# Music Classification: Beyond Supervised Learning, Towards Real-world Applications

Minz Won, Janne Spijkervet, Keunwoo Choi

# CONTENTS

This is a web book written for a tutorial session of the 22nd International Society for Music Information Retrieval Conference, Nov 8-12, 2021 in an online format. The ISMIR conference is the world's leading research forum on processing, searching, organising and accessing music-related data.

# The scope

The history of music classification dates back to at least 1996 [WBKW96]. The motivation of music classification remains the same since then.

> *The rapid increase in speed and capacity of computers and networks has allowed the inclusion of audio as a data type in many modern computer applications.*

It was further clarified in [TC02].

> *..gaining importance as a way to structure and organize the increasingly large numbers of music files available digitally...*

In this book, we focus on the more modern history of music classification since the popularization of deep learning in mid 2010s. Please refer to [FLTZ10] for the earlier progress in 2000s, which was mainly the design of audio features and adoption of classifiers as well as the birth of many music classification problems. [NCL+18] includes detailed discussion of the transition to deep learning approaches. There also exist other existing tutorials, [SLBock20] and [CFCS17b], that include more general MIR topics with a special focus on deep learning.

## Motivation

**Lower the barrier**: As deep learning emerges, music classification research has entered a new phase, and many data-driven approaches have been proposed to solve the problem. However, researchers sometimes use jargon in various ways. Also, some implementation details and evaluation methods are ambiguously described in the papers, blocking access to the information without personal contact. These are tremendous obstacles when new researchers want to dive into this fascinating research area. Through this book, we would like to lower the barrier for newcomers and reduce miscommunication between researchers by sharing the secrets.

**Cope with data issue**: Another issue that we are facing under the deep learning era is the exhaustion of labeled data. Labeling musical attributes requires strong domain knowledge and a significant amount of time for listening; hence expensive. Because of this, deep learning researchers started actively utilizing large-scale unlabeled data. This book introduces the recent advances in semi- and self-supervised learning that enables music classification models to step further beyond supervised learning.

**Narrow the gap**: Music classification has been applied to solve real-world problems successfully. However, some important procedures and considerations for real-world applications are rarely discussed as research topics. In this book, based on the various industry experiences of the authors, we try our best to raise awareness of these questions and provide answers and perspectives. We hope this helps academia and industries harmonize better together.

## About the authors

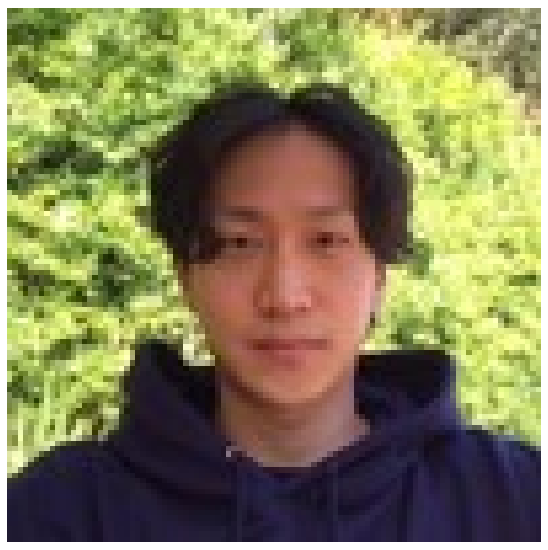

**Minz Won** is a Ph.D candidate at the Music Technology Group (MTG) of Universitat Pompeu Fabra in Barcelona, Spain. His research focus is music representation learning. Along with his academic career, he has put his knowledge into practice with industry internships at Kakao Corp., Naver Corp., Pandora, Adobe, and he recently joined ByteDance as a research scientist. He contributed to the winning entry in the WWW 2018 Challenge: Learning to Recognize Musical Genre.

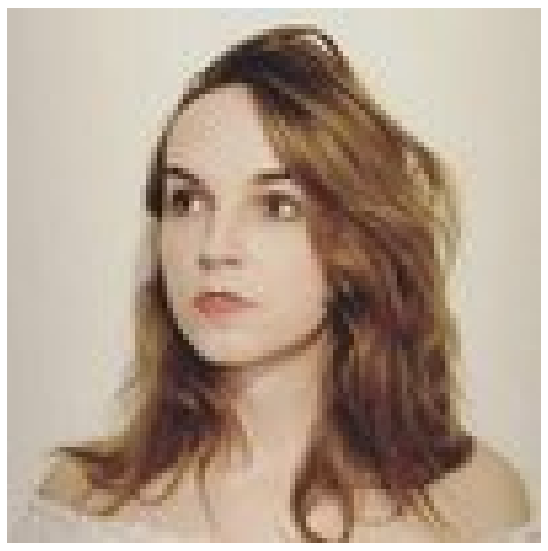

**Janne Spijkervet** graduated from the University of Amsterdam in 2021 with her Master's thesis titled "Contrastive Learning of Musical Representations". The paper with the same title was published in 2020 on self-supervised learning on raw audio in music tagging. She has started at ByteDance as a research scientist (2020 - present), developing generative models for music creation. She is also a songwriter and music producer, and explores the design and use of machine learning technology in her music.

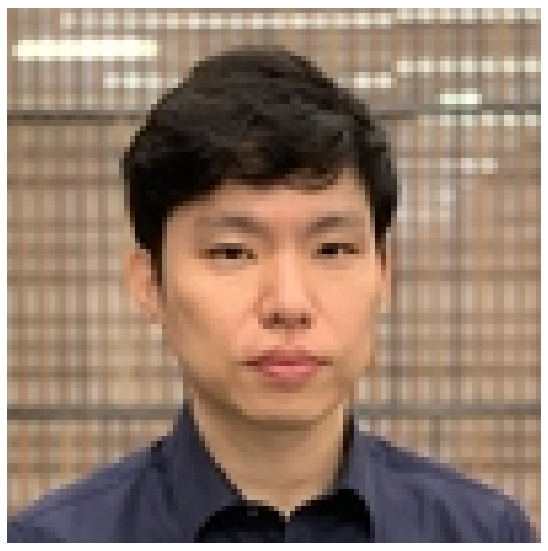

**Keunwoo Choi** is a senior research scientist at ByteDance, developing machine learning products for music recommendation and discovery. He received a Ph.D degree from Queen Mary University of London (c4dm) in 2018. As a researcher, he also has been working at Spotify (2018 - 2020) and several other music companies as well as open-source projects such as `Kapre`, `librosa`, and `torchaudio`. He also writes some music.

## Software

We use Jupyter Book[Com20], Librosa 0.8.1[MRL+15] [MMM+21], Pytorch[PGM+19], Torchaudio[YHN+21], Matplotlib[Hun07], and Numpy[HMvdW+20].

## Citing this book

```
@book{musicclassification:book,
    Author = {Won, Minz and Spijkervet, Janne and Choi, Keunwoo},
    Month = Nov.,
    Publisher = {https://music-classification.github.io/tutorial},
    Title = {Music Classification: Beyond Supervised Learning, Towards Real-world Applications},
    Year = 2021,
    Url = {https://music-classification.github.io/tutorial},
    doi = {10.5281/zenodo.5703779}
}
```

## Note

- You can download a pdf of this book from zenodo. If the pdf is not up-to-date, you can build it by yourself on your local machine.

# Part I

# The Basics

CHAPTER

ONE

# WHAT IS MUSIC CLASSIFICATION?

Music classification is a music information retrieval (MIR) task whose objective is the computational understanding of music semantics. For a given song, the classifier predicts relevant musical attributes. Based on the task definition, there are a nearly infinite number of classification tasks – from genres, moods, and instruments to broader concepts including music similarity and musical preferences. The retrieved information can be further utilized in many applications including music recommendation, curation, playlist generation, and semantic search.

## 1.1 Single-label classification

Let’s say there are two record stores in your town. ‘ABC Records’ curates all the records in alphabetic order, while ‘MIR Records□’ categorizes their stocks based on musical genres. When you already know what you want to buy, ‘ABC Records’ is a good place to go as you can search by the alphabetic index. However, when you want to browse and discover new music, ‘MIR Records’ will be preferable as you can visit the section with your favorite genre. Like this, well-designed categorization (i.e., music classification) helps customers browse music more efficiently. This record store scenario can be interpreted as a single-label classification task. One item can be in a single section; hence categories (genres in this example) are exclusive.

**Warning:** Genres are not always exclusive to each other. One song can belong to multiple genres.

## 1.2 Multi-label classification

Different from the example above, one item may belong to multiple categories. For example, one song can be Disco and K-Pop simultaneously, and these categories are not exclusive to each other. Also, listeners would like to browse music by instruments, languages, moods, or context, not only musical genres. We can handle these multiple musical attributes with multi-label classification. The multi-label classification is often referred to as “music tagging” since it puts various music tags for a given song.

Multi-label classification is handled as a binary classification for each musical attribute. For each label, the system determines whether a given song is positive to the label or not. In contrast with single-label classification, labels are not exclusive, and multiple tags can exist together.

## 1.3 Music classification tasks

There can be an almost infinite number of music classification tasks based on product requirements. Among them, the most explored music classification tasks in MIR research are listed as follow:

- Genre classification [TC02]
- Mood classification [KSM+10]
- Instrument identification [HBPD03]
- Music tagging [Lam08]

We discuss music classification tasks in more detail later in this chapter.

**Note:** Music tagging subsumes all other classification tasks as any class (label) can be musical tags.

## 1.4 Applications

The explosion of digital music has dramatically changed our music consumption behavior. Massive music libraries are available through streaming platforms, and it is impossible to browse the entire collections item-by-item. As a result, we need robust knowledge management systems more than ever. Music classification is a technique that supports knowledge management. Music classification models enhance users' music experience through many applications, including recommendation, curation, playlist generation, semantic search, and analysis of listening behavior.

- Recommendation: Once we have labeled or predicted musical attributes, a system can recommend music to users based on frequently consumed musical attributes of the users. Unlike collaborative filtering, a prevalent recommender system using user-item interactions, this content-based recommendation does not suffer from cold-start problems and popularity bias [Cel10].
- Curation: As we checked from the previous record store example, well-designed music curation helps users browse enormous music libraries efficiently. Hence, music streaming services curate music by genres, subgenres, or moods. Human agents can manually do this process, but music classification models can replace human efforts.
- Playlist generation: The usage of music classification models in playlist generation is similar to the use in music recommendation. But playlist generation needs to consider the order of the songs and more user context.
- Listening behavior analysis: Most modern streaming services provide annual reports of personal listening trends. This report helps users to understand their taste better and is basically fun!

CHAPTER

TWO

# INPUT REPRESENTATIONS

In this notebook, we'll study various audio input representations that are used for music classification. Choosing the right input representation is crucial to successful training of neural networks. This may sound against the spirit of deep learning – assume minimally and let the model learn.

This is because the optimal choice of audio input representations is more difficult than in other domains as you'll see here. In other words, this is an important design choices that music/audio researchers should make as opposed to people in natural language processing or computer vision.

---

**Note:** In this section, you will be introduced to various input representations as well as their relationship with our perception of sound. The connection is rarely mentioned, but it provides rigorous explainations about why we choose some design choices.

---

## 2.1 Biological Plausibility

Neural networks are inspired by biological neural networks. However, that doesn't mean we have to follow every detail of them. It is a classic debate topic where airplanes are often mentioned as a counterexample.

How about audio representations? Does biological plausibility matter when designing it?

The answer would depend on the problem we solve as well as the empirical evidence. For music classifcation, the answer seems to be "Yes". The way we perceive sounds decides what we care about and how we label music. Our understanding of music, then, defines music classification tasks. For example, no one cares about pattern recognition of inaudible frequency ranges.

In this section, I will help the readers to connect many choices we make regarding input representations to related concepts in psychoacoustics, a study of our perception of sound.

---

Alright, let's get started! Let me prepare some modules and variables first.

```
import numpy as np
import matplotlib.pyplot as plt
import librosa
import librosa.display
import IPython.display as ipd

plt.rcParams.update({'font.size': 16, 'axes.grid': True})

SR = 22050  # sample rate of audio
```

(continues on next page)

(continued from previous page)

```
wide = (18, 3)   # figure size
big = (18, 8)   # figure size

print(f"{librosa.__version__=}")
```

```
librosa.__version__='0.8.1'
```

```
src, sr = librosa.load('are-you-here-with-me(mono).mp3', sr=SR, mono=True, duration=5.
↪0)

print(f'{src.shape=}, {sr=}')
```

```
/Users/admin/miniconda3/lib/python3.8/site-packages/librosa/core/audio.py:165:␣
↪UserWarning: PySoundFile failed. Trying audioread instead.
  warnings.warn("PySoundFile failed. Trying audioread instead.")
```

```
src.shape=(110250,), sr=22050
```

## 2.2 Waveforms

The first representation we'll discuss is waveforms.

**Note:** Waveforms are records of amplitudes of audio signals.

```
plt.figure(figsize=wide)  # plot using matplotlib
plt.title('Waveform of the example signal')
plt.plot(src);plt.ylim([-1, 1]);
```

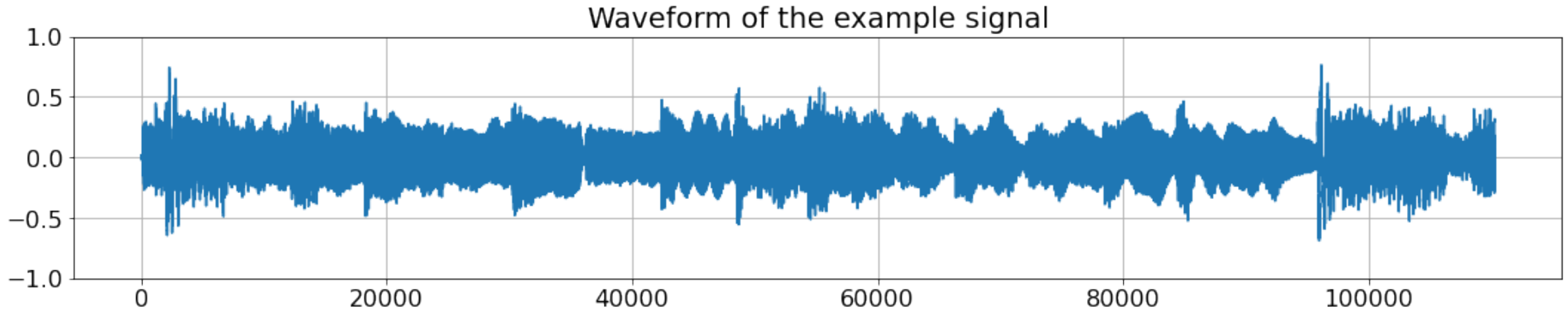

```
ipd.Audio(src, rate=sr) # load a NumPy array
```

```
<IPython.lib.display.Audio object>
```

Effective visualization of audio representations is trickier than you think. FYI, you can ask `librosa` to take care of it as below. We'll use both `matplotlib` directly and `librosa` depending on what I want to display.

```
plt.figure(figsize=wide)  # plot using librosa
plt.title('Waveform of the example signal')
librosa.display.waveshow(src, sr=22050);plt.ylim([-1, 1]);
```

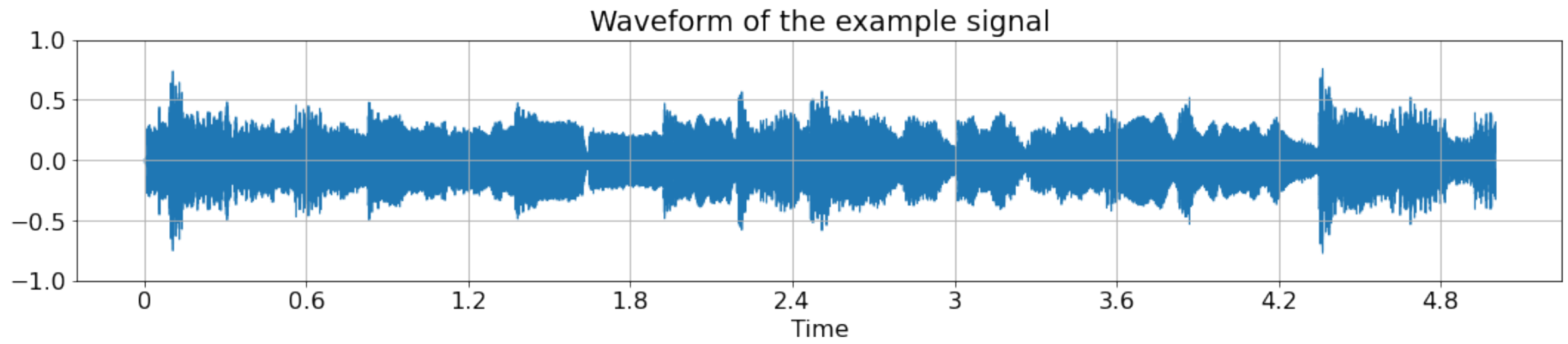

This 5-second 22,050-Hz sampled mono audio has a shape of `(110250, )`. So, what is this long 1-dimensional array? It is a representation for (diaphragms of) speakers, whose goal is to produce the sound (the change of pressure in the air) correctly.

□ Let's zoom into the waveform

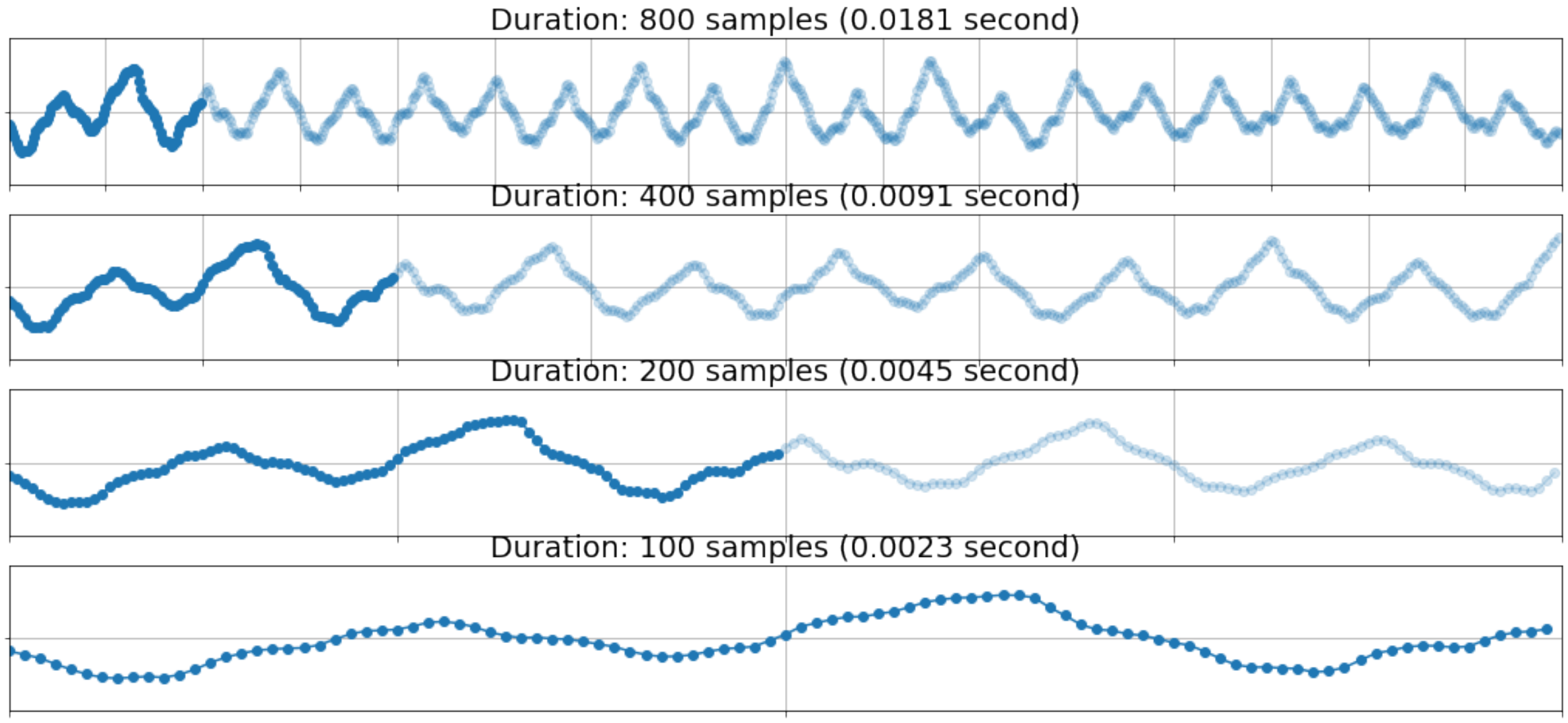

As the raw end of audio signals, waveforms were used as input of music classification models in [DS14], [LPKN17], etc. Is waveform the best representation for neural networks? It depends, but usually it's not.

- If you have a lot of data, it's worth trying with the minimally assuming, waveform-based models such as Sample-CNN ([LPKN17]). Beware though, it's requires a large memory, lots of computation, and large-scale data.

What are the alternatives then? There is no single answer to the question. A bunch of types of **spectrograms** could be the answer depending on what you're looking for.

But - first of all, what are spectrograms?

## 2.3 Spectrograms: time-frequency representations

**Note:** Spectrogram refers to a (2D) visualization of sound.

```
Text(0.5, 1.0, 'A Spectrogram')
```

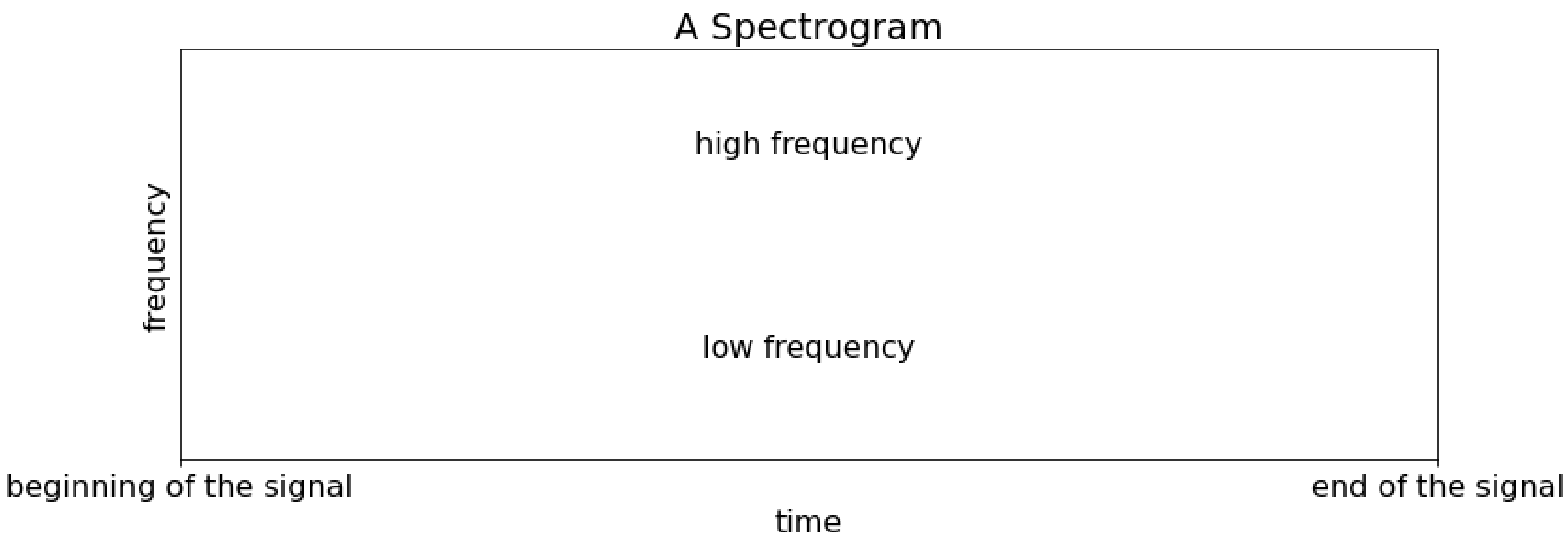

Why is spectrogram a good representation?

Let's think about how we perceive sound. The amplitudes of acoustic waves – waveforms – are what our eardrums respond to. The sound travels through our auditory system. At some point, it is then converted to a (sort of) 2-dimensional representation in the cochlear to perform some frequency analysis. This is done by the basilar membrane physically responding to a certain frequency component as below.

(Image from Wikipedia)

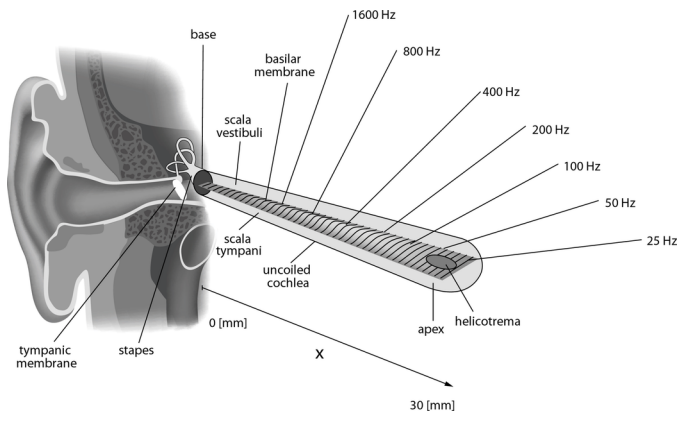

This means that 2-dimensioal representations are biologically plausible, which is not bad.

Additionally, spectrograms are neural network-friendly because they fit well with properties of some popular architectures. For example, local correlation and the shift invariance of CNNs can be utilized nicely when we're using spectrograms. One would argue that the harmonic relationship along the frequency axis is hardly considered in CNNs. But i) we can modify the structure to take it into account [WCNCS19], and its performance was on par with typical (non-harmonic) CNNs [WFBS20].

For the rest of this section, we'll focus on three types of spectrogram. They are STFT, melspectrograms, or constant-Q transform (CQT). Let me first summarize their property from a deep learning point of view.

---

**Note:**

- **STFT** has some good properties but its size is usually bigger than Melspectrograms or CQT. That means more computation and memory usage, so it is less desirable.
- **Melspectrogram** has been very popular for practical reasons. The performance is strong, its memory usage is small, and the computation is simple. This is probably everyone's go-to choice if you're not sure.
- **Constant-Q Transform (CQT)** is quite similar to Melspectrogram for ML models. However, it is more computation heavy and is less available in softwares we use.

---

## 2.4 STFT

STFT (short-time Fourier transform) is the most "raw" kind of spectrograms. It has two axes - time and frequency.

- It has a linear frequency resolution. Its frequency axis spans from 0 Hz (DC component) to `sample_rate / 2` Hz (aka Nyquist frequency).
- We can fully reconstruct the audio signal from a STFT.
- STFT consists of complex numbers.

```
n_fft = 512  # STFT parameter. Higher n_fft, higher frequency resolution you get.
hop_length = n_fft // 4  # STFT parameter. Smaller hop_length, higher time resolution.
stft_complex = librosa.stft(y=src, n_fft=n_fft, hop_length=hop_length)

print(f"{src.shape=}\n{stft_complex.dtype=}\n{stft_complex.shape=}\n{stft_complex[3,␣
↪3]=}\n")
```

```
src.shape=(110250,)
stft_complex.dtype=dtype('complex64')
stft_complex.shape=(257, 862)
stft_complex[3, 3]=(-0.9134862+0.22103079j)
```

The shape of `(257, 862)` means there are 257 frequency bands and 862 frames.

- 257 = (n_fft / 2) + 1. Originally, there are `n_fft` number of frequency bins. But, they are mirror image for real signals (such as audio signals) so we can discard the half. These bins include the boundaries, hence there is one more bin.
- 862 = ceil(signal_length / hop_length) = ceil(110250 / 128) = 862. There could be a few more frames depending on how you handel the boundaries.

But, we rarely use `stft_complex` as it is.

- **Modification 1**: For analysis purposes, we usually use the magnitudes of STFT only. This is not only convenient but also biologically plausible since the human auditory system is insensitive to phase information. (Nevertheless, this doesn't mean it is always better to discard the phase information.)

In MIR literatures, STFT usually refers to the magnitude of STFT while the original, complex-numbered STFT is referred as "complex STFT".

**Note:** STFT has a linear frequency resolution and we often use magnitude of it.

� Let's see how a magnitude-STFT looks like.

```
stft = np.abs(stft_complex)
print(f"{stft.dtype=}\n{stft.shape=}\n{stft[3, 3]=}\n")

plt.figure(figsize=wide)
img = plt.imshow(stft)
plt.colorbar(img)
plt.ylabel('bin index\n(lower index -> low freq)');plt.xlabel('time index')
plt.title('Manitude spectrogram (abs(STFT))');plt.grid(False);
```

```
stft.dtype=dtype('float32')
stft.shape=(257, 862)
stft[3, 3]=0.9398466
```

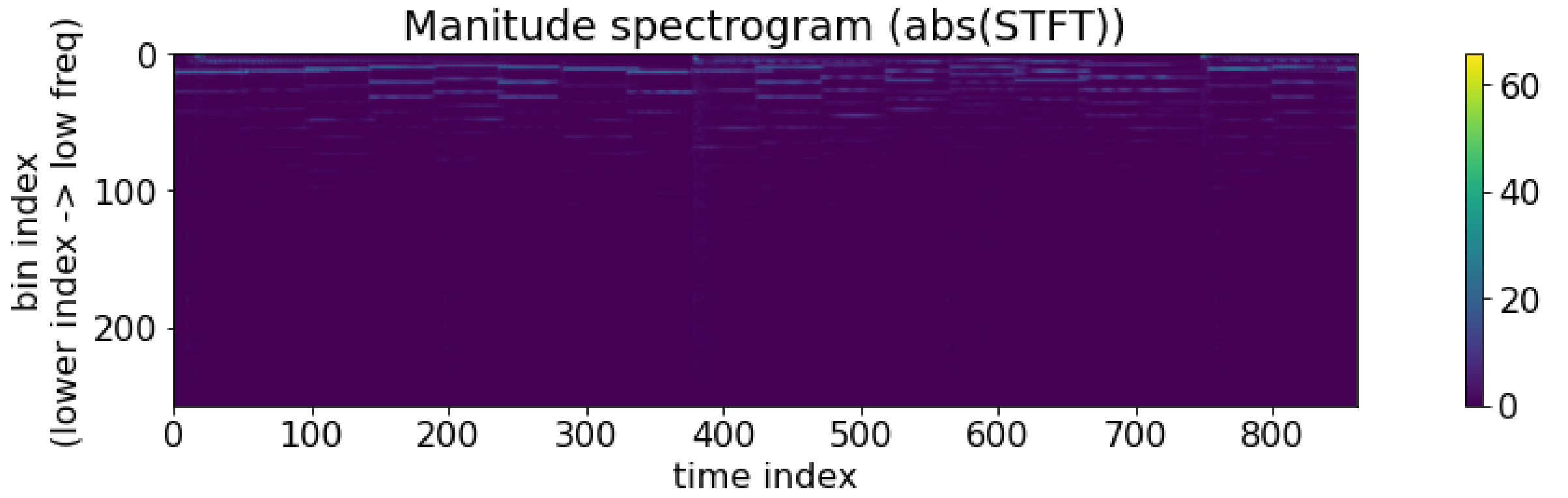

Can you see the slight activations on the lower frequency (near the upper boundary)? That's the magnitudes of the STFT of our example signal. But this image looks pretty sparse.

Let's time-average the frequency distribution and plot it.

```
plt.figure(figsize=wide); plt.subplot(1, 2, 1)
stft_freq_distritubion = np.mean(stft, axis=1)  # axis=1 is the time axis.
plt.plot(stft_freq_distritubion)
plt.xlabel('bin index\n(lower index -> low freq)')
plt.title('Frequency magnitude (linear scale)');
```

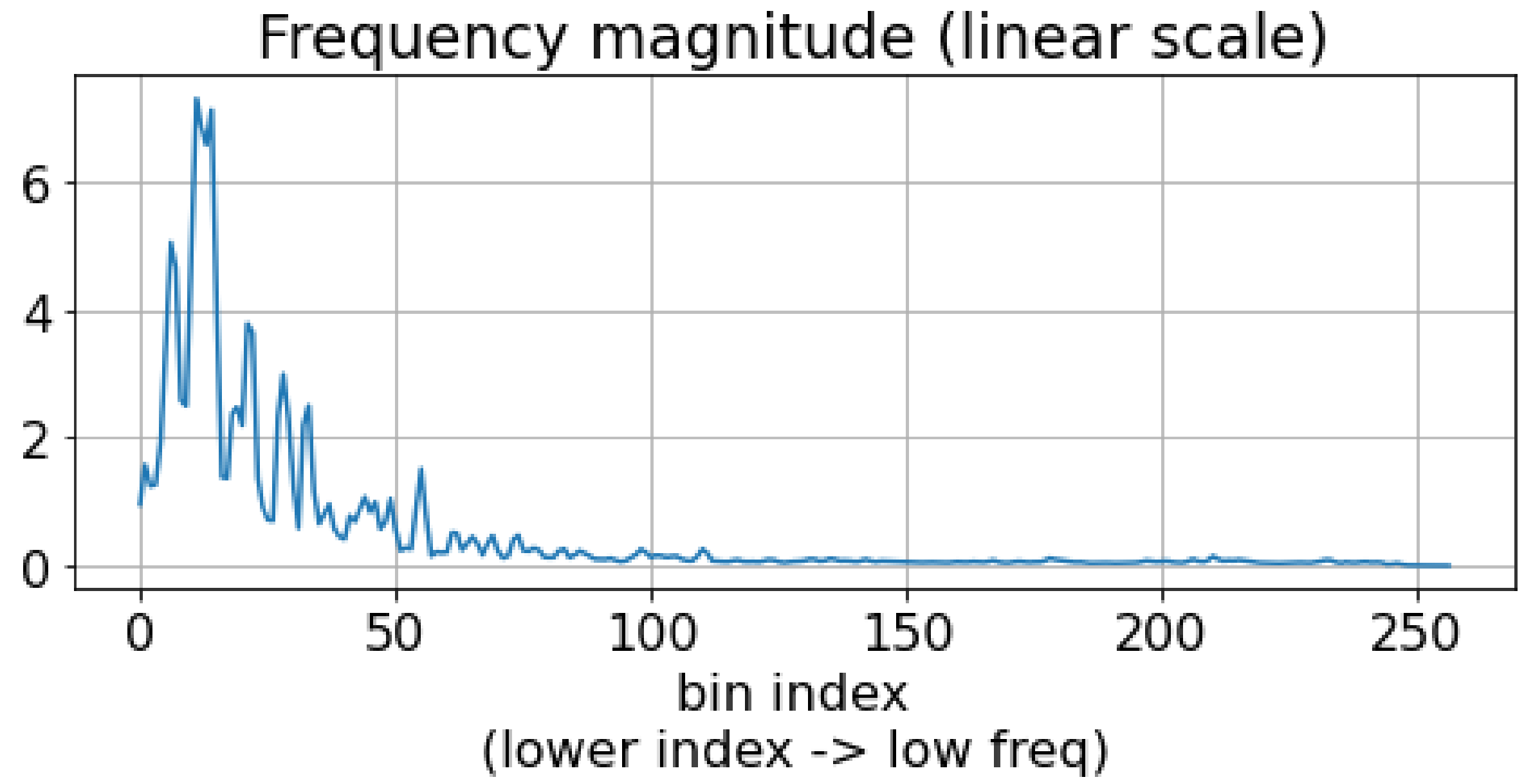

Seems like the magnitude is much larger in the low frequency region. This is very common due to our nonliear perception of loudness per frequency, and this is why the spectrogram above didn't seem very clear to us. And even worse, this kind of an extreme distribution is not good for neural networks. You can find more discussion on this in [CFCS17a].

The solution is to take `log()` to magnitude spectrograms.

- **Modification 2** After `abs()`, we compress the magnitude with `log()`. This is also biologically plausible - the

human perception of loudness is much closer to a logarithmic scale than a linear scale (i.e., it follows Weber–Fechner law).

Both of the modifications 1 and 2 are so common that people often omit them in papers.

**Note:** By "STFT" in deep learning-related articles, people often mean log-magnitude STFT.

Remember that these modification have nothing to do with the total shape – they are element-wise operations.

```
eps = 0.001
log_stft = np.log(np.abs(stft_complex) + eps)
print(f"{log_stft.dtype=}\n{log_stft.shape=}")

plt.figure(figsize=wide)
img = plt.imshow(log_stft)
plt.colorbar(img)
plt.ylabel('bin index\n(lower index -> low freq)');plt.xlabel('time index')
plt.title('Log-manitude spectrogram log(abs(STFT))');plt.grid(False);

ipd.Audio(src, rate=sr) # load a NumPy array
```

```
log_stft.dtype=dtype('float32')
log_stft.shape=(257, 862)
```

```
<IPython.lib.display.Audio object>
```

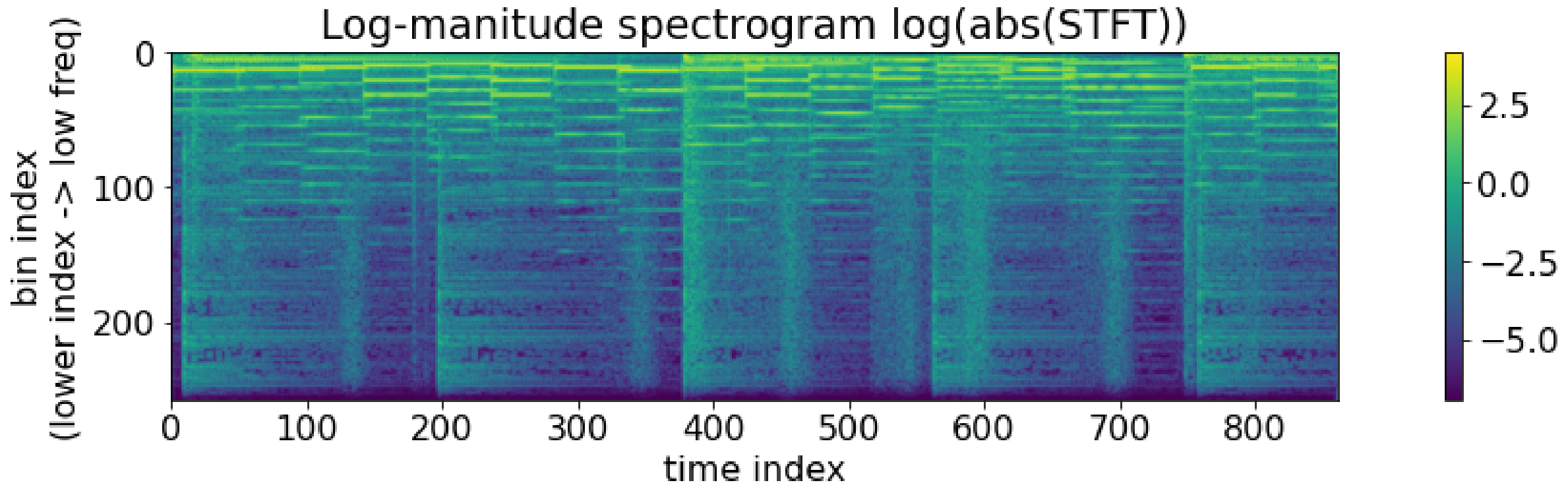

This is exactly what's happening in Decibel scaling. Decibel scaling is also logarithm mapping but with a few different choices of the constants (e.g., log10 vs log, etc) so that 0 dB becomes absolute silence and 130 dB becomes a really really loud sound. Check out the implementations in `librosa.amplitude_to_db()` and `librosa.power_to_db()` for more correct and numerically stable and decibel scaling.

Finally, it doesn't look so right when low-frequency is at the top of the image. Let's flip up-down to correct it.

```
eps = 0.001
log_stft = np.log(np.abs(stft_complex) + eps)
log_stft = np.flipud(log_stft)  # <-- Here! The rest of the code is hidden.
```

```
log_stft.dtype=dtype('float32')
log_stft.shape=(257, 862)
```

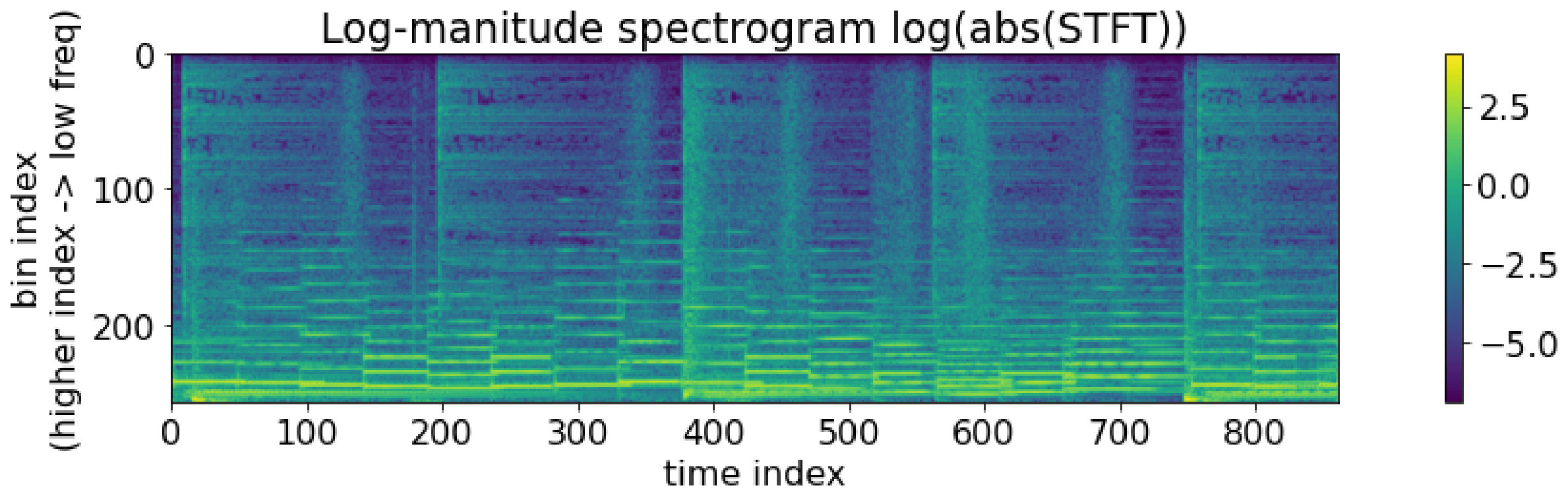

OK! This is the famous log-magnitude STFT. Quite often, people call it simply a STFT or a log-STFT.

**Note:** People usually use log-magnitude STFT.

From here, I'll use `libros` more actively because i) the implementation of other representations is not trivial and ii) so that the x- and y-axes are nicely displayed in more convenient units.

Additionally, it's always safe to make your data zero-centered. That's also done quite nicely with the default parameters of `librosa`.

```
plt.figure(figsize=wide)
img = librosa.display.specshow(librosa.amplitude_to_db(stft), sr=SR, x_axis='s', y_
↪axis='linear', hop_length=hop_length)
plt.colorbar(img, format="%+2.f dB")
plt.title('decibel scaled STFT, i.e., log(abs(stft))');
```

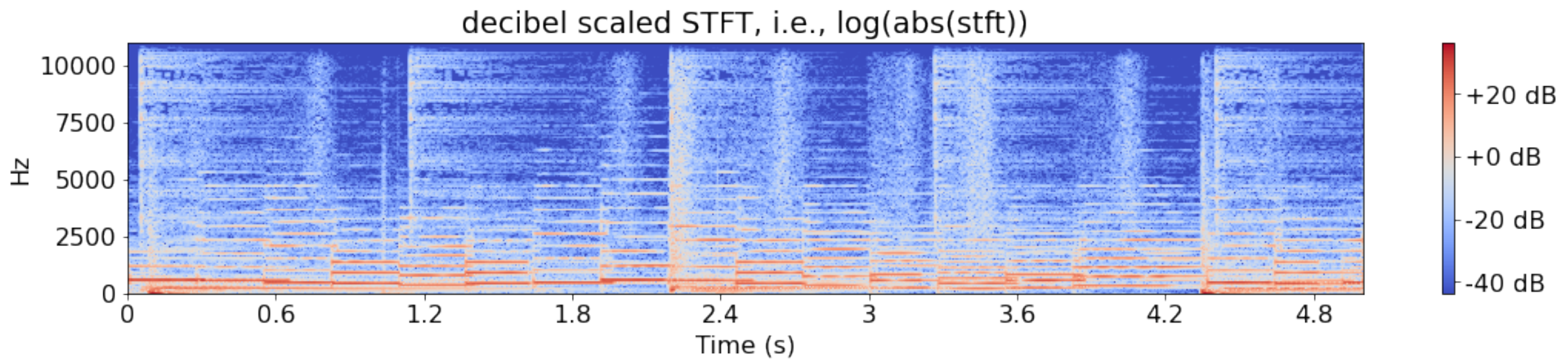

## 2.5 Even more modifications

By computing log(abs(STFT)), we get a nice image of the sound. But, it is just a beginning!

**Reasons for modifying the frequency scale**

- Our perception of frequency is also nonlinear (approximately.. (drum rolls!..) logarithmic).
- Similarly, we defined pitches in the octave (=logarithmic) scale

**Reasons why it's fine to remove some high frequency bands**

- The highest frequency of original audio signals is usually 22kHz which is pretty far beyond our range.
- Similarly, the information in high-frequency ranges (e.g., f > 10kHz) is i) sparse, ii) not that necessary for most of MIR tasks, and iii) barely audible for us.

**Reason why, it's *better* to remove some high frequency bands**

- Because we want to remove any redundant memory and computation especially in deep leanring.

Because of these, researchers have been using time-frequency representations that are even more modified than log(abs(STFT)).

## 2.6 Melspectrograms

Melspectrograms have been the top choices for music tagging and classification. But what is a melspectrogram?

Melspectrogram is a result of converting the linear frequency scale into the *mel scale*. Mel scale is invented to mimic the our perception of pitch. The popular implementation of these days assumes a linearity under 1 kHz and a logarithmic curve above 1 kHz.

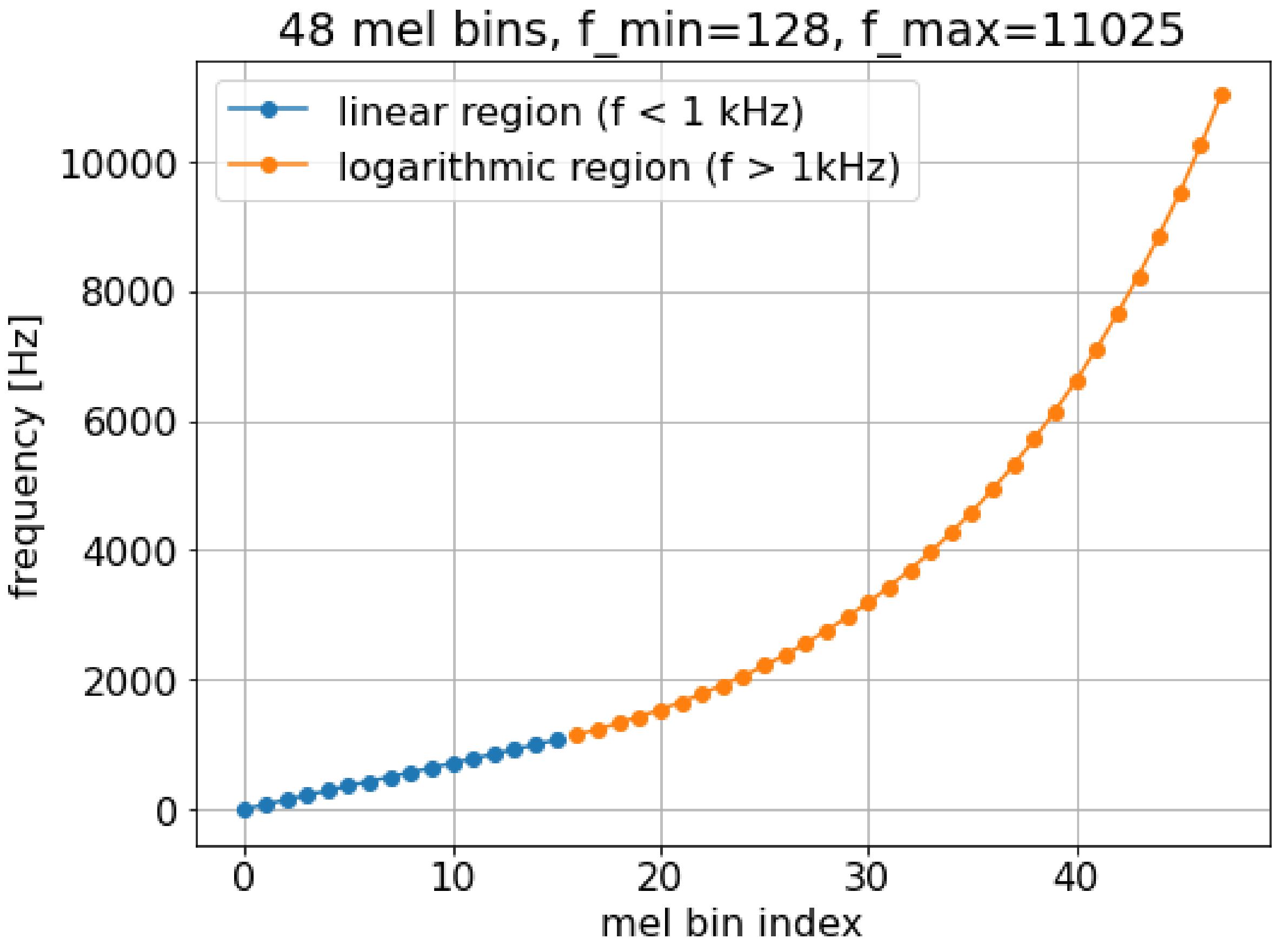

**Note:** Melspectrogram is based on a mel-scale, which is nonlinear and approximates human perception.

**Benefits**

- It's reduces the number of frequency band greatly. For example, 1028 -> 128. What a deep learning-plausible number it becomes!
- It's simple and fast - the computation is a matrix multiplication of a pre-computed filterbank matrix.
- It's effective - the model works even better in many cases.

```
log_melgram = librosa.power_to_db(
    np.abs(
        librosa.feature.melspectrogram(src, sr=SR, n_fft=n_fft, hop_length=hop_length,
 ↪ power=2.0,
                                      n_mels=128)
    )
)
print(log_melgram.shape, stft.shape)  # 257 frequency bins became 128 mel bins.
```

```
(128, 862) (257, 862)
```

We can directly compare STFT and melspectrogram.

```
plt.figure(figsize=(wide))
plt.subplot(1, 2, 1)
img = plt.imshow(np.flipud(librosa.amplitude_to_db(stft)))
plt.colorbar(img, format="%+2.f dB")
plt.title('log(abs(stft))'); plt.grid(False);
plt.yticks([0, n_fft//2], [str(SR // 2), '0']); plt.ylabel('[Hz]'); plt.xlabel('time␣
 ↪[index]')

plt.subplot(1, 2, 2)
img = plt.imshow(np.flipud(log_melgram))
plt.colorbar(img, format="%+2.f dB")
plt.title('log(melspectrogram)');  plt.grid(False);plt.yticks([]);
plt.yticks([0, 128], [str(SR // 2), '0']); plt.ylabel('[Hz]'); plt.xlabel('time␣
 ↪[index]');
```

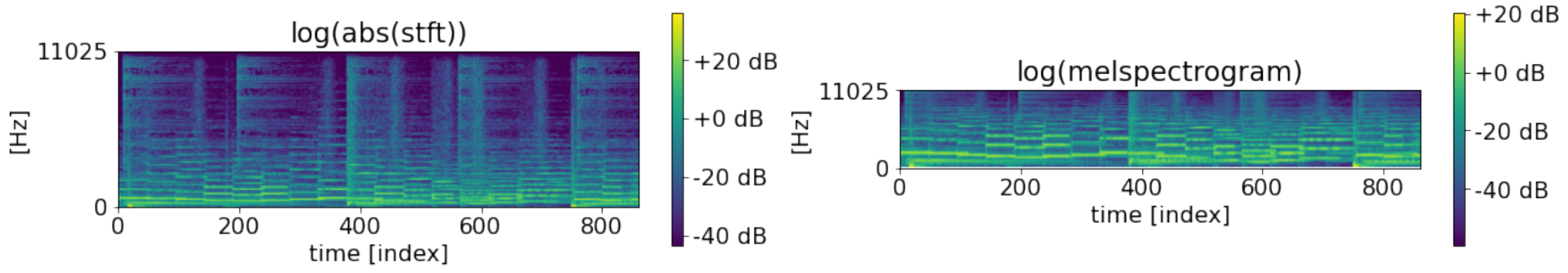

A few observation:

- Melspectrogram is smaller than STFT.
- Even if it's smaller, the low frequency region has allocated more bins than it does in STFT.
- The frequency range is the same.

## 2.7 Constant-Q Transform

As we've seen, melspectrograms are great! But it is just a simple aggregation of high-frequency bins into one. This means the frequency resolution of melspectrogram is bound by that of STFT. And there is always a trade-off between time and frequency resolutions in STFT.

Constant-Q Transform is more radical. Why not having accurately octave scale representation?

Its implementation is not trivial, but the idea is to use time-varying windows for different center frequency. Let's see the result.

```
log_cqt = librosa.amplitude_to_db(
    np.abs(
        librosa.cqt(src, sr=SR, hop_length=hop_length, n_bins=24*7, bins_per_
 ↪octave=24, fmin=librosa.note_to_hz('C1'))
    )
)
```

```
plt.figure(figsize=(18, 8))
plt.subplot(1, 2, 1)
img = librosa.display.specshow(log_melgram, y_axis='mel', sr=SR, hop_length=hop_
 ↪length)
plt.colorbar(img, format="%+2.f dB")
plt.title('log(melspectrogram)')

plt.subplot(1, 2, 2)
img = librosa.display.specshow(log_cqt, y_axis='cqt_hz', sr=SR, hop_length=hop_length,
 ↪ bins_per_octave=24)
plt.colorbar(img, format="%+2.f dB")
plt.title('log(cqt)')
```

```
Text(0.5, 1.0, 'log(cqt)')
```

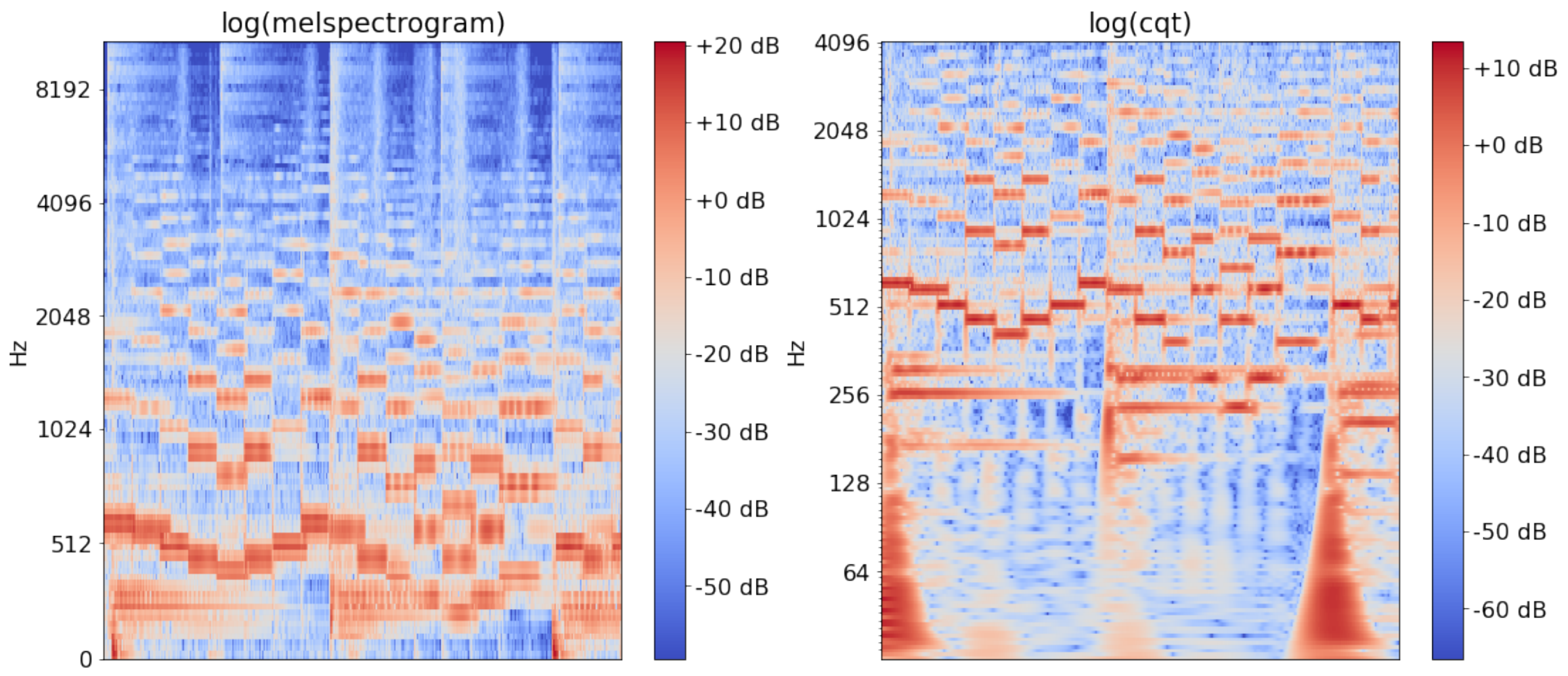

The difference is pretty obvious here - we get a much better pitch resolution!

## 2.8 Practical Issue: How to compute them?

There are several softwares that computes these representations. Take this information with a grain of salt because it will be outdated after releasing this book (2021 Nov).

### 2.8.1 Waveforms

First of all, you get waveforms by simply loading the audio.

- You'd need an audio codec (e.g., FFMPEG) if your audio comes in formats such as mp3, m4a, aac, or ogg.
  - Usually, your audio file loader returns a floating-point data array where the amplitude is in `[-1.0, 1.0]`. But, the original `wav` file usually stores the amplitudes in `int16` format. (No worries though, `Float32` is precise enough to represent them.)
- You can use `scipy.io.wavfile.read` to load PCM audio.

### 2.8.2 Spectrograms

There are various softwares and approaches to compute STFT, melspectrogram, and CQT.

- On CPU
  - **librosa**: librosa.stft internally uses `stft` but with a more tailored and carefully chosen API with default values. It also includes functions to compute a melspectrogram and a CQT [MRL+15].
  - **scipy**: scipy.signal.stft is a cpu-based implementation of STFT using FFT [VGO+20].
  - **Essentia** provides C++-based implementation of STFT and melspectrogram as well as their python bindings [BWGomezGutierrez+13].
- On CPU & GPU
  - **Torchaudio**: One can compute STFT using `torchaudio.functional.spectrogram`. It also includes a wide variety of functions and utilities such as `amplitude_to_DB` and `resampling` [YHN+21].
  - **nnAudio**: It has multiple versions of CQT computation functions as well as others e.g., STFT and melspectrogram. Its STFT computation is based on Conv1D, not FFT [CAAH20].
  - **Tensorflow**: It has a native support of stft. However, it is slow on cpu, which might be critical if your model is going to be deployed on cpu-only machines. It, by the way, even has a function to convert an STFT to melspectrogram [ABC+16].
  - **kapre**: kapre.time_frequency includes `tf.keras` layers such as STFT and melspectrogram as well as utilities as decibel conversion. Its STFT layer is a wrapper of `tf.signal.stft` [CJK17].

### 2.8.3 Consistency between softwares

Not all the implementations are equal! Especially.. nothing.

- Default behavior such as padding may be different
- There is no single canonical reference implementation of some concepts such as mel scale.
- CQT is an approximation yet. There is no method to compute CQT with perfect reconstruction.

One should make sure that all the data are processed consistently. The easiest way is to have a single method that process all the data during training and after deployment. This may be tricky, but possible for most of the cases. See the comparisons and suggestions linked below for more information.

- Librosa and Scipy
- Librosa and Tensorflow
- Librosa and Torchaudio

CHAPTER

THREE

# DATASETS

## 3.1 Overview

There already exists a great comprehensive list of MIR datasets. In this book, we focus on some important datasets and *discussion* of them including some secrets that worth spreading.Over time, researchers have adopted different strategies to create and release datasets. This resulted in various pros and cons, and traps.

These are some important but rarely discussed aspects.

### 3.1.1 Availabilities of audio signal

- This basic and fundamental requirement is already difficult. This is because, well, music is usually copyright-protected. The solutions are i) just-do-it, ii) distribute the features, iii) use copyright-free music, iv) distribute the IDs.

### 3.1.2 Hidden traps!

..Because some of the dataset creation procedure was not perfect.

- How shall we split them
  - Many datasets don't have a official dataset split, and this caused many problems. Usually, wrong split gets us an incorrectly optimistic result, which incentives us to overlook the problem.
- How noisy the labels are?
  - No annotation is perfect, but on a varying level. Why? How?
  - Regarding the inherent noisiness of the label (subjectivity, fuzzy definition, etc), what is the practical/meaningful best performance?
- How realistic the audio signals are?
  - We want our research (and the resulting models) to be practical. A lot of this depends on how similar the dataset is to the real, target data.

---

Now, let's look into actual datasets.

## 3.2 Gtzan Music Genre (2002)

**Note:**

- Audio is directly available
- 100 items x 10 genres x 30-second mp3 files.
- Single-label genre classification

The famous GTZAN dataset [TC01] deserves to be the MNIST for music. The first paper using this dataset [TC02] remains a foundational work in the modern music classification. The dataset was used in more than 100 papers already in 2013 according to a survey ([Stu13]). It is popular since the concept of music genres and single-label classification is easy, simple, and straightforward. 30-second mp3 is small and short. With a lot of features and a power classifier, researchers these days can quickly achieve 90+% (or even 95+% (or even 100%!)) accuracy.

However, now we know that there are way too many issues with the dataset. This is summarized very well in the aforementioned survey by Bob L. Sturm's survey [Stu13]. We'll list a few.

**Warning:**

- The audio quality varies by samples (though it was intended) and it is not annotated
- There are heavy artist repetition, which are very often ignored during dataset split
- The labels don't seem to be 100% correct (which makes the 100%-accuracy models questionable)

Because of these known issues, GTZAN doesn't seem to be as popular as it used to be in published research. Still, one may find it a simple benchmark dataset. In that case, please refer to this repo made by Corey Kereliuk and Bob Sturm and use a fault-filtered split.

**Tip:**

- Use other, bigger and better datasets
- Use a cleaned version and split

A tremendous number of following researchers owe its creator, George Tzanetakis, for the dataset release. Here's a quote from the website, where you can simply one-click-download the dataset.

> ..Unfortunately the database was collected gradually and very early on in my research so I have no titles (and obviously no copyright permission etc)..

This is not a viable option these days anymore. Let's see more modern approaches.

## 3.3 MagnaTagATune (2009)

**Note:**

- Designed for tagging problem
- Audio is directly available. They're from magnatune.com, a marketplace of indie music. John Buckman, the founder of magnatune contributed these files.
- 5,405 tracks (25,863 x 29-second clips), 230 artists, 446 albums, 188 tags.

MagnaTagATune [LWM+09] has played a significant role since its release until even now (2021). It was used in pioneering research such as [DS14], [CFS16], [LPKN17], [WCS21], [SB21], etc.

"Tagging" is a specific kind of classification, and MagnaTagATune is one of the earliest tagging datasets that is in this scale and that comes with audio. The songs are all indie music, so use this dataset at your own risk - the property of the music/audio might not be as realistic as you want.

The gamification of the annotation process is worth mentioning. In this game called "Tag a Tune", two players were asked to tag a clip, then shown the other player's tagging results to finally judge if they were listening to the same clip or not. This constraint-free annotation process has pros and cons; it is realistic, which is good; but it makes the label noisy, which is bad as a benchmark dataset.

There are various approaches how to split and whether to include below top-50 tags during training and/or testing. This hidden difference makes the comparison silently noisy. Finally, The authors of [WFBS20] decided to include items with top-50 tags only, both in training and testing. They then trained various types of models and shared the result in the paper. We recommend follow-up researchers to use the same split for a correct comparison.

**Warning:**

- Tags are weakly labeled and have synonyms
- DO NOT RANDOM SPLIT 25,863 clips! They're from the same track!
- Researchers used slightly different splits.
- The score on this dataset is still improving, but only slightly. It means we might be near the glass ceiling.

This dataset turned out to be big enough to train some early deep neural network models such as 1D and 2D CNNs. Until late 2010s, MagnaTagATune was probably the most popular dataset in music tagging.

**Tip:**

- Follow the split and refer to the numbers in [WFBS20].
- If you split by yourself, do it by tracks (instead of clips) |
- Know you're dealing with an indie music|
- Know you're dealing with a weakly labeled dataset|

## 3.4 Million Song Dataset (2011)

**Note:**

- Audio is not directly available.
  - As of 2021, only a crawled version that contains ~99% of the preview clips is available by word of mouth.
- Literally a million tracks: By far the biggest dataset
- The provided last.fm tags are realistic

The million song dataset (MSD, [BMEWL11]) is a monumental music dataset. It was ahead of time in every aspect – size, quality, reliability, and various complementary features.

MSD has been *the* music dataset since the beginning of deep learning era. It enabled the first deep learning-based music recommendation system [VdODS13] and the first large-scale music tagging [CFS16].

Researchers usually formulate the music tagging on MSD as a top-50 prediction task. This may be partially due to the convention of MagnaTagATune and earlier research, but it makes sense considering the sparsity of the tags. The tags in MSD are in an extremely long tail.

> in the MSD, .. there are 522,366 tags. This is outnumbering the 505,216 unique tracks..
>
> .. the most popular tag is 'rock' which is associated with 101,071 tracks. However, 'jazz', the 12th most popular tag is used for only 30,152 tracks and 'classical', the 71st popular tag is used 11,913 times only. ..

(from [CFCS18])

**Warning:**

- Some splits have artist leakage
- It might be difficult to get the mp3s

The dataset split used in [CFS16] was based on simple random sampling. However, this resulted in potentially allowing information between splits as same artists appear in different split.To avoid this issue, the authors of [WCS21] introduced CALS split - a cleaned and artist-level stratified split. This includes [TRAIN, VALID, TEST, STUDENT, NONE] sets where STUDENT set is a set of unlabeled items with respect to top-50 tags and can be used for semi-supervised and unsupervised learning. (NONE is a subset of discarded items since their artists appeared in TRAIN.)

One critical downside of MSD is the availability of the audio. The creators of MSD adopted a very modern approach on this - while only distributing audio features and metadata, they released a code snippet for fetching 30-second audio previews from 7digital. (Recently, people have reported the audio preview API does not work anymore. This means the audio is available only by word of mouth.)

**Tip:**

- ? Ask around for the audio!
- Use the recent split
- No music after 2011

## 3.5 FMA (2017)

**Note:**

- Rigorously processed metadata and split. Maintained nicely on Github.
- More than 100k full tracks of copyright-free indie music
- Artist-chosen genres in a hierarchy defind by the website (free music archive)

Free Music Archive dataset (FMA, [DBVB16]) is a modern, large-scale dataset that contains full-tracks, instead of short preview clips. Along with MTG-Jamendo, it enables interesting research towards fully utilizing the information of the whole audio signal.

**Warning:**

- Audio quality varies, and the music is quite “indie”.
- Genre labels are i) from a pre-defined 163-genre hierarchy and ii) chosen by the artist.

From a machine learning point of view, the second item in Warning is an advantage. However, it limits the development of realistic models.

**Tip:**

- Good for genre classification/hierarchical classification.
- A full-track is available, which is rare in the community

## 3.6 MTG-Jamendo (2019)

**Note:**

- 55,000 full audio tracks (320kbps MP3)
- 195 tags from genre, instrument, and mood/theme
- Pre-defined split based on the target tasks (genre, instrument, mood/theme, top-50, overall.)

MTG-Jamendo [BWT+19] is a modern dataset that shares some pros of MSD and FMA. Its audio is readily and legally available, the audio is full-track and high-quality, contains various and realistic tags, and comes with properly defined splits.

There are some interesting properties of this dataset, too. Pop and rock is the top genres in most of the genre datasets, and that could be the same for your target test set. In MTG-Jamendo, the genre distribution is skewed towards some other genres: The most popular genres are Electronic (16,480 items), soundtrack (~8k), pop, ambient, and rock. For mood alone, MTG-Jamendo is still great but there are alternatives (more information is under Resources section).

**Warning:**

- Genre distribution is slightly unusual

The distribution mismatch between training, validation, and testing sets is a classic yet critical problem. This wouldn't be a problem if all the testing and evaluation is within the provided split of MTG-Jamendo. Otherwise, one would want to have a different sampling strategy to alleviate the issue. (To be fair, this is not only applicable for MTG-Jamendo.)

## 3.7 AudioSet (2017)

- Preview of AudioSet.

**Note:**

- Large scale (2.1 million in total), 1 million under music
- Fairly strongly labeled in terms of temporal resolution (labeled for 10-second segment)
- High-quality annotation
- Official and reliable split is provided

AudioSet [GEF+17] is made for general audio understanding and not specifically for music. But, in their well-designed taxonomy, there is a high-level category 'music' that includes 'musical instrument', 'music genre', 'musical concepts', 'music role', and 'music mood'. In total, there are more than 1M items, each of which corresponds to a specific 10-second of YouTube video.

The annotation is considered to be more than quite reliable. Also, for each category, AudioSet provides the estimated accuracy of the annotation.

**Warning:**

- It includes music with a low audio quality
- Only the video URLs are provided
- The exact version would vary by people

The varying audio quality might be a downside depending on the target application. The dataset includes a live session, a noisy and amateur recording, music with a low SNR, etc.

To use AudioSet, one has to crawl the audio signal by themselves. Downloading YouTube video/audio is in a grey zone in terms of copyright, let alone the use of them.

Another issue is that the availabilities of the items in AudioSet are time-varying and country-dependent! Once the videos are taken down, that's it. Depending on the setting, some videos are just not available in some countries. Given the large size, this issue might not be critical in practice – so far.

## 3.8 NSynth (2017)

**Note:**

- '305,979 musical notes, each with a unique pitch, timbre, and envelope' as well as five different velocities
- 16 kHz, 4-second, monophonic.

NSynth [ERR+17] is 'a dataset of musical notes'. Yes, it is a music dataset. But is it a music *classification* dataset? Yes, in a sense that MNIST is an image dataset. We suggest using this dataset only as a simple proof of concept.

**Warning:**

- This dataset is great for a lot of purposes, not exactly for music classification

## 3.9 Summary

We showed that many popular datasets are different (and flawed) in many aspects. This is applied to the datasets we did not discussed above. But that is a part of reality. In general, we strongly recommend investigating the dataset you use closely - audio, labels, split, etc. It is always helpful to talk to the other researchers – the creators and the users of the dataset.

There's good news as well. The research community is learning lessons from the mistakes and adopting better data science practices. Recently, as a result, we witness the quality of datasets increases significantly. At the end of this book, we will revisit this in more detail and discuss what to consider when creating datasets.

## 3.10 Resources

- We barely cover mood-related datasets in this section. We would like to refer to this repo[GCCE+21] which provides great information about music/mood datasets.
- `mirdata` [BFR+19] is handy Python package that helps researchers handle MIR datasets easily and correctly. Many classification datasets are included e.g., the AcousticBrainz genre dataset [BPS+19].

CHAPTER

# FOUR

# PROBLEM FORMULATION

In this section, we share our hands-on experiences of and details about music classification tasks. There are various types of music classification tasks, and a task can be formulated in different ways. The ideas and considerations are reflected when one constructs a new dataset. After then, users of the dataset follow what was assumed in the dataset.

## 4.1 Genre Classification

Music genre is one of the first things that come to people's mind when they talk about music. Every music listener knows at least some genre names. When talking about musical preferences, people assume they're supposed to talk about their favorite genres.

The simplest problem formulation of genre classification is to define a genre taxonomy that is **flat** and **mutually exclusive** (single-label classification). This is how the pioneering Gtzan genre classification dataset was constructed [TC02]. The authors set 10 high-level genres:

> "blues, classical, country, disco, hip hop, jazz, metal, pop, reggae, rock".

Are they really flat? Musicologists can argue about it for a whole night. Are they mutually exclusive? Probably.. not. One can always find (or write) hybrid music by combining some important features of various genres. However, this simplification works to some extent (Every problem formulation is wrong, but some are useful.) We also suspect people naturally have the idea of mutual exclusiveness when they think of music genres. If that's true, the simplification is not only a bad thing. It was also adopted in Ballroom dataset [CGomezG+06], FMA-small and FMA-medium [DBVB16], ISMIR 2004 genre [CGomezG+06], etc.

We can find a different problem formulation in more modern datasets. The mutual exclusiveness assumption was loosened in Million Song Dataset [BMEWL11] (with tagtraum genre annotations [Sch15]). This allows a track to have more than one genre labels (=**multi-label classification**), which is probably more correct. This is also the usual case where genre classification is treated as a part of a tagging problem (since tags usually include various types of labels including genres) such as MTG-Jamendo [BWT+19].

Finally, a hierarchical genre taxonomy is considered in datasets such as FMA-Full [DBVB16] and AcousticBrainz-Genre [BPS+19].

## 4.2 Mood classification

The genre boundaries are already fuzzy, but perhaps not as much as those of mood. By definition, mood is 100% subjective; and then there is the difference between perceived mood (the mood of music) and induced mood (the mood one would feel when listening to the music) – let alone a time-varying nature of it. In practice, MIR researchers have been brave enough to be ignorant about those details and formulate mood classification problems in various ways.

In general, the whole scene is similar to that of genre classification. Some early datasets are based on flat hierarchy (e.g., MoodsMIREX). When being a part of tagging problem, it's allowed to have multi-labeling (MSD, MTG-Jamendo) [BMEWL11], [BPS+19].

But, there is something special in mood classification. Not all the researchers in mood understanding agreed to make a compromise and formulate it as a classification problem. As a result, some continuity was allowed when annotating mood of music.

The most common method is to annotate it in a **two-dimensional plane** where the axes represents arousal and valence. DEAP and Emomusic are examples [KMS+11], [SCS+13]. Sometimes, researchers even went further. For example, the music can be annotated in a three-dimensional space – valence, arousal, and dominance. In another direction, there is time-varying annotation (every 1 second) in DEAM/Mediaeval [SAY16].

## 4.3 Instrument identification

Instrument identification is another interesting problem that is a bit different from all the others. When pop music is the target, researchers have no control in the range of the existing instruments - when sampling some tracks, it is not possible to limit the instruments to be in a pre-defined taxonomy. One can manually sample items so that there only exist some selected, target instruments. But what's the point when reality ignores the constraint?

That was, though, not a problem in the early days since researchers didn't dare to annotate all the instruments. In fact, in the very early days, the target of instrument identification model was not even music tracks – instrument **samples** (e.g., 1-second clips that contains only a single note of one instrument) were the items to classify.

The problem became more realistic with datasets such as IRMAS [BJFH12]. It has annotations of a single **'predominant' instrument** of an item. This means mutual exclusiveness is assumed and the problem becomes a single-label classification. It is subjective and noisy, but again - we always approximate anyway.

More recently, instrument identification was treated as **multi-label classification** in dataset such as OpenMIC-2018 [HDM18]. Like other tasks, it is also a multi-label classification when being solved as a part of music tagging in MSD or MTG-Jamendo [BMEWL11], [BPS+19].

We can (relatively) safely assume that Instrument annotation is, compared to others such as genres or mood, objective. This may sound good, but this makes it difficult for researchers to accept the noise in the label. When it comes to mood or genre, when the label doesn't seem quite right, researchers may still accept that as a result of inherent subjectivity. However, there are many, many unannotated existing instruments in OpenMIC-2018, MSD, and MTG-Jamendo – in other words, we really know with high confidence that they are wrong! For this reason, we think that the future of instrument identification dataset might be a synthetic(ally mixed) dataset with 100% correct instrument labels.

## 4.4 Music tagging

We already mentioned music tagging above, but what is it exactly? The progress of the computer and internet has given the privilege of labeling music to every single music listener – the **democratization of annotation**. We call this process music tagging. Social music services such as Last.fm gathered these tags, and predicting these tags from the audio content became a task named music (auto-)tagging.

There is no constraint on which tag to use as long as the text field UI allows. Because of this freedom, tagging datasets are extremely **messy, noisy, and in a long-tail**. For example, million song dataset has 505,216 tracks with at least one last.fm tags and the total number of unique tags is.. 522,366 [BMEWL11]. There are more tags than the number of the tracks! Out of them, the 7th popular tag is.. "favorite". The 18th is "Awesome" (yes, it distinguishes lower/uppercases). 33th is "seen live". 37th "Favorite" (I told you). 41th is "Favourite".

Surprisingly, we can still solve this to some extent! How? Well, actually, many other tags – especially the top ones – are relevant to the music content. These are the top-15 tags after removing those mentioned fuzzy tags.

> 'rock', 'pop', 'alternative', 'indie', 'electronic', 'female vocalists', 'dance', '00s', 'alternative rock', 'jazz', 'beautiful', 'metal', 'chillout', 'male vocalists', 'classic rock',

There are genre, mood, and instruments – each of which has been treated as a target category for automatic classification.

What is the exact point of solving a tagging problem if the tag taxonomy is merely a superset of other labels? Musically, the taxonomy and the occurrence of tags reflects what listeners care about. This means the knowledge a model learns can be more universally useful than that from other tasks. Practically, it is easier to collect music tags than collecting (expert-annotated) genre, mood, or instrument labels. This enabled researchers to train and evaluate deep learning models, and this is why tagging remains to be the most popular music classification problem.

CHAPTER

FIVE

# EVALUATION

Evaluation of models is one of the most crucial parts of music classification. No matter how many state-of-the-art models are available, the practical performance of the application can be different depending on which model we choose. Hence, proper evaluation metrics that are fit for purpose are essential in the model selection. In this section, we explore widely used evaluation metrics of music classification. Along with the concepts and definitions of evaluation metrics, their implementation using scikit-learn library is provided together.

Let's explore different evaluation metrics with an example of a binary classification task. We want to assess a classifier that detects vocals in music. Our dataset has ten songs with vocal (blue) and ten songs without vocal (orange). The green circle is a decision boundary of the model. The model predicts that the items in the green circle are vocal music, and the items at the outside circle are instrumental music.

```
import numpy as np
y_true = np.array([False, False, False, False, False, False, False, False, False ,
↪False, True, True, True, True, True, True, True, True, True, True])
y_pred = np.array([False, False, False, False, False, False, False, True, True, True,␣
↪False, False, True, True, True, True, True, True, True, True])
```

As shown in the figure below, we can separate the predictions into four categories.

- True positives (TP): Correctly predicted vocal music (upper left).
- False positives (FP): Predicted as vocal music but they are non-vocal music (upper right).
- False negatives (FN): Predicted as non-vocal music but they are vocal music (lower left).
- True negatives (TN): Correctly predicted non-vocal music (lower right).

```
TP = (y_true & y_pred).sum()
FP = (~y_true & y_pred).sum()
FN = (y_true & ~y_pred).sum()
TN = (~y_true & ~y_pred).sum()
print('True Positive: %d' % TP)
print('False Positive: %d' % FP)
print('False Negative: %d' % FN)
print('True Negative: %d' % TN)
```

```
True Positive: 8
False Positive: 3
False Negative: 2
True Negative: 7
```

## 5.1 Accuracy

Accuracy is an intuitive evaluation metric to assess classification models. It measures how many items are correctly predicted. The formula of accuracy is:

```
accuracy = (TP + TN) / (TP + TN + FP + FN)
print('Accuracy: %.4f' % accuracy)

from sklearn.metrics import accuracy_score
sklearn_accuracy = accuracy_score(y_true, y_pred)
print('Accuracy (sklearn): %.4f' % sklearn_accuracy)
```

```
Accuracy: 0.7500
Accuracy (sklearn): 0.7500
```

## 5.2 Precision

Precision measures how many retrieved items are truly relevant. Among 11 retrieved items in the green circle, 8 of them are vocal music, and 3 of them are not. The formula of precision is:

```
precision = TP / (TP + FP)
print('Precision: %.4f' % precision)

from sklearn.metrics import precision_score
sklearn_precision = precision_score(y_true, y_pred)
print('Precision (sklearn): %.4f' % sklearn_precision)
```

```
Precision: 0.7273
Precision (sklearn): 0.7273
```

## 5.3 Recall

Recall measures how many relevant items are correctly retrieved. Among 10 songs with vocal, 8 of them are correctly predicted as vocal music. The formula of recall is:

```
recall = TP / (TP + FN)
print('Recall: %.4f' % recall)

from sklearn.metrics import recall_score
sklearn_recall = recall_score(y_true, y_pred)
print('Recall (sklearn): %.4f' % sklearn_recall)

sensitivity = TP / (TP + FN)
specificity = TN / (FP + TN)
print('Sensitivity: %.4f' % sensitivity)
print('Specificity: %.4f' % specificity)
```

```
Recall: 0.8000
Recall (sklearn): 0.8000
Sensitivity: 0.8000
Specificity: 0.7000
```

---

**Tip:**

- High precision is directly related to user experience. When retrieved items are truly relevant, users can trust the system.
- However, a high precision / low recall system only retrieves a few positive items, which end up with low diversity. A lot of relevant items (False negatives) will be discarded.

---

## 5.4 F-measure

F-measure or F-score is an evaluation metric of binary classification. The traditional F-measure (F1-score) is defined as the harmonic mean of precision and recall. The maximum value is 1.0, and the lowest is 0 (either precision or recall is zero).

```
F1 = 2 * precision * recall / (precision + recall)
print('F1-score: %.4f' % F1)

from sklearn.metrics import f1_score
sklearn_F1 = f1_score(y_true, y_pred)
print('F1-score (sklearn): %.4f' % sklearn_F1)
```

```
F1-score: 0.7619
F1-score (sklearn): 0.7619
```

---

**Tip:** Depending on system requirements, either precision or recall may be more critical. Fbeta-measure controls the balance of precision and recall using a coefficient beta.

---

## 5.5 High precision vs high recall?

The model outputs the likelihood of the input to have vocal between 0 and 1. Hence, to make a final decision, we need to set a threshold. With a high threshold, the model becomes more strict, which means the green circle becomes smaller. The retrieved results by the model for a given query "vocal music" will be reliable. However, the model only retrieves a few songs among the entire vocal tracks (i.e., high precision and low recall). This can be observed from the precision-recall curve below. As the threshold gets closer to 1.0, precision goes higher while recall goes lower.

On the other hand, if the threshold gets lower, it results in high recall and low precision, which means the system returns any item to be positive. Like this, appropriate decision making of threshold is crucial.

## 5.6 Area under receiver operating characteristic curve (ROC-AUC)

As we checked from the precision-recall curve, the model's performance varies by a decision boundary (threshold). The receiver operating characteristic curve (ROC curve) reflects the model's threshold-varying characteristics. The ROC curve is created by plotting true positive rate (TPR) against false positive rate (FPR), where TPR is also known as sensitivity or recall, and FPR is calculated as (1 - specificity).

In the figure above, a dotted black line indicates the ROC curve of a random classifier, a blue line indicates a better classifier, and an orange line shows a perfect classifier. As a classifier gets better, the area under the curve (AUC) gets wider. We call this area under the ROC curve as ROC-AUC score.

```
y_true = np.array([0, 0, 0, 0, 0, 0, 0, 0, 0, 0, 1, 1, 1, 1, 1, 1, 1, 1, 1, 1])
y_pred_random = np.array([0.5, 0.5, 0.5, 0.5, 0.5, 0.5, 0.5, 0.5, 0.5, 0.5, 0.5, 0.5,␣
↪0.5, 0.5, 0.5, 0.5, 0.5, 0.5, 0.5, 0.5])
y_pred_blue = np.array([0.1, 0.3, 0.8, 0.6, 0.1, 0.4, 0.5, 0.1, 0.2, 0.2, 0.4, 0.4, 0.
↪5, 0.6, 0.7, 0.8, 0.9, 0.6, 0.8, 0.7])
y_pred_orange = np.array([0, 0, 0, 0, 0, 0, 0, 0, 0, 0, 1, 1, 1, 1, 1, 1, 1, 1, 1, 1])

from sklearn.metrics import roc_auc_score
roc_auc_random = roc_auc_score(y_true, y_pred_random)
roc_auc_blue = roc_auc_score(y_true, y_pred_blue)
roc_auc_orange = roc_auc_score(y_true, y_pred_orange)
print('ROC-AUC (random): %.4f' % roc_auc_random)
print('ROC-AUC (blue): %.4f' % roc_auc_blue)
print('ROC-AUC (orange): %.4f' % roc_auc_orange)
```

```
ROC-AUC (random): 0.5000
ROC-AUC (blue): 0.8450
ROC-AUC (orange): 1.0000
```

## 5.7 Area under precision-recall curve (PR-AUC)

It is known that ROC-AUC may report overly optimistic results with imbalanced data. Therefore, the area under the precision-recall curve (PR-AUC) is often provided together with ROC-AUC. The precision-recall curve is created by plotting precision against recall at different thresholds. Unlike the ROC-AUC score, which has 0.5 as its lowest value, the lowest bound of PR-AUC differs by data. When a model predicts every item to be positive regardless of threshold, the recall will always be 1.0, and precision will be a ratio of positive items w.r.t. all items. Hence, the lowest value of PR-AUC is the ratio of positive items.

```
y_true = np.array([0, 0, 0, 0, 0, 0, 0, 0, 0, 0, 0, 0, 0, 0, 0, 0, 0, 0, 1, 1])
y_pred_random = np.array([0.5, 0.5, 0.5, 0.5, 0.5, 0.5, 0.5, 0.5, 0.5, 0.5, 0.5, 0.5,␣
↪0.5, 0.5, 0.5, 0.5, 0.5, 0.5, 0.5, 0.5])
y_pred = np.array([0.1, 0.3, 0.8, 0.6, 0.1, 0.4, 0.5, 0.1, 0.2, 0.2, 0.4, 0.4, 0.5, 0.
↪6, 0.7, 0.8, 0.9, 0.6, 0.8, 0.7])


from sklearn.metrics import roc_auc_score, average_precision_score
roc_auc = roc_auc_score(y_true, y_pred)
roc_auc_random = roc_auc_score(y_true, y_pred_random)
pr_auc = average_precision_score(y_true, y_pred)
pr_auc_random = average_precision_score(y_true, y_pred_random)
print('ROC-AUC (random): %.4f' % roc_auc_random)
```

(continues on next page)

(continued from previous page)

```
print('PR-AUC (random): %.4f' % pr_auc_random)
print('ROC-AUC: %.4f' % roc_auc)
print('PR-AUC: %.4f' % pr_auc)
```

```
ROC-AUC (random): 0.5000
PR-AUC (random): 0.1000
ROC-AUC: 0.8472
PR-AUC: 0.2917
```

**Warning:** The average precision (`sklearn.metrics.average_precision_score`) is one method for calculating PR-AUC. There are other methods such as trapezoid estimates and the interpolated estimates.

**Tip:** When the classification task has multiple labels, we need to aggregate multiple ROC-AUC scores and PR-AUC scores. In scikit-learn library, there is an option called `average`. Most automatic music tagging research uses the option `average='macro'`, which averages tag-wise metrics. For more details, check their documentation (roc_auc_score, average_precision_score).

# Part II

# Supervised Learning

CHAPTER

SIX

# INTRODUCTION

In this chapter, we discuss a learning paradigm, supervised learning, which fully relies on ground truth to solve music classification tasks.

The state-of-the-art in music classification has been improved with various deep neural network architectures that are based on different goals and assumptions. We discuss the most successful and commonly used architectures in music classification.

We need some information to train a neural network model. We call this *ground truth*, *annotations*, or *labels* of the dataset. To reach state-of-the-art performance, deep neural networks often need many different labeled ground truth examples.

Creating a large dataset is costly and tricky. As a solution, in this chapter, we introduce data augmentation – a technique we use to increase the size of dataset. Data augmentation is deeply domain specific, and we discuss the methods for musical data.

Along this chapter, we implement a practical example of a supervised music classification model using GTZAN dataset. This will help the readers to put everything in perspective,

# CHAPTER SEVEN

# ARCHITECTURES

## 7.1 Overview

This tutorial mainly covers deep learning approaches for music classification. Before we jump into the details of different deep architectures, let's check some essential attributes of music classification models.

As shown in the figure above, a music classification model can be broken into preprocessing, front end, and back end modules. In the previous section, we have already covered the preprocessing steps where the model extracts different input representations. The front end of the music classification model usually captures local acoustic characteristics such as timbre, pitch, or existence of a particular instrument. Then the back end module summarizes a sequence of the extracted features, which are the output of the front end module. The boundary between the front and back end may be ambiguous, sometimes.

Another important attribute of music classification is song-level training vs instance-level training. Although our goal is to make song-level predictions, music classification models often use only short audio segments during the training. This is called instance-level training. Instance-level training is justified by our intuition; that humans can predict music tags (e.g., rock music) with just a few-second snippet. As shown in the figure above, when we train a model with an instance level, we end up having more training examples. The task may become more difficult because the model is given a less amount of information. In practice, sometimes this ends up obtaining a more robust music tagging model, probably due to the higher stochasticity. After training an instance-level model, if we need a song-level prediction, the instance-level predictions can be aggregated. Max-pooling, average-pooling, or majority vote is the common operations used for the aggregation.

We summarize important attributes of music classification models as follow:

| Model | Preprocessing | Input length | Front end | Back end | Training | Aggregation |
|---|---|---|---|---|---|---|
| FCN | STFT | 29.1s | 2D CNN | . | song-level | . |
| VGG-ish / Short-chunk CNN | STFT | 3.96s | 2D CNN | Global max pooling | instance-level | Average |
| Harmonic CNN | STFT | 5s | 2D CNN | Global max pooling | instance-level | Average |
| MusiCNN | STFT | 3s | 2D CNN | 1D CNN | instance-level | Average |
| Sample-level CNN | . | 3s | 1D CNN | 1D CNN | instance-level | Average |
| CRNN | STFT | 29.1s | 2D CNN | RNN | song-level | . |
| Music tagging transformer | STFT | 5s-30s | 2D CNN | Transformer | instance-level | Average |

## 7.2 Fully Convolutional Networks (FCNs)

Motivated by the huge success of convolutional neural networks (CNN) in computer vision, MIR researchers adopted the successful architectures to solve automatic music tagging problems. The fully convolutional network (FCN) is one of the early deep learning approaches for music tagging, which comprises four convolutional layers [CFS16]. Each layer is followed by batch normalization, rectified linear unit (ReLU) non-linearity, and a max-pooling layer. 3x3 convolutional filters are used to capture spectro-temporal acoustic characteristics of an input melspectrogram.

## 7.3 VGG-ish / Short-chunk CNNs

The VGG-ish model [HCE+17] and Short-chunk CNNs [WFBS20] are very similar to FCN except for their inputs. Instead of learning song-level representation, they utilize instance-level (chunk-level) training.

Since their input length is shorter than FCN's, the VGG-ish model and Short-chunk CNN do not need to increase the size of receptive fields with sparse strides. Instead, Short-chunk CNN, for example, consists of 7 convolutional layers with dense max-pooling (2, 2), which fits a 3.69s audio chunk. When its input becomes longer, the model summarizes the features using global max pooling.

## 7.4 Harmonic CNNs

The convolutional modules of Harmonic CNNs are identical to those of Short-chunk CNNs, but they use slightly different inputs [WCNS20]. Harmonic CNNs take advantage of trainable band-pass filters and harmonically stacked time-frequency representation inputs. In contrast with fixed mel filterbanks, trainable filters bring more flexibility to the model. And harmonically stacked representation preserves spectro-temporal locality while keeping the harmonic structures through the channel of the input tensor in the first convolutional layer.

## 7.5 MusiCNN

Instead of using 3x3 filters, the authors of MusiCNN proposed to use manually designed filter shapes for music tagging [PS19]. Let's first assume that x- and y-axes correspond to time and frequency. Vertically long filters are designed to capture timbral characteristics, while horizontally long filters are designed to capture temporal energy flux that is probably related to rhythmic patterns and tempo.

## 7.6 Sample-level CNNs

Sample-level CNNs and its variant tackle automatic music tagging in an end-to-end manner by directly using raw audio waveforms as their inputs [LPKN17]. In this architecture, 1x2 or 1x3 (1D) convolution filters are used. Each layer consists of 1D convolution, batch normalization, and ReLU non-linearity. Strided convolution is used to increase the size of the receptive field.

## 7.7 Convolutional Recurrent Neural Networks (CRNNs)

Unlike the previously introduced instance-level models, the convolutional recurrent neural networks (CRNNs) are designed to represent music as a long sequence of multiple instances [CFSC17a]. CRNNs can be described as a combination of CNN and RNN. The CNN front end captures local acoustic characteristics (instance-level), and the RNN back end summarizes the sequence of instance-level features.

## 7.8 Music tagging transformer

The motivation of the convolutional neural network with self-attention (CNNSA) [WCS19] and Music tagging transformer [WCS21] is identical to that of the CRNN model. The front end captures local acoustic characteristics, and the back end summarizes the sequence. In the field of natural language processing, Transformer has shown its suitability in long sequence modeling by using self-attention layers. Both CNNSA and Music tagging transformer use the CNN front end and the Transformer back end. The back end summarizes the instance-level features effectively.

## 7.9 Which model should we use?

After exploring these many different architectures, the first natural question would be about the *best* model to use. In previous work [WFBS20], experimental results in three datasets (MagnaTagATune, Million Song Data, MTG-Jamendo) are reported as follows.

---

**Note:** Summary:

- For the best performance, use the Music tagging transformer.
- VGG-ish and Short-chunk CNN are simple but powerful choices.
- When your training dataset is small, try with a reduced search space by using MusiCNN or Harmonic CNN.
- Sample-level CNN achieves strong performance with the increase of the size of the dataset. Still, spectrogram-based models are showing state-of-the-art results.

---

**Tip:**

- PyTorch implementation of introduced models are available online [Github]
- You can try an online demo of pretrained models [Replicate.ai]

---

CHAPTER

# EIGHT

# AUDIO DATA AUGMENTATIONS

In this chapter, we will discuss common transformations that we can apply to audio signals in the **time domain**. We will refer to these as "audio data augmentations".

Data augmentations are a set of methods that add modified copies to a dataset, from the existing data. This process creates many variations of natural data, and can act as a regulariser to reduce the problem of overfitting. It can also help deep neural networks become robust to complex variations of natural data, which improves their generalisation performance.

In the field of computer vision, the transformations that we apply to images are often very self-explanatory. Take this image, for example. It becomes clear that we are zooming in and removing the color of the image:

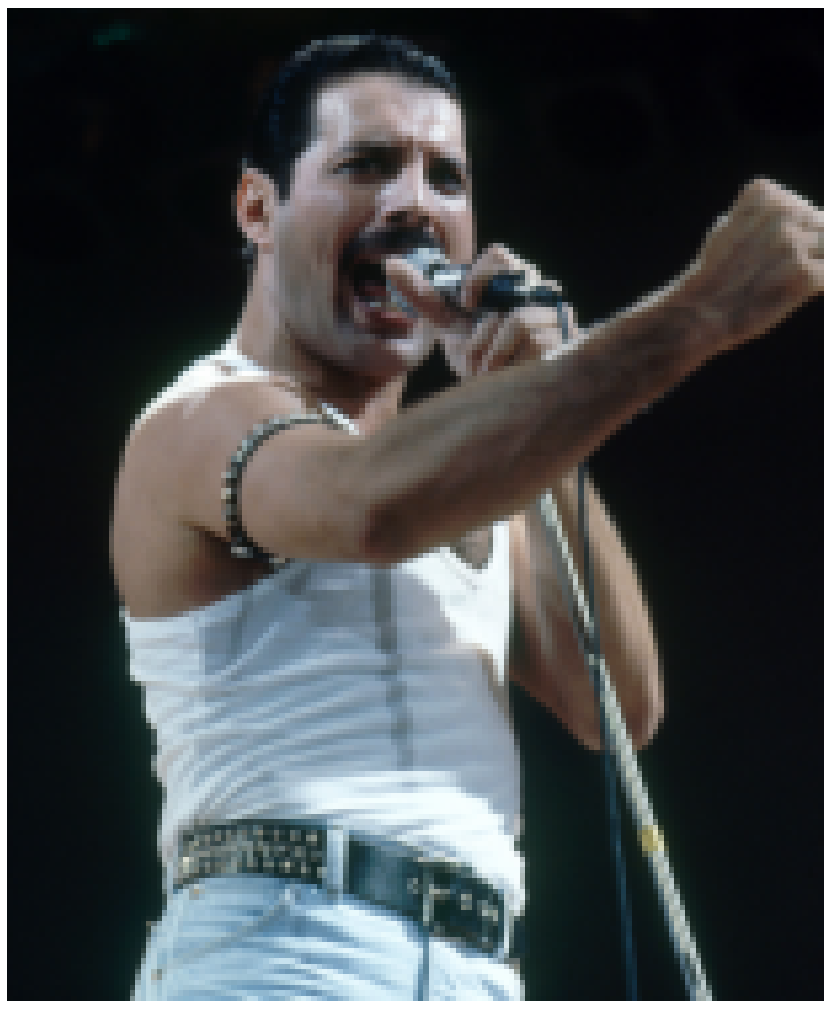

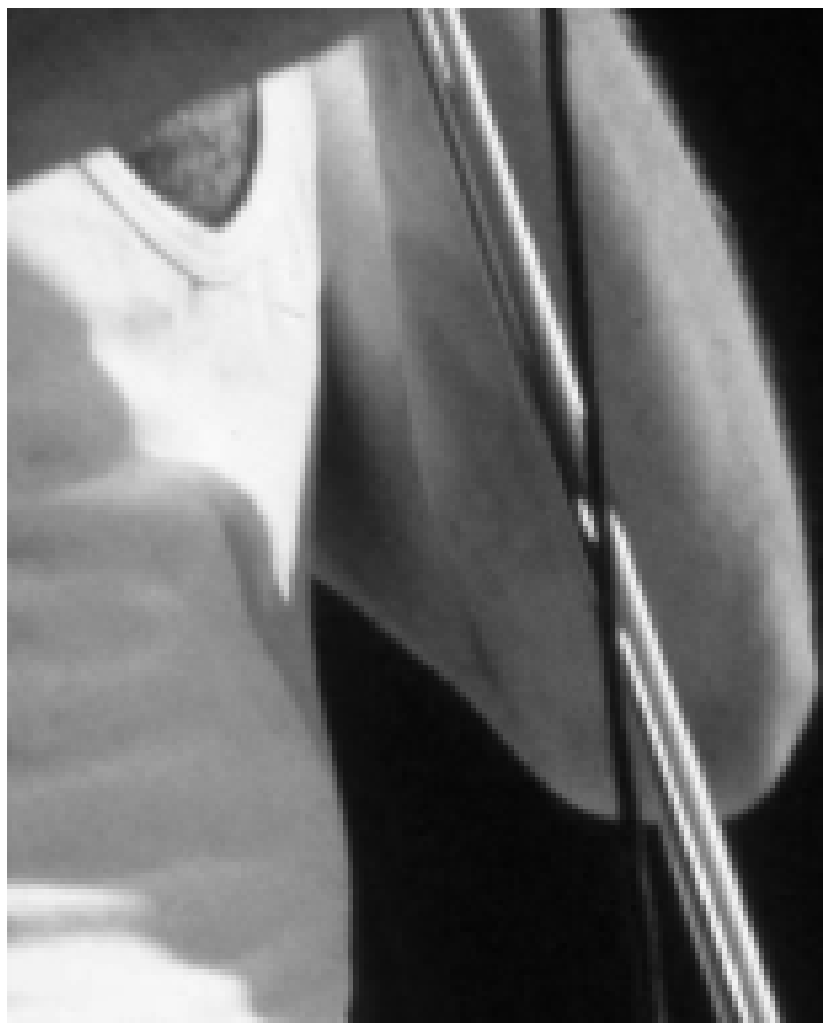

Naturally, we cannot translate transformations from the vision domain directly to the audio domain. Before we explore a battery of audio data augmentations, we now list the currently available code libraries:

## 8.1 Code Libraries

| Name | Author | Framework | Language | License | Link |
|---|---|---|---|---|---|
| Muda | B. McFee et al. (2015) | General Purpose | Python | ISC License | source code |
| Audio Degradation Toolbox | M. Mauch et al. (2013) | General Purpose | MATLAB | GNU General Public License 2.0 | source code |
| rubberband | - | General Purpose | C++ | GNU General Public License (non-commercial) | website, pyrubberband |
| audiomentations | I. Jordal (2021) | General Purpose | Python | MIT License | source code |
| tensorflow-io | tensorflow.org | TensorFlow | Python | Apache 2.0 License | tutorial |
| torchaudio | pytorch.org | PyTorch | Python | BSD 2-Clause "Simplified" License | source code |
| torch-audiomentations | Asteroid (2021) | PyTorch | Python | MIT License | source code |
| torchaudio-augmentations | J. Spijkervet (2021) | PyTorch | Python | MIT License | source code |

## 8.2 Listening

One of the most essential, and yet overlooked, parts of music research is exploring and observing the data. This also applies to data augmentation research: one has to develop a general understanding of the effect of transformations that can be applied to audio. Even more so, when transformations are applied sequentially.

For instance, we will understand why a reverb applied *before* a frequency filter will sound different than when the reverb is applied *after* the frequency filter. Before we develop this intuition, let's listen to a series of audio data augmenations.

```
Number of datapoints in the GTZAN dataset: f442

Selected track no.: 5
Genre: 0
Sample rate: 22050
Channels: 1
Samples: 639450
```

```
<IPython.lib.display.Audio object>
```

### 8.2.1 Random Crop

Similar to how we can crop an image, so that only a subset of the image is represented, we can 'crop' a piece of audio by selecting a fragment between two time points $t_0 - t_1$.

Various terms for this exist, e.g.,: slicing, trimming,

### 8.2.2 Frequency Filter

---

**Note:** In these examples and the accompanying code, we assume the shape of audio ordered in our array is follows: (channel, time)

---

A frequency filter is applied to the signal. We can process the signal with either the LowPass or HighPass algorithm [47]. In a stochastic setting, we can determine which one to apply by, for example, a coin flip. Another filter parameter we can control stochastically is the *cutoff* frequency: the frequency at which the filter will be applied. All frequencies above the cut-off frequency are filtered from the signal for a low-pass filter (i.e., we let the *low* frequencies *pass*). Similarly for the high-pass filter, all frequencies below the cut-off frequency are filtered from the signal (i.e., we let the *high* frequencies *pass*).

```
Original
```

```
<IPython.lib.display.Audio object>
```

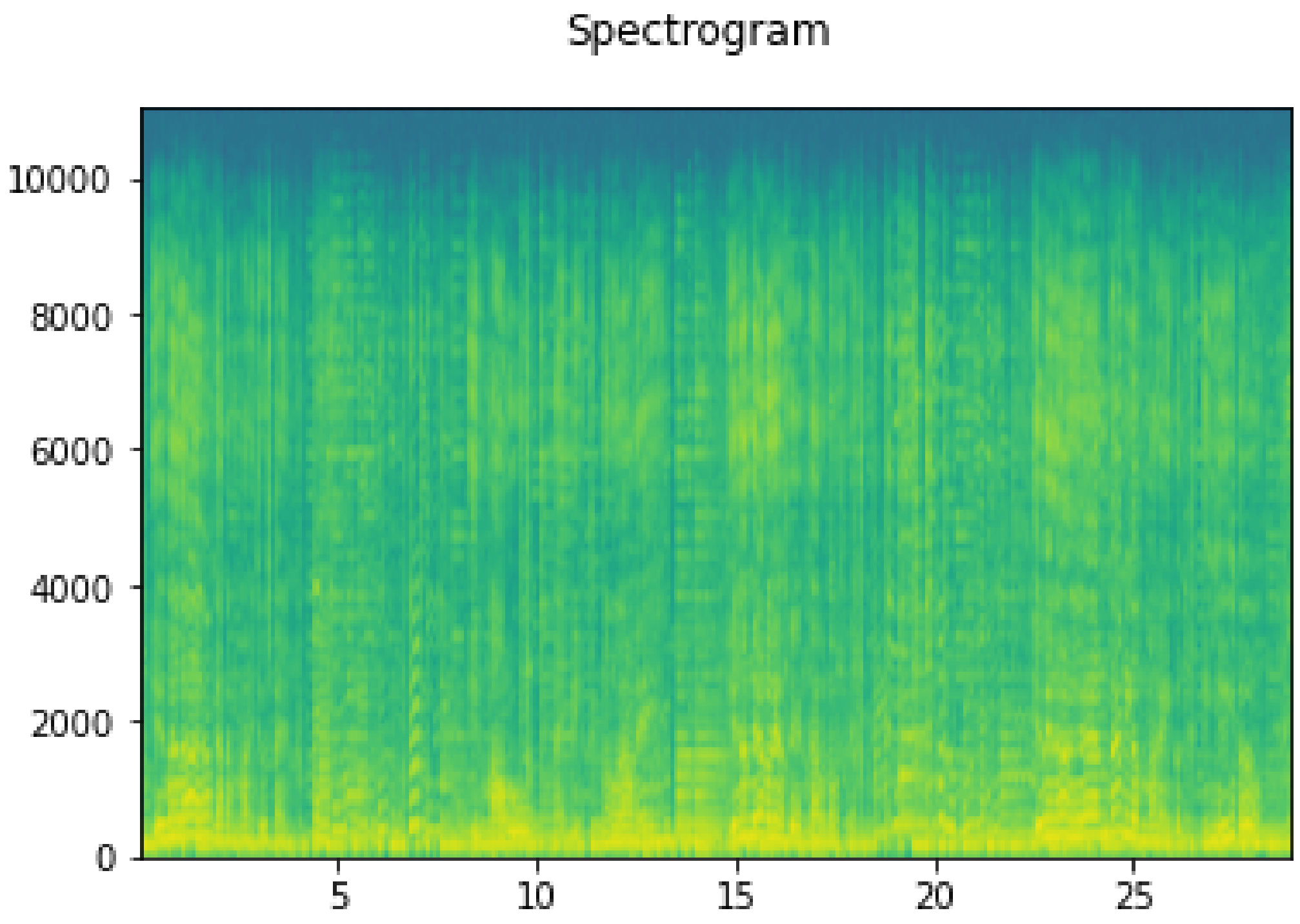

```
LowPassFilter
```

```
<IPython.lib.display.Audio object>
```

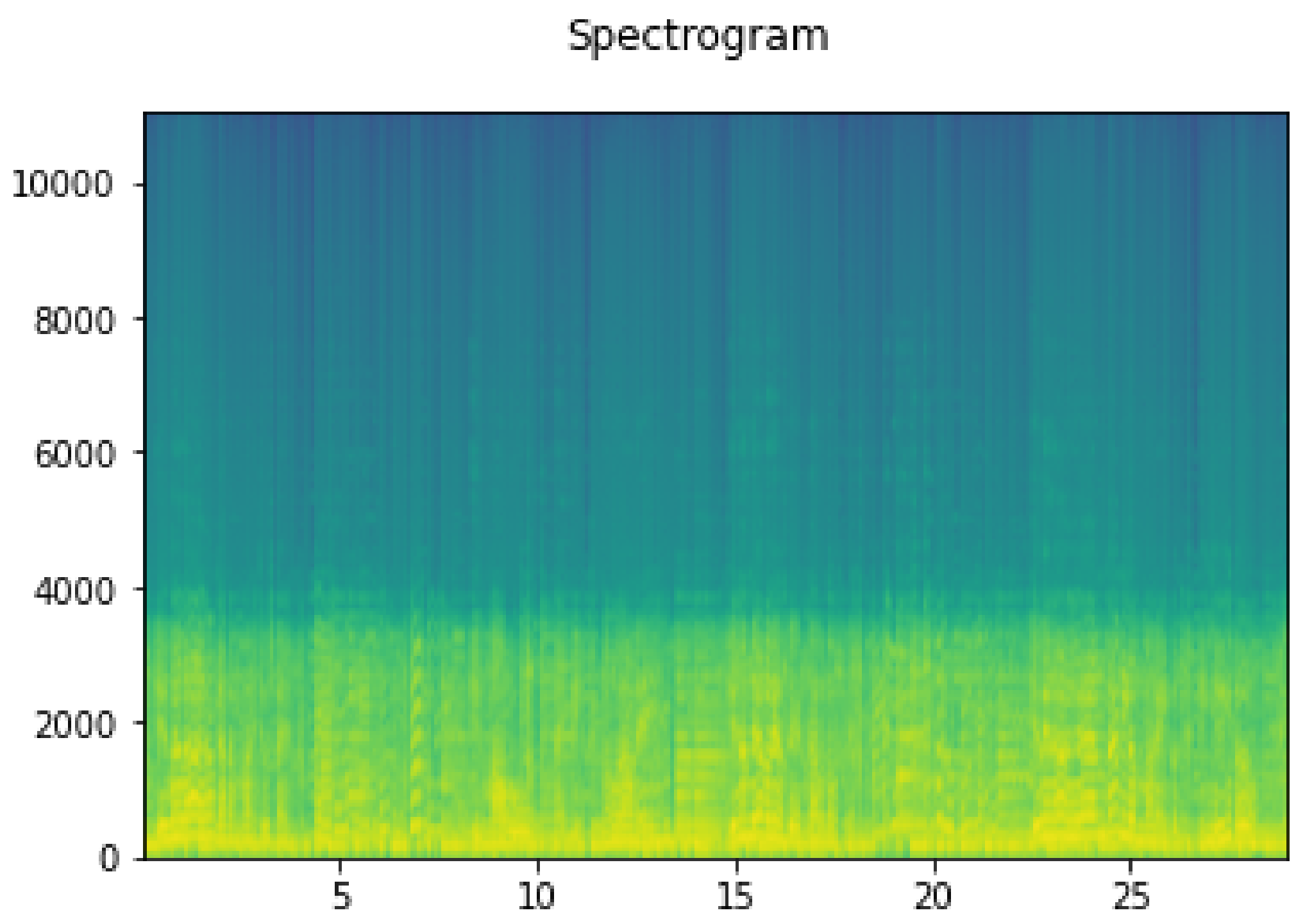

### 8.2.3 Delay

The signal is delayed by a value that can be chosen arbitrarily. The delayed signal is added to the original signal with a volume factor, e.g.,, we can multiply the signal's amplitude by 0.5.

```
Original
```

```
<IPython.lib.display.Audio object>
```

```
Delay of 200ms
```

```
<IPython.lib.display.Audio object>
```

#### Comb filter

When we apply a delayed signal to the original with a short timespan and a high volume factor, it will cause interferences. These audible interferences are called a "comb filter".

```
Original
```

```
<IPython.lib.display.Audio object>
```

```
Delay of 61ms
```

```
<IPython.lib.display.Audio object>
```

### 8.2.4 Pitch Shift

The pitch of the signal is shifted up or down, depending on the pitch interval that is chosen beforehand. Here, we assume a 12-tone equal temperament tuning that divides a single octave in 12 semitones.

```
Original
```

```
<IPython.lib.display.Audio object>
```

```
Pitch shift of 4 semitones
```

```
<IPython.lib.display.Audio object>
```

### 8.2.5 Reverb

To alter the original signal's acoustics, we can apply a Schroeder reverberation effect. This gives the illusion that the sound is played in a larger space, in which it takes longer for the sound to reflect.

Applying a reverberation of a "small" room on a signal that was recorded in a larger room does not have the opposite effect: the process of reverberation is an additive process. The reverse process is called "dereverberation".

```
Original
```

```
<IPython.lib.display.Audio object>
```

```
Reverb
```

```
<IPython.lib.display.Audio object>
```

### 8.2.6 Gain

> **Warning:** In Jupyter notebook's Audio() object, we have to set `normalize=False` so that we can hear an unnormalized version of the audio. This is important to reflect the true audio transformation output.

We can apply a volume factor to the signal, so that it is perceived as louder. It is generally accepted that a loudness gain of 10 decibels is perceived as twice as loud, and similarly 10 decibels of gain reduction is perceived half as loud.

```
Original
```

```
<IPython.lib.display.Audio object>
```

```
Gain
```

```
<IPython.lib.display.Audio object>
```

### 8.2.7 Noise

White Gaussian noise is added to the complete signal with a signal-to-noise ratio (SNR) that can be specified. A uniform distribution between the minimum and maximum SNR boundaries is made so that, for example, we can draw a different SNR value for each example in a mini-batch during training.

```
Original
```

```
<IPython.lib.display.Audio object>
```

```
Noise
```

```
<IPython.lib.display.Audio object>
```

### 8.2.8 Polarity Inversion

While this does not have an effect on a time-frequency representation of audio, e.g., a spectrogram, encoders that are trained on raw waveforms can benefit from an audio data augmentation that flips the phase of an audio signal: Polarity Inversion. Simply put, the signal is multipled by $-1$, which causes the phase to invert.

Interestingly, when we add the original signal to the phase-inverted signal, all phases will cancel out. This will naturally result in silence. This is the core principle behind noise-cancelling headphones, which record the sound of your surroundings and apply a polarity inversion as to reduce unwanted noise.

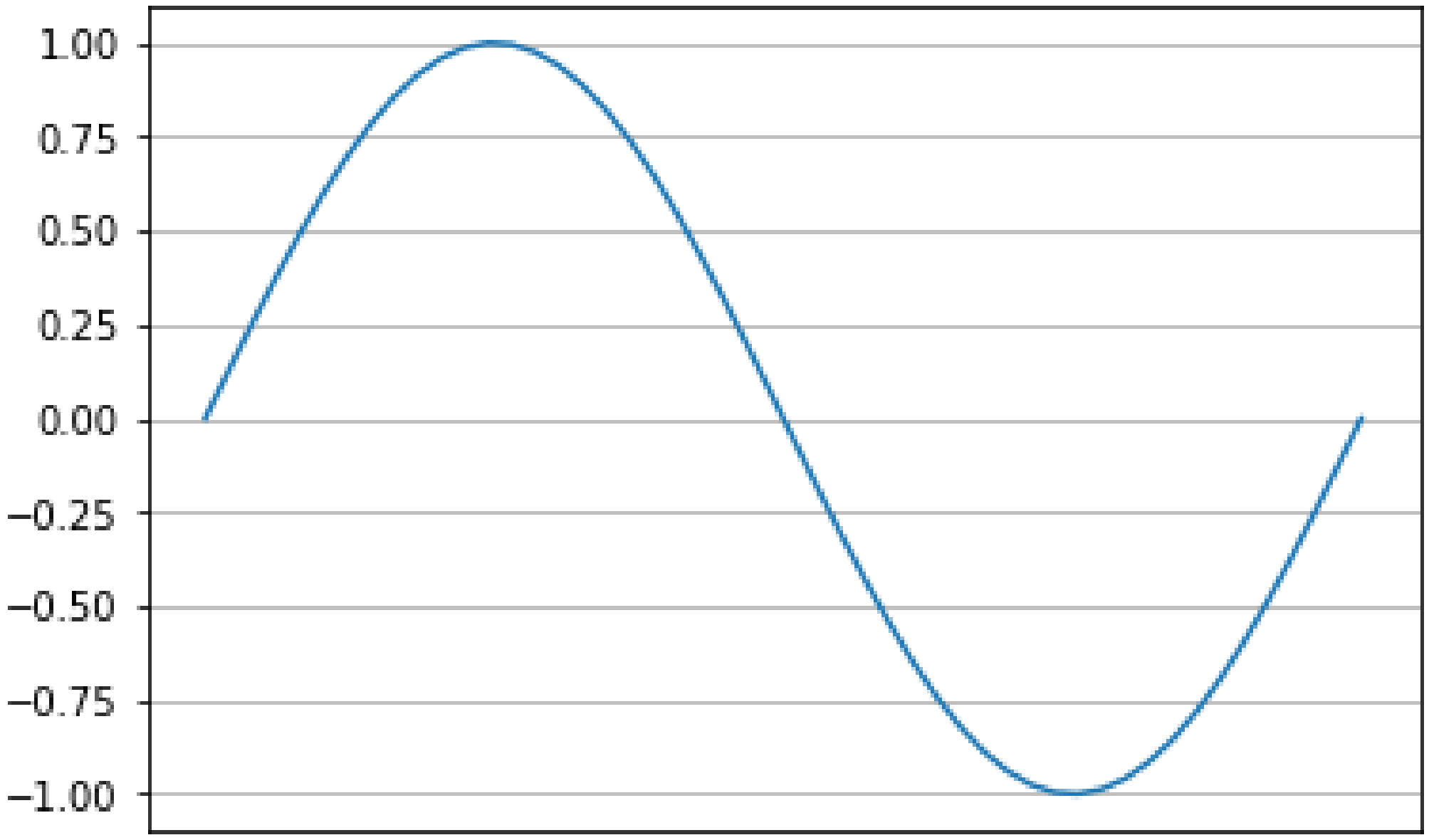

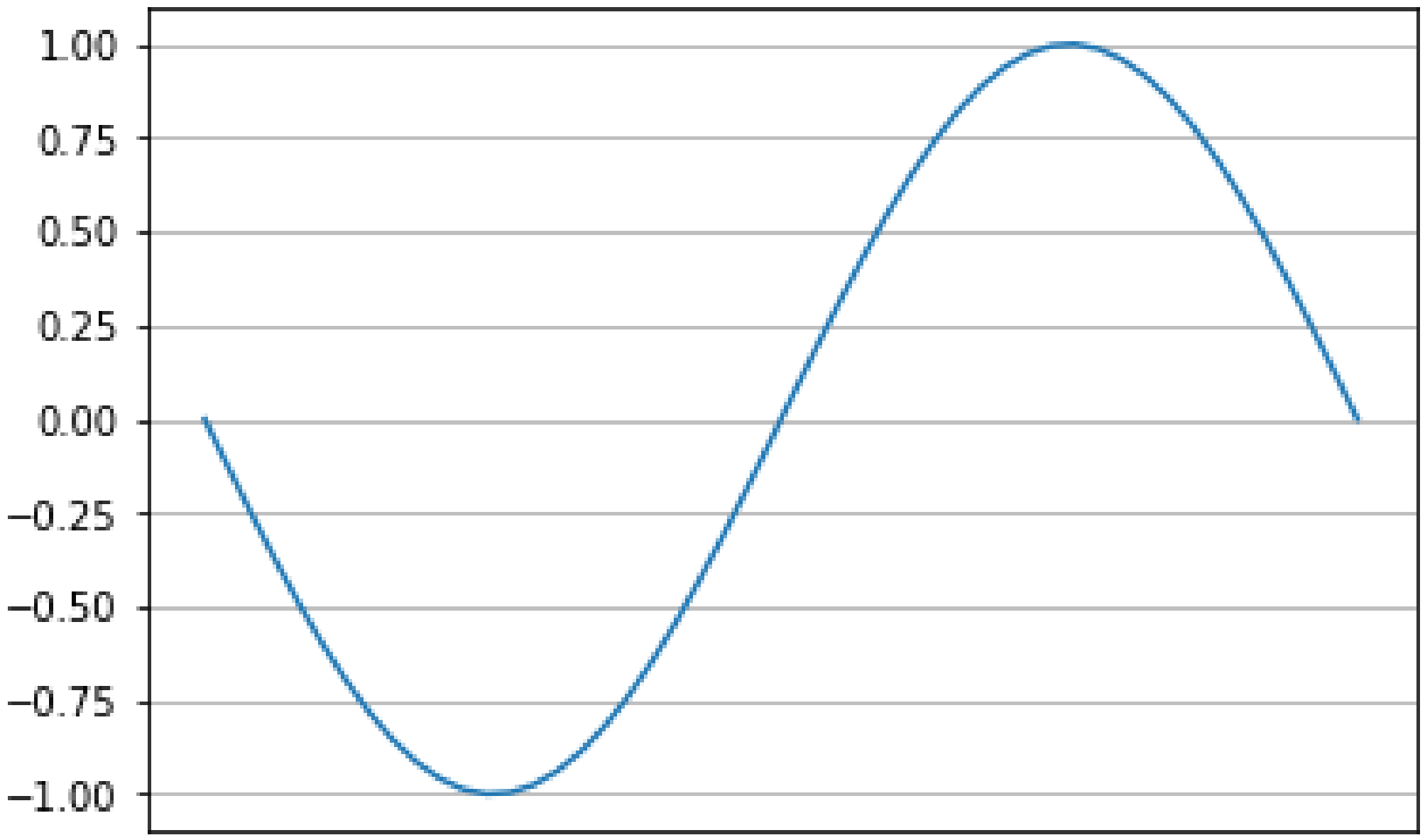

```
Original
```

```
<IPython.lib.display.Audio object>
```

```
Polarity Inversion
```

```
<IPython.lib.display.Audio object>
```

```
Original + Polarity Inversion
```

```
<IPython.lib.display.Audio object>
```

## 8.3 Sequential Audio Data Augmentations

Now that we have built up some intuition of some of the audio transformations, let us observe how they can be applied sequentially. More importantly, to develop an understanding on how different audio transformations interact when we apply them before, or after each other.

For this, we can use a `Compose` module, which takes as input a list of audio transformations. These will be applied in the order they appear in the supplied list. This interface is similar to `torchvision.transforms` and `torchaudio.transforms`' `Compose` modules.

```
Original:
```

```
<IPython.lib.display.Audio object>
```

```
Transform: Compose(
        Delay()
        HighLowPass()
)
```

```
<IPython.lib.display.Audio object>
```

Now that we have listened to what a sequential audio transformation sounds like, let's observe how two different transforms interact when they are applied in a different sequential order.

Let's take the following two transforms:

- `Noise`
- `Reverb`

A signal that does not have any reverberation added, is commonly called a *dry* signal. A signal that is reverberated is called a *wet* signal.

When we first apply the `Noise` transform, the `Reverb` transform will apply the reverberation to the dry signal **and** the added noise signal. This will result in a completely *wet* signal.

Conversely, when we first apply the `Reverb` transform, the `Noise` signal will be added **after** the reverberated signal. The noise is thus *dry*, i.e., it is not reverberated.

```
Transform 1: Compose(
        Noise()
        Reverb()
)
```

```
<IPython.lib.display.Audio object>
```

```
Transform: Compose(
        Reverb()
        Noise()
)
```

```
<IPython.lib.display.Audio object>
```

### 8.3.1 More Sequential Audio Data Augmentations

Let's continue to develop our intuition for sequential audio transformations a bit more in the following examples:

```
Transform: Compose(
        RandomResizedCrop()
        HighLowPass()
        Delay()
)
```

```
<IPython.lib.display.Audio object>
```

Instead of retrieving a single augmented example, let's return `4` different views of the original sound:

```
Transform: ComposeMany(
        RandomResizedCrop()
        HighLowPass()
        Delay()
)
```

```
<IPython.lib.display.Audio object>
```

```
<IPython.lib.display.Audio object>
```

```
<IPython.lib.display.Audio object>
```

```
<IPython.lib.display.Audio object>
```

## 8.4 Stochastic Audio Data Augmentations

We can also apply audio data augmentations stochastically, in which each data augmentation is applied with a random probability $p$. This will increase the number of natural examples the model can learn - and generalize - from:

```
from torchaudio_augmentations import RandomApply

# we want 4 augmented samples from ComposeMany
num_augmented_samples = 4

# 4 seconds of audio
num_samples = sr * 4

stochastic_transforms = [
    RandomResizedCrop(n_samples=num_samples),
    # apply with p = 0.3
    RandomApply(
        [
            PolarityInversion(),
            HighLowPass(
                sample_rate=sr,
                lowpass_freq_low=2200,
                lowpass_freq_high=4000,
                highpass_freq_low=200,
                highpass_freq_high=1200,
            ),
            Delay(
                sample_rate=sr,
                volume_factor=0.5,
                min_delay=100,
                max_delay=500,
                delay_interval=1,
            ),
        ],
        p=0.3,
    ),
    # apply with p = 0.8
```

(continues on next page)

(continued from previous page)

```
    RandomApply(
        [
            PitchShift(sample_rate=sr, n_samples=num_samples),
            Gain(),
            Noise(max_snr=0.01),
            Reverb(sample_rate=sr),
        ],
        p=0.8,
    ),
]
transform = ComposeMany(
    stochastic_transforms, num_augmented_samples=num_augmented_samples
)

print("Transform:", transform)
transformed_audio = transform(audio)

for ta in transformed_audio:
    display(Audio(ta, rate=sr))
plt.show()
```

```
Transform: ComposeMany(
        RandomResizedCrop()
        RandomApply(
    p=0.3
    PolarityInversion()
    HighLowPass()
    Delay()
)
        RandomApply(
    p=0.8
    <torchaudio_augmentations.augmentations.pitch_shift.PitchShift object at␣
↪0x7f874c4d07f0>
    Gain()
    Noise()
    Reverb()
)
)
```

```
<IPython.lib.display.Audio object>
```

```
<IPython.lib.display.Audio object>
```

```
<IPython.lib.display.Audio object>
```

```
<IPython.lib.display.Audio object>
```

### 8.4.1 Single stochastic augmentations

```
# we want 4 augmented samples from ComposeMany
num_augmented_samples = 4

# 4 seconds of audio
num_samples = sr * 4


# define our stochastic augmentations
transforms = [
    RandomResizedCrop(n_samples=num_samples),
    RandomApply([PolarityInversion()], p=0.8),
    RandomApply([HighLowPass(sample_rate=sr)], p=0.6),
    RandomApply([Delay(sample_rate=sr)], p=0.6),
    RandomApply([PitchShift(sample_rate=sr, n_samples=num_samples)], p=0.3),
    RandomApply([Gain()], p=0.6),
    RandomApply([Noise(max_snr=0.01)], p=0.3),
    RandomApply([Reverb(sample_rate=sr)], p=0.5),
]


transform = ComposeMany(transforms, num_augmented_samples=num_augmented_samples)

print("Transform:", transform)
transformed_audio = transform(audio)

for ta in transformed_audio:
    # plot_spectrogram(ta, sr, title=e="")
    display(Audio(ta, rate=sr))
plt.show()
```

```
Transform: ComposeMany(
        RandomResizedCrop()
        RandomApply(
    p=0.8
    PolarityInversion()
)
        RandomApply(
    p=0.6
    HighLowPass()
)
        RandomApply(
    p=0.6
    Delay()
)
        RandomApply(
    p=0.3
    <torchaudio_augmentations.augmentations.pitch_shift.PitchShift object at␣
↪0x7f874b7973a0>
)
        RandomApply(
    p=0.6
    Gain()
)
        RandomApply(
```

(continues on next page)

(continued from previous page)

```
    p=0.3
    Noise()
)
        RandomApply(
    p=0.5
    Reverb()
)
)
```

```
<IPython.lib.display.Audio object>
```

```
<IPython.lib.display.Audio object>
```

```
<IPython.lib.display.Audio object>
```

```
<IPython.lib.display.Audio object>
```

## 8.5 Conclusion

Hopefully, this chapter on audio data augmentations has given you an intuition of what transformations we can apply to audio signals. We will be using these audio data augmentations in the other code tutorials, to see how they can be applied effectively to improve training of deep neural networks in the task of music classification.

# CHAPTER NINE

# PYTORCH TUTORIAL

## 9.1 Data collection

In this PyTorch tutorial, we use GTZAN dataset which consists of 10 exclusive genre classes. Please run the following script in your local path.

```
!wget http://opihi.cs.uvic.ca/sound/genres.tar.gz
!tar -zxvf genres.tar.gz
!wget https://raw.githubusercontent.com/coreyker/dnn-mgr/master/gtzan/train_filtered.
↪txt
!wget https://raw.githubusercontent.com/coreyker/dnn-mgr/master/gtzan/valid_filtered.
↪txt
!wget https://raw.githubusercontent.com/coreyker/dnn-mgr/master/gtzan/test_filtered.
↪txt
```

## 9.2 Data loader

```
import os
import random
import torch
import numpy as np
import soundfile as sf
from torch.utils import data
from torchaudio_augmentations import (
    RandomResizedCrop,
    RandomApply,
    PolarityInversion,
    Noise,
    Gain,
    HighLowPass,
    Delay,
    PitchShift,
    Reverb,
    Compose,
)


GTZAN_GENRES = ['blues', 'classical', 'country', 'disco', 'hiphop', 'jazz', 'metal',
↪'pop', 'reggae', 'rock']
```

(continues on next page)

(continued from previous page)

```
class GTZANDataset(data.Dataset):
    def __init__(self, data_path, split, num_samples, num_chunks, is_augmentation):
        self.data_path =  data_path if data_path else ''
        self.split = split
        self.num_samples = num_samples
        self.num_chunks = num_chunks
        self.is_augmentation = is_augmentation
        self.genres = GTZAN_GENRES
        self._get_song_list()
        if is_augmentation:
            self._get_augmentations()

    def _get_song_list(self):
        list_filename = os.path.join(self.data_path, '%s_filtered.txt' % self.split)
        with open(list_filename) as f:
            lines = f.readlines()
        self.song_list = [line.strip() for line in lines]

    def _get_augmentations(self):
        transforms = [
            RandomResizedCrop(n_samples=self.num_samples),
            RandomApply([PolarityInversion()], p=0.8),
            RandomApply([Noise(min_snr=0.3, max_snr=0.5)], p=0.3),
            RandomApply([Gain()], p=0.2),
            RandomApply([HighLowPass(sample_rate=22050)], p=0.8),
            RandomApply([Delay(sample_rate=22050)], p=0.5),
            RandomApply([PitchShift(n_samples=self.num_samples, sample_rate=22050)],␣
↪p=0.4),
            RandomApply([Reverb(sample_rate=22050)], p=0.3),
        ]
        self.augmentation = Compose(transforms=transforms)

    def _adjust_audio_length(self, wav):
        if self.split == 'train':
            random_index = random.randint(0, len(wav) - self.num_samples - 1)
            wav = wav[random_index : random_index + self.num_samples]
        else:
            hop = (len(wav) - self.num_samples) // self.num_chunks
            wav = np.array([wav[i * hop : i * hop + self.num_samples] for i in␣
↪range(self.num_chunks)])
        return wav

    def __getitem__(self, index):
        line = self.song_list[index]

        # get genre
        genre_name = line.split('/')[0]
        genre_index = self.genres.index(genre_name)

        # get audio
        audio_filename = os.path.join(self.data_path, 'genres', line)
        wav, fs = sf.read(audio_filename)

        # adjust audio length
        wav = self._adjust_audio_length(wav).astype('float32')
```

(continues on next page)

(continued from previous page)

```
        # data augmentation
        if self.is_augmentation:
            wav = self.augmentation(torch.from_numpy(wav).unsqueeze(0)).squeeze(0).
↪numpy()

        return wav, genre_index

    def __len__(self):
        return len(self.song_list)

def get_dataloader(data_path=None,
                   split='train',
                   num_samples=22050 * 29,
                   num_chunks=1,
                   batch_size=16,
                   num_workers=0,
                   is_augmentation=False):
    is_shuffle = True if (split == 'train') else False
    batch_size = batch_size if (split == 'train') else (batch_size // num_chunks)
    data_loader = data.DataLoader(dataset=GTZANDataset(data_path,
                                                       split,
                                                       num_samples,
                                                       num_chunks,
                                                       is_augmentation),
                                  batch_size=batch_size,
                                  shuffle=is_shuffle,
                                  drop_last=False,
                                  num_workers=num_workers)
    return data_loader
```

Let's check returned data shapes.

```
train_loader = get_dataloader(split='train', is_augmentation=True)
iter_train_loader = iter(train_loader)
train_wav, train_genre = next(iter_train_loader)

valid_loader = get_dataloader(split='valid')
test_loader = get_dataloader(split='test')
iter_test_loader = iter(test_loader)
test_wav, test_genre = next(iter_test_loader)
print('training data shape: %s' % str(train_wav.shape))
print('validation/test data shape: %s' % str(test_wav.shape))
print(train_genre)
```

```
training data shape: torch.Size([16, 639450])
validation/test data shape: torch.Size([16, 1, 639450])
tensor([9, 3, 4, 2, 2, 5, 2, 5, 7, 1, 1, 7, 8, 7, 4, 0])
```

**Note:**

- A data loader returns a tensor of audio and their genre indice at each iteration.
- Random chunks of audio are cropped from the entire sequence during the training. But in validation / test phase, an entire sequence is split into multiple chunks and the chunks are stacked. The stacked chunks are later input to a trained model and the output predictions are aggregated to make song-level predictions.

## 9.3 Model

We are going to build a simple 2D CNN model with Mel spectrogram inputs. First, we design a convolution module that consists of 3x3 convolution, batch normalization, ReLU non-linearity, and 2x2 max pooling. This module is going to be used for each layer of the 2D CNN.

```
from torch import nn


class Conv_2d(nn.Module):
    def __init__(self, input_channels, output_channels, shape=3, pooling=2, dropout=0.
↪1):
        super(Conv_2d, self).__init__()
        self.conv = nn.Conv2d(input_channels, output_channels, shape, padding=shape//
↪2)
        self.bn = nn.BatchNorm2d(output_channels)
        self.relu = nn.ReLU()
        self.maxpool = nn.MaxPool2d(pooling)
        self.dropout = nn.Dropout(dropout)

    def forward(self, wav):
        out = self.conv(wav)
        out = self.bn(out)
        out = self.relu(out)
        out = self.maxpool(out)
        out = self.dropout(out)
        return out
```

Stack the convolution layers. In a PyTorch module, layers are declared in `__init__` and they are built up in `forward` function.

```
import torchaudio


class CNN(nn.Module):
    def __init__(self, num_channels=16,
                       sample_rate=22050,
                       n_fft=1024,
                       f_min=0.0,
                       f_max=11025.0,
                       num_mels=128,
                       num_classes=10):
        super(CNN, self).__init__()

        # mel spectrogram
        self.melspec = torchaudio.transforms.MelSpectrogram(sample_rate=sample_rate,
                                                            n_fft=n_fft,
                                                            f_min=f_min,
                                                            f_max=f_max,
                                                            n_mels=num_mels)
        self.amplitude_to_db = torchaudio.transforms.AmplitudeToDB()
        self.input_bn = nn.BatchNorm2d(1)
```

(continues on next page)

(continued from previous page)

```
        # convolutional layers
        self.layer1 = Conv_2d(1, num_channels, pooling=(2, 3))
        self.layer2 = Conv_2d(num_channels, num_channels, pooling=(3, 4))
        self.layer3 = Conv_2d(num_channels, num_channels * 2, pooling=(2, 5))
        self.layer4 = Conv_2d(num_channels * 2, num_channels * 2, pooling=(3, 3))
        self.layer5 = Conv_2d(num_channels * 2, num_channels * 4, pooling=(3, 4))

        # dense layers
        self.dense1 = nn.Linear(num_channels * 4, num_channels * 4)
        self.dense_bn = nn.BatchNorm1d(num_channels * 4)
        self.dense2 = nn.Linear(num_channels * 4, num_classes)
        self.dropout = nn.Dropout(0.5)
        self.relu = nn.ReLU()

    def forward(self, wav):
        # input Preprocessing
        out = self.melspec(wav)
        out = self.amplitude_to_db(out)

        # input batch normalization
        out = out.unsqueeze(1)
        out = self.input_bn(out)

        # convolutional layers
        out = self.layer1(out)
        out = self.layer2(out)
        out = self.layer3(out)
        out = self.layer4(out)
        out = self.layer5(out)

        # reshape. (batch_size, num_channels, 1, 1) -> (batch_size, num_channels)
        out = out.reshape(len(out), -1)

        # dense layers
        out = self.dense1(out)
        out = self.dense_bn(out)
        out = self.relu(out)
        out = self.dropout(out)
        out = self.dense2(out)

        return out
```

**Note:** In this example, we performed preprocessing on-the-fly using torchaudio. This process can be done offline outside of the network using other libraries such as librosa and essentia.

**Tip:**

- There is no activation function at the last layer since `nn.CrossEntropyLoss` already includes softmax in it.
- If you want to perform multi-label binary classification, include `out = nn.Sigmoid()(out)` at the last layer and use `nn.BCELoss()`.

## 9.4 Training

Iterate training. One epoch is defined as visiting all training items once. This definition can be modified in `def __len__` in data loader.

```
from sklearn.metrics import accuracy_score, confusion_matrix


device = torch.device('cuda:0' if torch.cuda.is_available() else 'cpu')
cnn = CNN().to(device)
loss_function = nn.CrossEntropyLoss()
optimizer = torch.optim.Adam(cnn.parameters(), lr=0.001)
valid_losses = []
num_epochs = 30

for epoch in range(num_epochs):
    losses = []

    # Train
    cnn.train()
    for (wav, genre_index) in train_loader:
        wav = wav.to(device)
        genre_index = genre_index.to(device)

        # Forward
        out = cnn(wav)
        loss = loss_function(out, genre_index)

        # Backward
        optimizer.zero_grad()
        loss.backward()
        optimizer.step()
        losses.append(loss.item())
    print('Epoch: [%d/%d], Train loss: %.4f' % (epoch+1, num_epochs, np.
↪mean(losses)))

    # Validation
    cnn.eval()
    y_true = []
    y_pred = []
    losses = []
    for wav, genre_index in valid_loader:
        wav = wav.to(device)
        genre_index = genre_index.to(device)

        # reshape and aggregate chunk-level predictions
        b, c, t = wav.size()
        logits = cnn(wav.view(-1, t))
        logits = logits.view(b, c, -1).mean(dim=1)
        loss = loss_function(logits, genre_index)
        losses.append(loss.item())
        _, pred = torch.max(logits.data, 1)

        # append labels and predictions
        y_true.extend(genre_index.tolist())
        y_pred.extend(pred.tolist())
    accuracy = accuracy_score(y_true, y_pred)
```

(continues on next page)

(continued from previous page)

```
    valid_loss = np.mean(losses)
    print('Epoch: [%d/%d], Valid loss: %.4f, Valid accuracy: %.4f' % (epoch+1, num_
↪epochs, valid_loss, accuracy))

    # Save model
    valid_losses.append(valid_loss.item())
    if np.argmin(valid_losses) == epoch:
        print('Saving the best model at %d epochs!' % epoch)
        torch.save(cnn.state_dict(), 'best_model.ckpt')
```

```
Epoch: [1/30], Train loss: 2.4078
Epoch: [1/30], Valid loss: 2.3558, Valid accuracy: 0.1117
Saving the best model at 0 epochs!
Epoch: [2/30], Train loss: 2.3422
Epoch: [2/30], Valid loss: 2.2748, Valid accuracy: 0.1218
Saving the best model at 1 epochs!
Epoch: [3/30], Train loss: 2.2830
Epoch: [3/30], Valid loss: 2.2013, Valid accuracy: 0.1929
Saving the best model at 2 epochs!
Epoch: [4/30], Train loss: 2.2026
Epoch: [4/30], Valid loss: 2.0716, Valid accuracy: 0.2487
Saving the best model at 3 epochs!
Epoch: [5/30], Train loss: 2.1279
Epoch: [5/30], Valid loss: 1.9948, Valid accuracy: 0.2640
Saving the best model at 4 epochs!
Epoch: [6/30], Train loss: 2.1007
Epoch: [6/30], Valid loss: 1.9407, Valid accuracy: 0.3249
Saving the best model at 5 epochs!
Epoch: [7/30], Train loss: 2.0670
Epoch: [7/30], Valid loss: 1.9217, Valid accuracy: 0.3096
Saving the best model at 6 epochs!
Epoch: [8/30], Train loss: 2.0387
Epoch: [8/30], Valid loss: 1.9618, Valid accuracy: 0.2893
Epoch: [9/30], Train loss: 2.0034
Epoch: [9/30], Valid loss: 1.7882, Valid accuracy: 0.3604
Saving the best model at 8 epochs!
Epoch: [10/30], Train loss: 1.9669
Epoch: [10/30], Valid loss: 1.7608, Valid accuracy: 0.3807
Saving the best model at 9 epochs!
Epoch: [11/30], Train loss: 1.9212
Epoch: [11/30], Valid loss: 1.7428, Valid accuracy: 0.3604
Saving the best model at 10 epochs!
Epoch: [12/30], Train loss: 1.9497
Epoch: [12/30], Valid loss: 1.7381, Valid accuracy: 0.3401
Saving the best model at 11 epochs!
Epoch: [13/30], Train loss: 1.8578
Epoch: [13/30], Valid loss: 1.7946, Valid accuracy: 0.3350
Epoch: [14/30], Train loss: 1.8934
Epoch: [14/30], Valid loss: 1.6822, Valid accuracy: 0.3959
Saving the best model at 13 epochs!
Epoch: [15/30], Train loss: 1.8459
Epoch: [15/30], Valid loss: 1.6475, Valid accuracy: 0.4416
Saving the best model at 14 epochs!
Epoch: [16/30], Train loss: 1.8433
Epoch: [16/30], Valid loss: 1.6429, Valid accuracy: 0.3503
```

(continues on next page)

(continued from previous page)

```
Saving the best model at 15 epochs!
Epoch: [17/30], Train loss: 1.8358
Epoch: [17/30], Valid loss: 2.0232, Valid accuracy: 0.3046
Epoch: [18/30], Train loss: 1.8106
Epoch: [18/30], Valid loss: 1.6712, Valid accuracy: 0.3655
Epoch: [19/30], Train loss: 1.7393
Epoch: [19/30], Valid loss: 2.2497, Valid accuracy: 0.2741
Epoch: [20/30], Train loss: 1.7158
Epoch: [20/30], Valid loss: 1.5637, Valid accuracy: 0.4162
Saving the best model at 19 epochs!
Epoch: [21/30], Train loss: 1.7603
Epoch: [21/30], Valid loss: 1.4845, Valid accuracy: 0.5178
Saving the best model at 20 epochs!
Epoch: [22/30], Train loss: 1.7305
Epoch: [22/30], Valid loss: 1.6282, Valid accuracy: 0.3503
Epoch: [23/30], Train loss: 1.7213
Epoch: [23/30], Valid loss: 1.4270, Valid accuracy: 0.5381
Saving the best model at 22 epochs!
Epoch: [24/30], Train loss: 1.7064
Epoch: [24/30], Valid loss: 1.6344, Valid accuracy: 0.3655
Epoch: [25/30], Train loss: 1.6306
Epoch: [25/30], Valid loss: 1.3873, Valid accuracy: 0.5330
Saving the best model at 24 epochs!
Epoch: [26/30], Train loss: 1.7458
Epoch: [26/30], Valid loss: 1.4194, Valid accuracy: 0.5076
Epoch: [27/30], Train loss: 1.6578
Epoch: [27/30], Valid loss: 1.7264, Valid accuracy: 0.3604
Epoch: [28/30], Train loss: 1.6247
Epoch: [28/30], Valid loss: 1.4872, Valid accuracy: 0.5076
Epoch: [29/30], Train loss: 1.6642
Epoch: [29/30], Valid loss: 1.3975, Valid accuracy: 0.4772
Epoch: [30/30], Train loss: 1.6681
Epoch: [30/30], Valid loss: 1.6023, Valid accuracy: 0.4213
```

## 9.5 Evaluation

Collect the trained model's predictions for the test set. Chunk-level predictions are aggregated to make song-level predictions.

```
# Load the best model
S = torch.load('best_model.ckpt')
cnn.load_state_dict(S)
print('loaded!')

# Run evaluation
cnn.eval()
y_true = []
y_pred = []

with torch.no_grad():
    for wav, genre_index in test_loader:
        wav = wav.to(device)
        genre_index = genre_index.to(device)
```

(continues on next page)

(continued from previous page)

```
        # reshape and aggregate chunk-level predictions
        b, c, t = wav.size()
        logits = cnn(wav.view(-1, t))
        logits = logits.view(b, c, -1).mean(dim=1)
        _, pred = torch.max(logits.data, 1)

        # append labels and predictions
        y_true.extend(genre_index.tolist())
        y_pred.extend(pred.tolist())
```

```
loaded!
```

Finally, we can assess the performance and visualize a confusion matrix for better understanding.

```
import seaborn as sns
from sklearn.metrics import confusion_matrix


accuracy = accuracy_score(y_true, y_pred)
cm = confusion_matrix(y_true, y_pred)
sns.heatmap(cm, annot=True, xticklabels=GTZAN_GENRES, yticklabels=GTZAN_GENRES, cmap=
↪'YlGnBu')
print('Accuracy: %.4f' % accuracy)
```

```
Accuracy: 0.5414
```

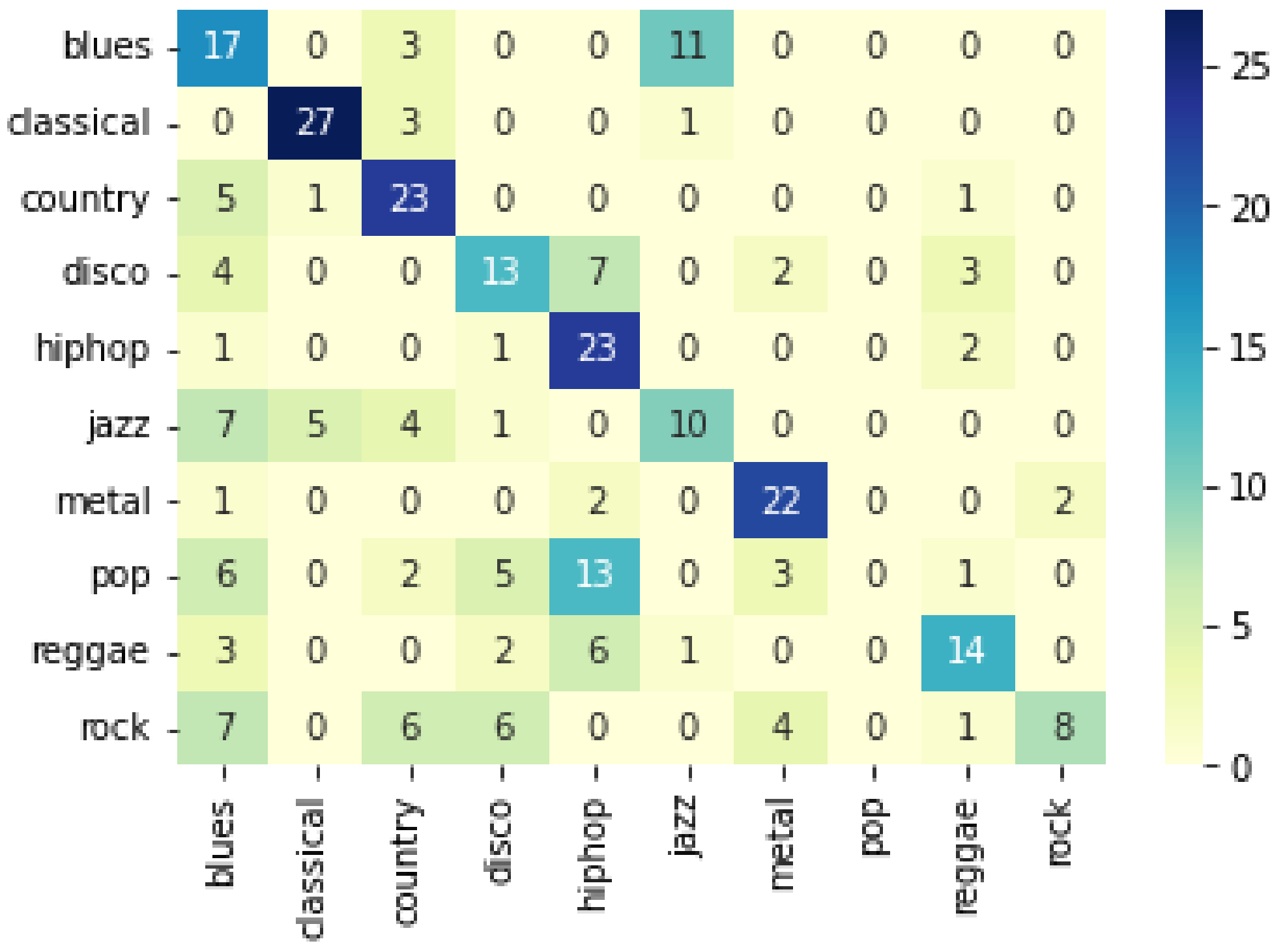

**Tip:** In this tutorial, we did not use any high-level library for more understandable implementation. We highly recom-mend checking the following libraries for simplified implementation:

- PyTorch Lightning
- Ignite
- Hydra

# Part III

# Semi-supervised Learning

CHAPTER

TEN

# BEYOND SUPERVISION

In previous sections, we learned how to facilitate music classification in a data-driven fashion when we have labeled music audio. The main motivation of building music classification models is to save human efforts of manually labeling musical attributes. However, modern deep learning models are data-hungry. As a result, ironically, we end up demanding a large amount of human effort again during dataset creation process. In the next two chapters, we explore training methods beyond supervised learning so that we can alleviate this irony.

In a real-world scenario, we may have a large-scale music library, but only a few of them might have manual labels. Also, sometimes, there is a discrepancy between the taxonomies of the existing training data and the target task. What can we do? As you'll learn from this chapter, one can adopt transfer learning which takes advantage of pretrained models. Semi- or self-supervised approaches can be another solution since they enable us to utilize abundant unlabeled data.

CHAPTER

ELEVEN

# TRANSFER LEARNING

The core idea of transfer learning is to i) learn knowledge from solving a problem (source task) and ii) apply the knowledge to solve other, relevant problems (target task) [TS10]. For example, if the model is able to perform instrument identification (source task), the learned knowledge would be useful to solve music genre classification (target task) since as the underlying concepts in music genres are related with instrumentation. The assumption is that although the source and target tasks are not identical, if the dataset for source task is much larger than target task, transferring could lead to a better performance.

In practice, i) we usually fit our model to solve a source task and ii) further optimize the pretrained model to solve the target task. In the second process, all or a part of learned parameters are updated. For example, the authors of [CFSC17b] pretrained a music tagging model on the Million Song Dataset. Then the model was transferred to solve downstream tasks such as genre classification, emotion recognition, audio event classification, etc .

Music information can be classified into three categories based on the nature of the metadata elaboration: editorial, cultural, and acoustic [Pac05]. The aforementioned transfer learning experiment takes advantage of music tags in the Million Song Dataset which are mostly related to acoustic information (e.g., genre, instrument). However, those music tags still relies on human effort of labeling. Instead of targeting the acoustic information, we can also design the source task to predict editorial or cultural metadata.

## 11.1 Pretext using editorial information

Editorial metadata is, by definition, obtained by the editor. Written information of the song such as artist names, album names, song titles, or released dates can be included. As we can distinguish artists by their acoustic characteristics, a previous work proposed to use artist classification as its pretext task for music representation learning and transferred the learned representation to solve downstream music genre classification tasks [PLP+17].

However, there are millions of artists which makes the pretext task to be unrealistic when with large-scale music libraries. To alleviate this issue, following researchers proposed to use clusters of artists as the prediction target of the source task [KWSL18].

## 11.2 Pretext using cultural information

Cultural information is generated by the way music is perceived and consumed in the society. One well-known approach is to use collaborative filtering. Collaborative filtering models the interests of users from user-item interaction data. As shown in the figure below, a user-item interaction matrix can be decomposed into two matrices with a lower dimensionality using matrix factorization. They represent the embeddings of items and users, respectively.

A previous work trained a pretext music representation model by targeting this item (song) embeddings [VdODS13]. The learned representation can include rich acoustic information if the original user-item interaction dataset is large enough. This pretext (source task) is especially beneficial in industry where such a type of data is accessible.

CHAPTER

TWELVE

# SEMI-SUPERVISED LEARNING

In many realistic scenarios, we have limited labeled data and abundant unlabeled data. For example, in the Million Song Dataset (MSD), only 24% of them are labeled with at least one of the top-50 music tags. As a consequence, most the existing MSD tagging research discarded the 76% of the audio included in MSD.

Semi-supervised learning is a broad concept of a hybrid approach of supervised learning and semi-supervised learning. In detail, many variants have been proposed.

- In self-training, a teacher model is first trained with labeled data. Then the trained teacher model predicts the labels of unlabeled data. A student model is optimized to predict both the labels of labeled data and the pseudo-labels (the prediction by teacher) of unlabeled data [YJegouC+19].
- Consistency training constrains models to generate noise invariant predictions [SSP+03]. Unsupervised loss of consistency training is formalized as follow:
- Entropy regularization minimizes the entropy of the model's predictions. A straightforward implementation is to directly minimize the entropy of the predictions for unlabeled data [GB+05]. But this can be also achieved in an implicit manner by training with one-hot encoded pseudo-labels [L+13]. In this case, the model first makes a prediction using unlabeled data. The prediction is then modified to be an one-hot vector.
- Some previous works incorporate multiple semi-supervised approaches together [BCG+19], [XLHL20].
- Other semi-supervised methods includes graph-based approaches [ZGL03] and generative modeling [KMRW14].

In this section, we explore a specific semi-supervised approach: Noisy student training [XLHL20]. Noisy student training is a self-training process that constrains the student model to be noise-invariant.

**Warning:** In some papers, SSL stands for semi-supervised learning, but others use the acronym to represent self-supervised learning. To avoid confusion, we do not use abbreviations of semi- and self-supervised learning in this book.

## 12.1 Noisy student training

Noisy student training is a kind of teacher-student learning (self-training) [XLHL20]. In the typical teacher-student learning, a teacher model is first trained with labeled data in a supervised scheme. Then, a student model is trained to resemble the teacher model by learning to predict the pseudo-labels, the prediction of the teacher model. What makes noisy student training special is to add noise to the input.

**Tip:** The pseudo-labels can be continuous (soft) vectors or one-hot encoded (hard) vectors. The original paper reported that both soft and hard labels worked, but soft labels worked slightly better for out-of-domain unlabeled data.

Now, a student model can be optimized using both labels (pseudocode line 10-11, follow orange lines) and pseudo-labels (pseudocode line 12-15, follow blue lines). In this process, strong data augmentation is applied for unlabeled data (pseudocode line 13) and this makes the student model perform beyond the teacher model. The current state-of-the-art music tagging models (short-chunk ResNet and Music tagging transformer) can be further improved by using the noisy student training.

**Tip:** A trained student model can be another teacher model to iterate the noisy student training process. However, different from the results in image classification, no significant performance gain was observed in music tagging with the MSD.

## 12.2 Knowledge expansion and distillation

In noisy student training, the size of the student model is not necessarily smaller than the size of the teacher model. As a student model is exposed with larger-scale data with more difficult environments (noise), it can learn more information than the teacher model. One can interpret this method as knowledge expansion [XLHL20].

On the other hand, we can also reduce the size of the student model for the sake of model compression. This process is called knowledge distillation and it is suitable for applications with less computing power [KR16].

As shown in the table, both Short-chunk ResNet and Music tagging transformer can be improved with data augmentation (DA). Then the models are further improved with noisy student training in both knowledge expansion (KE) and knowledge distillation (KD) manners [WCS21].

**Tip:**

- Tensorflow implementation of noisy student training [Github]
- PyTorch implementation of noisy student training for music tagging [Github]

# Part IV

# Self-supervised Learning

CHAPTER

# THIRTEEN

# INTRODUCTION

Supervised learning of deep neural networks has seen many breakthroughs in music information retrieval. Across tasks from music classification to source separation and music recommendation, large neural networks that use a supervised optimization scheme have reached state-of-the-art results by using large, human-annotated datasets.

These large parameterised networks are data-hungry; they require many independent and identically distributed (i.i.d.) data points to generalize well in the task they are trained on. Especially in music, it can be hard to manually annotate, and the annotations often suffer from a single source of truth. There is no single oracle: depending on the context, music theoretical background and cultural background, a song's analysis can yield different results. This was the motivation of the previous chapter, semi-supervised learning.

In this chapter, we introduce another training strategy that attempts to learn from unlabeled i.i.d. data points: self-supervised learning. We first consider the term pre-training, how it is linked to self-supervised learning and its inherent caveats. Then, we introduce the concept of self-supervised learning and review some key papers in this line of work at this moment of writing (November 2021).

## 13.1 Pre-training

The weights of neural networks have to be initialized before commencing the training on a target task. This initialization currently does not hold any apriori knowledge on the task at hand. Gaining the weights that belong to a global minimum in our task is not an easy feat currently, especially for smaller, annotated datasets.

One solution is to pre-train our network on a large annotated dataset to help it steer the optimization scheme in the right direction. This is called pre-training and gaining popularity more and more.

However, classical pre-training have many caveats. Music datasets can be biased towards certain concepts, which will be reflected in the trained model – and to its application to the target task. The source task and the target task won't be the same, and the negative effective of this difference is hard to predict.

Another challenge is that there are some musical concepts that rarely appear in labeled datasets. This means we would need a huge dataset to have enough representations of those rarity.

## 13.2 Self-supervised learning

In self-supervised learning, we obtain a supervisory signal from the data by leveraging its underlying structure. Generally, this can be done on the data itself, or in the space of the data representations. For example, we can predict part of the data from other parts of the data. Or, we can predict the future from events that occurred in the past. In short - we let the model predict the occluded from the visible while we control what to occlude. This has been a very popular approach in language modeling in the past 10 years.

Within the last two years, many different self-supervised methods have been proposed, in particular for vision tasks, and resulted in great improvements over supervised methods when labeled datapoints are scarce. Recently, they even started to perform better than equivalent networks trained in a supervised manner.

**Note: Learning scheme**

It is worth visiting the general learning scheme of this line of work:

1. First, we pre-train a neural network using a self-supervised objective (the pretext task).
2. In order to test the effectiveness of the learned representations, the pre-trained networks' weights are "frozen", and;
3. A linear evaluation on (part of) the supervised dataset is performed to compare against existing benchmarks.

The linear evaluation scheme involves training a supervised linear classifier (a fully-connected layer followed by a softmax) using the representations extracted from the self-supervised network, and (a subset of) the labels associated with the data.

## 13.3 Should I use self-supervised learning?

Self-supervised learning can be beneficial in the following situations:

- The amount of labeled data available is scarce
- You do not want to sacrifice the size and the expressivity of your model.
- You need general-purpose representations that are not less tightly coupled with a single use case.

You should take these considerations into account:

- A pre-trained model will have weights that reflect (and augment!) the biases embedded in a dataset.
- The pretext task used as the self-supervised learning objective is important to analyze and reflect on, as it can yield many assumptions for the downstream task.

CHAPTER

FOURTEEN

# METHODS FOR SELF-SUPERVISED LEARNING

## 14.1 Contrastive Learning

Contrastive learning is a method that describes learning representations by modeling similarity from natural variations of data. It is often presented in the following stages:

1. Encode different "views" from natural variations of a single example.
2. Train a model with metric learning using the representations from the encoder(s).
3. Use the representation of the encoder by applying another classifier for the downstream task. For evaluation, linear regression is used usually so that we can focus on the performance of the pre-trained encoder rather than the that of the added classifier.

## 14.2 Contrastive Predictive Coding

Contrastive predictive coding (CPC) was introduced by Aäron van den Oord et al. (2018) [OLV18] as a universal framework of representation learning. The data can be an image, in which neighbouring *patches* usually share spatial information locally. In the case of speech signals, it could be the phonemes that should be similar with the neighbors. Conversely, on a more global level, we expect a different pattern. For example, the chorus of a song is expected to repeat in another part of our audio signal.

In CPC, these related observations are mapped similarly as a representation in a latent space. Their hypothesis is that predictions of related observations are often conditionally dependent on similar, high-level pieces of latent information.

To test their hypothesis, they propose the following:

1. First, complex natural data, such as images and audio, are compressed into a latent embedding space. This makes it easier to model the predictions of related observations.
2. A more expressive *(read: larger, more powerful)* autoregressive model uses the representations in this latent space to make predictions for future observations. These observations are mapped to the corresponding representation.
3. The InfoNCE loss uses a cross-entropy loss to quantify how well the model can classify these future representations from a set of unrelated, *"negative"* examples [CITATION]. This loss is inspired by Noise Contrastive Estimation [CITATION].

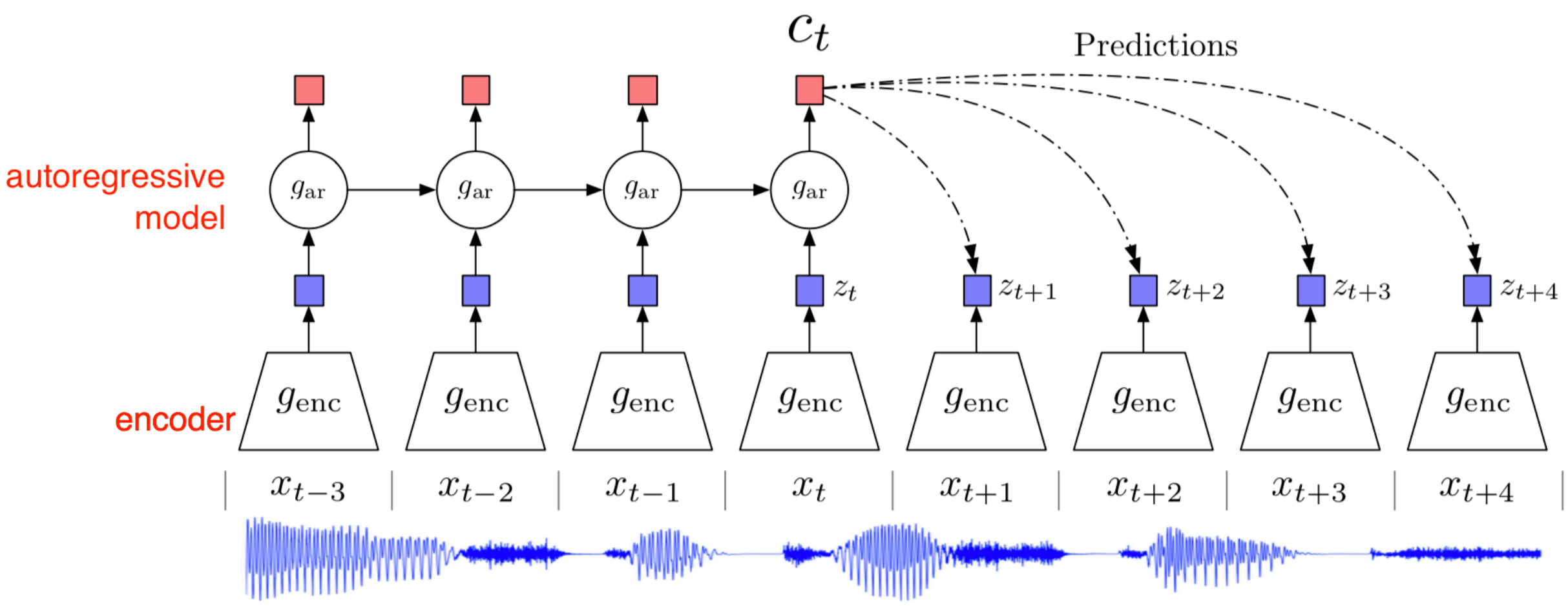

**Mini-batch composition**

In CPC, a mini-batch is made of $N$ examples that are chosen randomly from the full training dataset. Within this batch, a single positive example (also known as the *anchor* sample) and $N-1$ negative examples are used to compute the cross-entropy loss for classifying the correct positive example. In this way, the internal structure of the data is leveraged to obtain a loss signal that we can backpropagate.

## 14.3 Momentum Contrast (MoCO)

Momentum Contrast was proposed in [HFW+20]. It was one of the first papers to close the gap between unsupervised and supervised learning approaches on some vision tasks. In MoCO, a dictionary of examples in the data are maintained as a queue. Each example in the mini-batch is encoded, and put in front of the queue, while the last item in the dictionary is subsequently dequeued. The pretext task used in MoCO is to define a contrastive loss on the query and the keys of the dictionary: a query matches the key when the query is an embedding of a different *view* of the same datapoint. For example: if the query is an embedding of the bassoon solo in Stravinsky's "Rite of Spring", it should match with the key that corresponds to the "Rite of Spring". The encoded query should be similar to its corresponding key, and dissimilar to other keys in the dictionary.

Training a Momentrum Contrast encoder is done with positive and negative pairs of examples in a mini-batch. The positive example pairs are made of queries that correspond to keys of the current mini-batch. The negative pairs are queries of the current mini-batch and keys from past mini-batches.

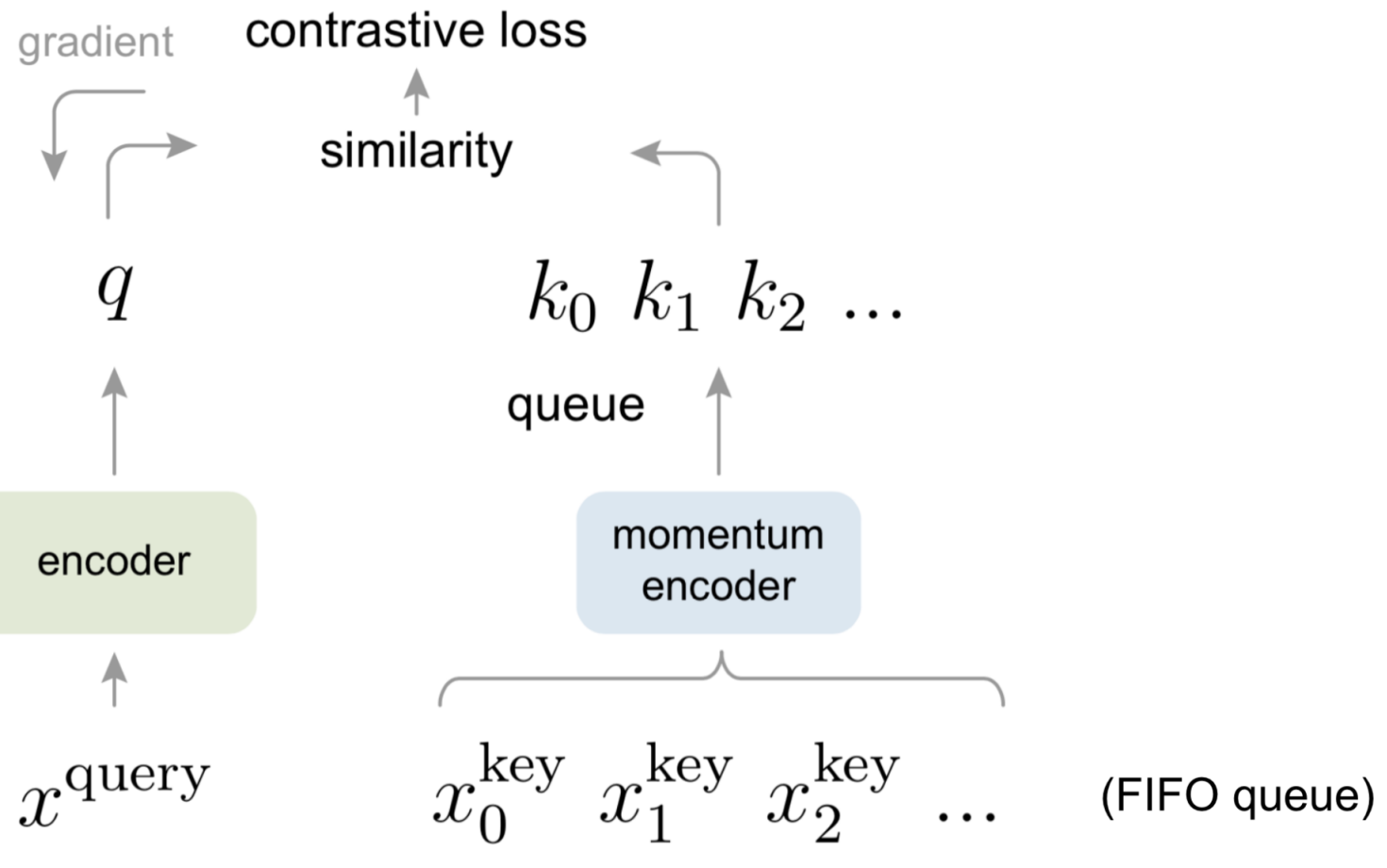

The keys are encoded by a *"slowly progressing"* encoder, because the dictionary's keys are drawn over multiple mini-batches. This encoder is implemented as a momentum-based moving average. We therefore have two encoders: an encoder for the queries and a momentum-encoder for the keys. The main difference between these two encoders is in the way they are updated. The query encoder is updated by backpropagation while the momentum encoder is updated by a linear interpolation of the query and the momentum encoder.

**Tip**

An advantage of Momentum Contrast is that the batch size is not related to the number of negative examples candidates in the dictionary lookup. Even for smaller batch sizes, the performance of Momentum Contrast is consistent. This can be especially useful for (raw) audio, for which it is often harder to compute with larger mini-batch sizes due to GPU memory constraints.

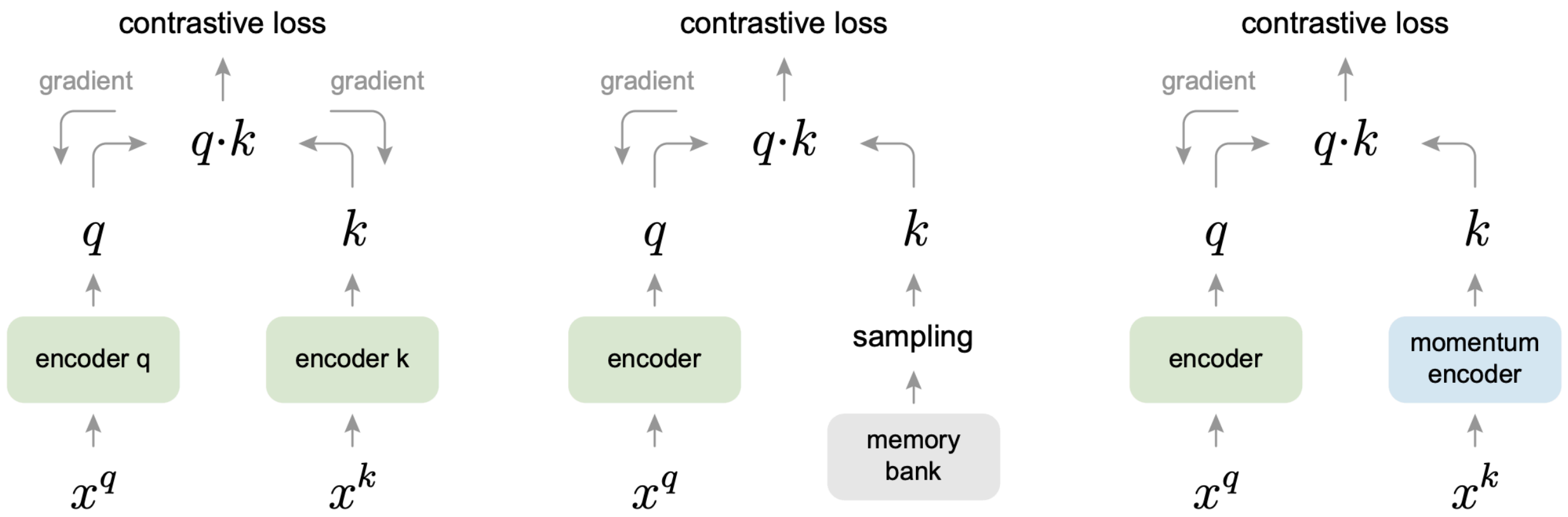

(a) end-to-end (b) memory bank (c) MoCo

## 14.4 SimCLR

SimCLR was introduced in [CKNH20] as a simple contrastive learning approach to learn strong visual representations. It leverages strong image data augmentations, large batch sizes, a single large encoder and a simple contrastive loss to pre-train an encoder that learns effective representations. These representations are used to train very effective linear classifiers in various downstream image classification tasks.

For each image example in the mini-batch, two augmented (but correlated!) views are taken. This is done by a series of data augmentations that are applied randomly to each example. This will naturally yield $2N$ datapoints per mini-batch. Each of these augmented views are then embedded using a standard ResNet encoder network. While these representations are used during linear evaluation, during the pre-training stage these representations are projected to a different latent space by a small linear layer on which the contrastive loss is computed.

During pre-training, the network only learns from the contrastive loss: the labels are only used during the linear evaluation phase (see the previous section for more details).

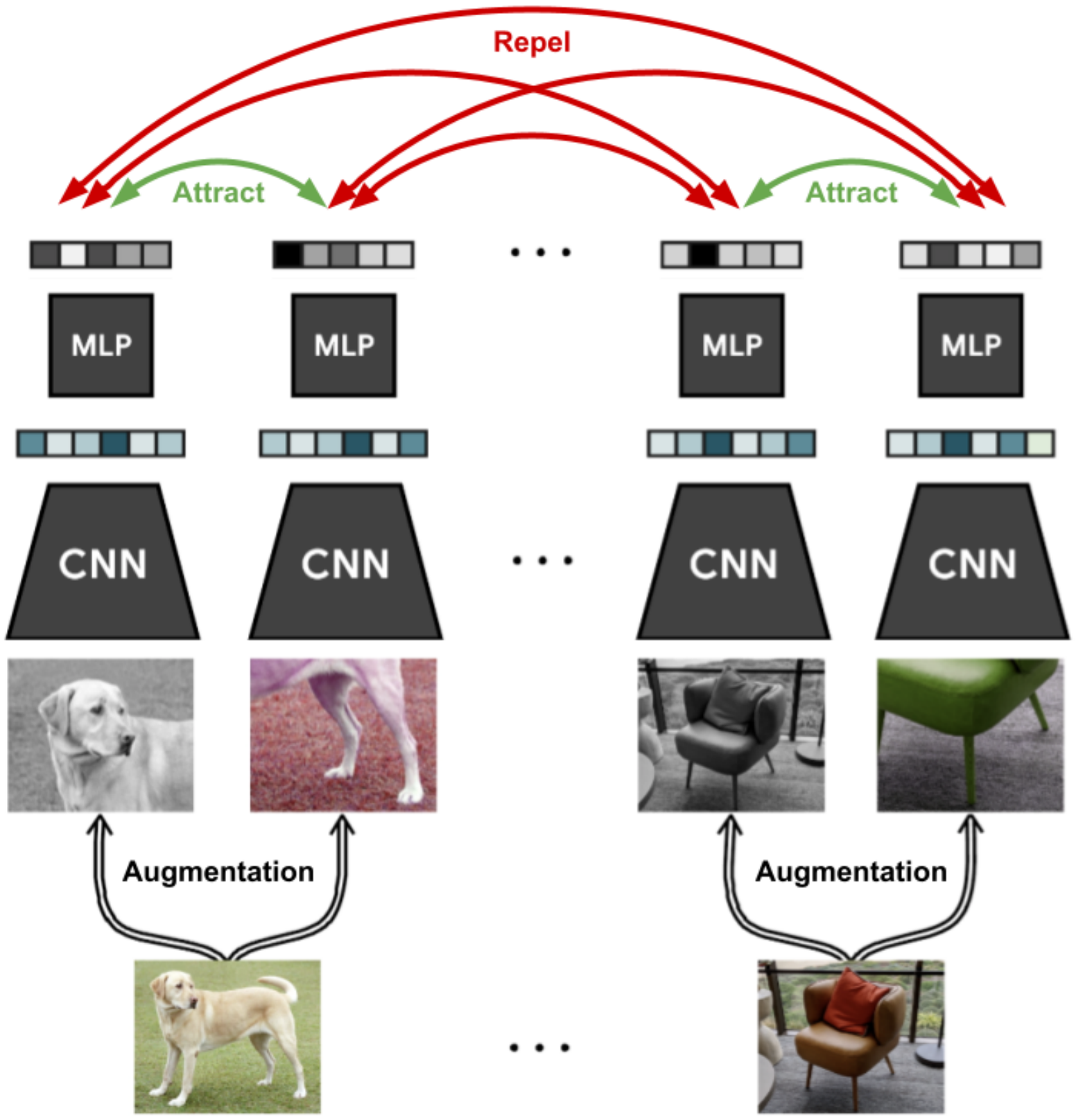

## 14.5 Contrastive Losses

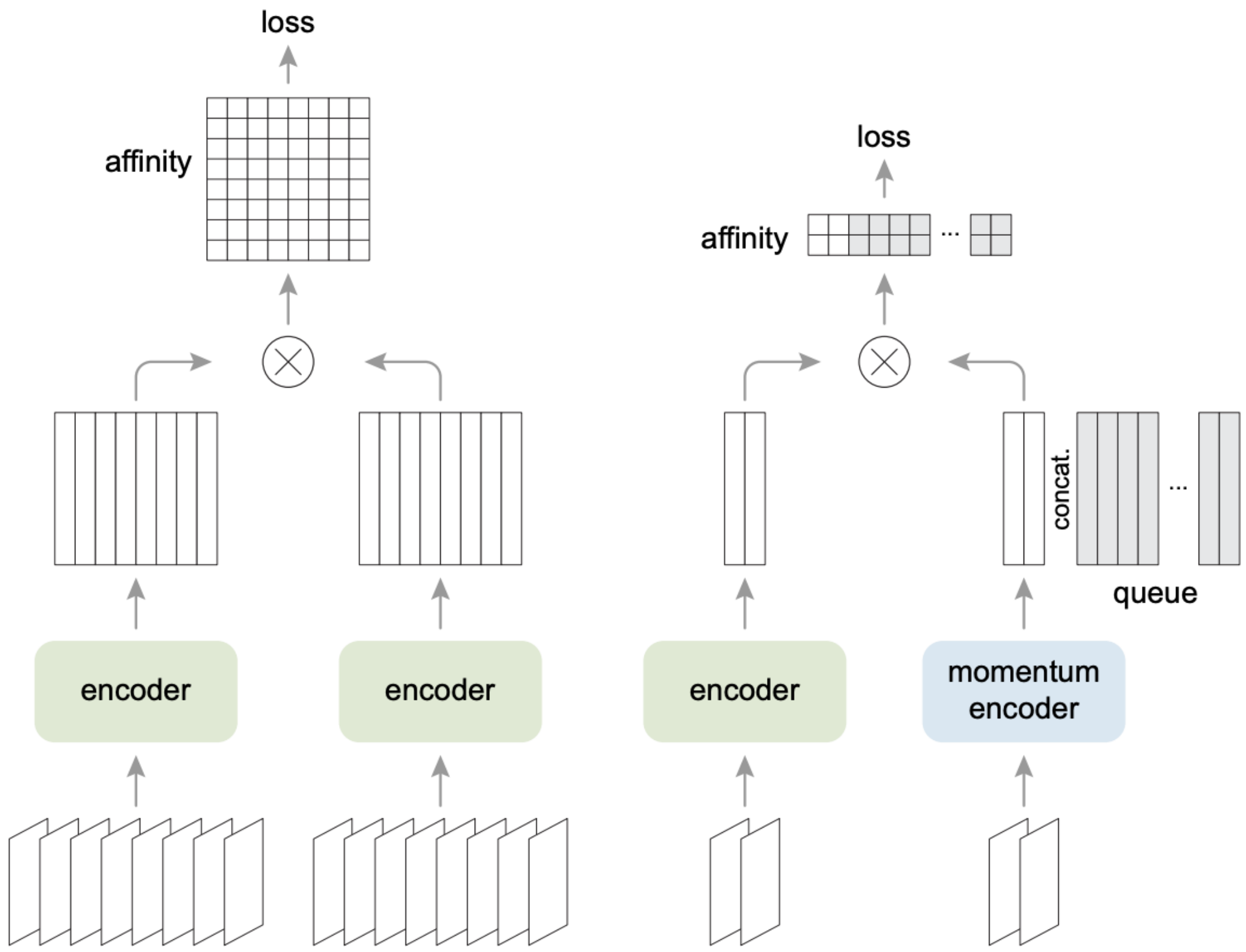

(a) end-to-end (b) Momentum Contrast

Many contrastive learning methods use a variant of a contrastive loss function. The contrastive loss function was first introduced in Noise Contrastive Estimation [GHyvarinen10] and subsequently the InfoNCE loss from Contrastive Predictive Coding [OLV18].

This loss can be minimized using a variety of methods, which mostly differ in the way they keep track of the keys of data examples. In SimCLR [CKNH20], a single batch consists both of "positive" and "negative" pairs, which act as "keys" to the original examples. These are updated end-to-end by back-propagation. To increase the complexity of the contrastive learning task, it requires a large batch size to contain more negative examples. In Momentum Contrast, the negative examples' keys are maintained in a queue. Note that only the queries and the positive keys in a single batch are encoded.

$$\mathcal{L}_{q,\, k^+,\, \{k^-\}} = -\log \frac{\exp(q \cdot k^+ / \tau)}{\exp(q \cdot k^+ / \tau) + \sum_{k^-} \exp(q \cdot k^- / \tau)}$$

## 14.6 PASE

PASE was proposed in [PRS+19]. It demonstrated that useful representations for speech recognition can be learned by defining multiple pretext tasks that jointly optimize an encoder neural network. The encoder distributes the representations of the input data to multiple, small feed-forward neural networks (called *workers*) that jointly solve different pretext tasks. Each worker is composed of a single hidden layer, and either solves a regression or binary classification task. These smaller feed-forward layers are chosen because the emphasis on learning the more expressive representations is put on the larger encoder, i.e., the encoder should learn more high-level features that can be used by the *worker* networks to help solve their tasks. After pre-training the network in a self-supervised manner, the learned representations are evaluated in the task of speaker recognition, emotion recognition and phoneme recognition.

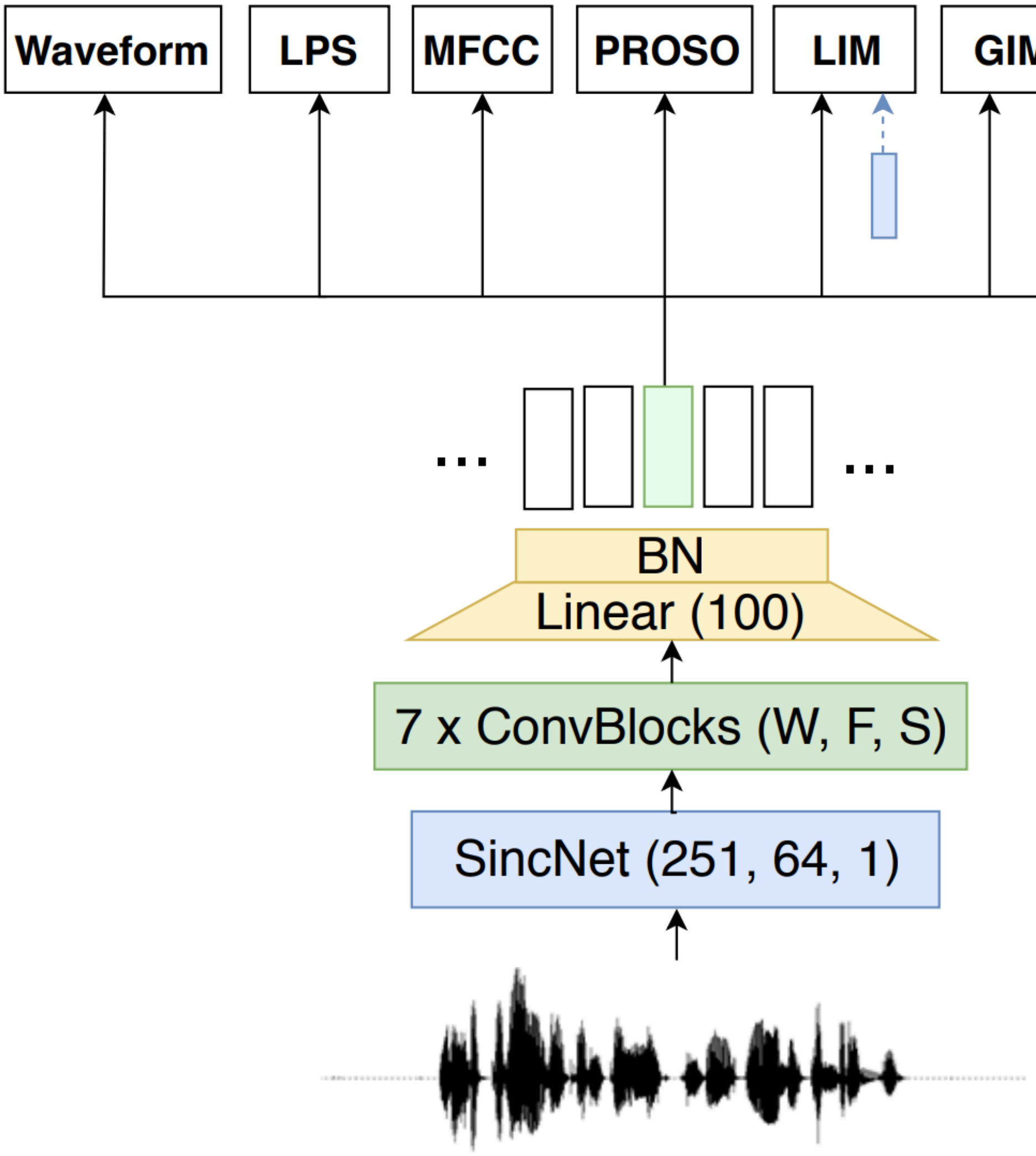

The improved version of PASE, which was called PASE+, uses a set of audio data augmentations to improve the robustness of the learned representations for the downstream task.

## 14.7 More papers on self-supervised learning

The following is a short list of important papers in self-supervised learning, of which a few are discussed more in-depth in this tutorial:

| Paper | Year | Tasks |
|---|---|---|
| Representation Learning with Contrastive Predictive Coding | 2018 | Speech, images, text, reinforcement learning |
| Noise-contrastive estimation: A new estimation principle for unnormalized statistical models | 2014 | Theoretical |
| Unsupervised Visual Representation Learning by Context Prediction | 2015 | Images |
| Momentum Contrast for Unsupervised Visual Representation Learning | 2019 | Images |
| wav2vec: Unsupervised Pre-training for Speech Recognition | 2019 | Speech recognition |
| Learning Problem-agnostic Speech Representations from Multiple Self-supervised Tasks | 2019 | Speech recognition |
| Bootstrap your own latent: A new approach to self-supervised Learning | 2020 | Images |
| A Simple Framework for Contrastive Learning of Visual Representations | 2020 | Images |
| Contrastive learning of general-purpose audio representations | 2020 | Sound classification |
| Contrastive Learning of Musical Representations | 2021 | Music classification |
| Vector Quantized Contrastive Predictive Coding for Template-based Music Generation | 2021 | Music generation |

CHAPTER

# FIFTEEN

# PYTORCH TUTORIAL

## 15.1 CLMR

In the following examples, we will be taking a look at how Contrastive Learning of Musical Representations (Spijkervet & Burgoyne, 2021) uses self-supervised learning to learn powerful representations for the downstream task of music classification.

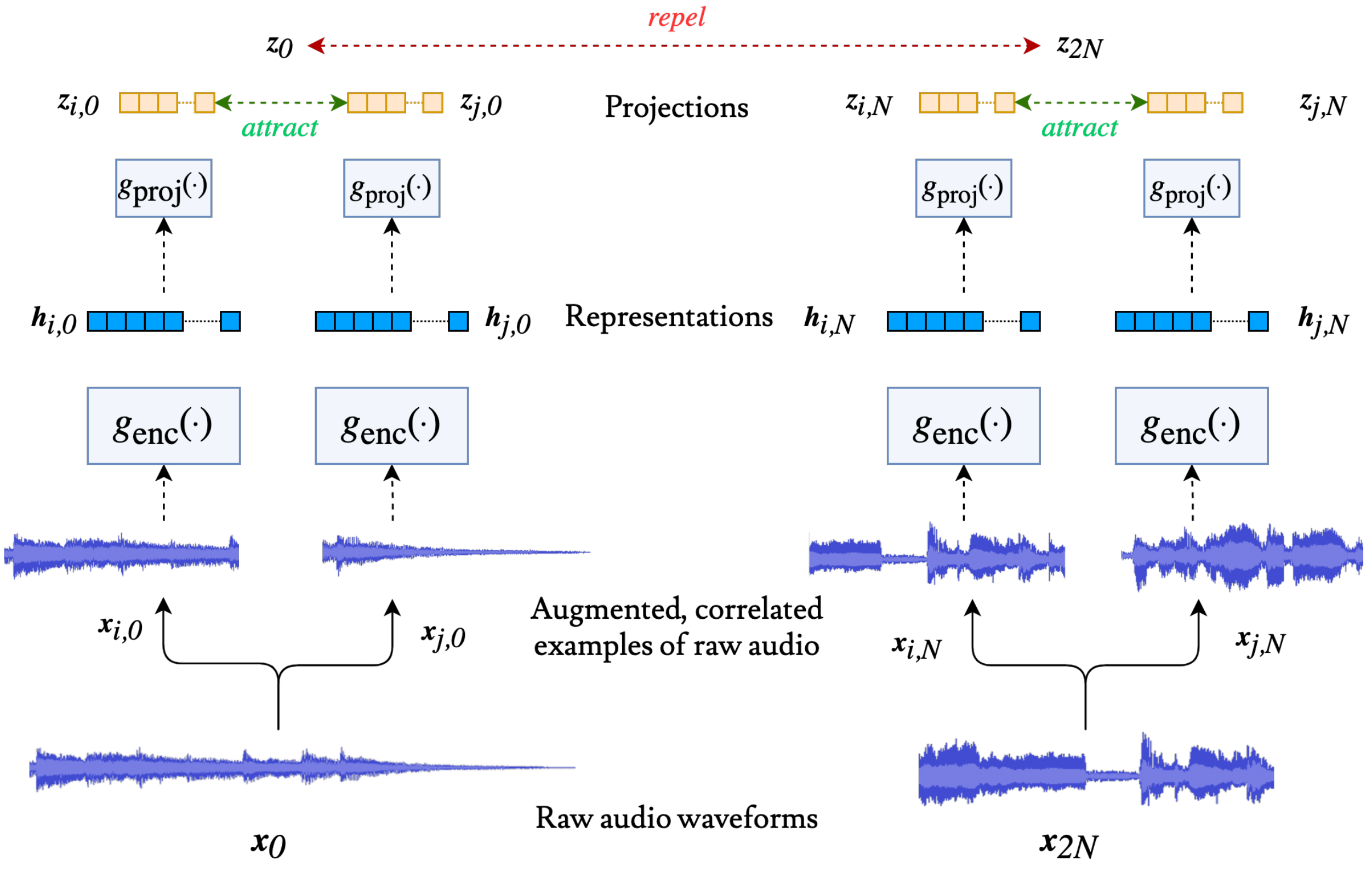

In the above figure, we transform a single audio example into two, distinct augmented views by processing it through a set of stochastic audio augmentations.

```
from argparse import Namespace

import torch
from tqdm import tqdm
```

(continues on next page)

(continued from previous page)

```
device = torch.device("cuda") if torch.cuda.is_available() else torch.device("cpu")
print(f"We are using the following the device to train: {device}")

# initialize an empty argparse Namespace in which we can store argumens for training
args = Namespace()


# every piece of audio has a length of 59049 samples
args.audio_length = 59049

# the sample rate of our audio
args.sample_rate = 22050
```

```
We are using the following the device to train: cpu
```

```
import os
import random

import numpy as np
import soundfile as sf
import torch
from torch.utils import data
from torchaudio_augmentations import (
    Compose,
    Delay,
    Gain,
    HighLowPass,
    Noise,
    PitchShift,
    PolarityInversion,
    RandomApply,
    RandomResizedCrop,
    Reverb,
)

GTZAN_GENRES = [
    "blues",
    "classical",
    "country",
    "disco",
    "hiphop",
    "jazz",
    "metal",
    "pop",
    "reggae",
    "rock",
]


class GTZANDataset(data.Dataset):
    def __init__(self, data_path, split, num_samples, num_chunks, is_augmentation):
        self.data_path = data_path if data_path else ""
        self.split = split
        self.num_samples = num_samples
        self.num_chunks = num_chunks
```

(continues on next page)

(continued from previous page)

```
        self.is_augmentation = is_augmentation
        self.genres = GTZAN_GENRES
        self._get_song_list()
        if is_augmentation:
            self._get_augmentations()

    def _get_song_list(self):
        list_filename = os.path.join(self.data_path, "%s_filtered.txt" % self.split)
        with open(list_filename) as f:
            lines = f.readlines()
        self.song_list = [line.strip() for line in lines]

    def _get_augmentations(self):
        transforms = [
            RandomResizedCrop(n_samples=self.num_samples),
            RandomApply([PolarityInversion()], p=0.8),
            RandomApply([Noise(min_snr=0.3, max_snr=0.5)], p=0.3),
            RandomApply([Gain()], p=0.2),
            RandomApply([HighLowPass(sample_rate=22050)], p=0.8),
            RandomApply([Delay(sample_rate=22050)], p=0.5),
            RandomApply(
                [PitchShift(n_samples=self.num_samples, sample_rate=22050)], p=0.4
            ),
            RandomApply([Reverb(sample_rate=22050)], p=0.3),
        ]
        self.augmentation = Compose(transforms=transforms)

    def _adjust_audio_length(self, wav):
        if self.split == "train":
            random_index = random.randint(0, len(wav) - self.num_samples - 1)
            wav = wav[random_index : random_index + self.num_samples]
        else:
            hop = (len(wav) - self.num_samples) // self.num_chunks
            wav = np.array(
                [
                    wav[i * hop : i * hop + self.num_samples]
                    for i in range(self.num_chunks)
                ]
            )
        return wav

    def get_augmentation(self, wav):
        return self.augmentation(torch.from_numpy(wav).unsqueeze(0)).squeeze(0).
↪numpy()

    def __getitem__(self, index):
        line = self.song_list[index]

        # get genre
        genre_name = line.split("/")[0]
        genre_index = self.genres.index(genre_name)

        # get audio
        audio_filename = os.path.join(self.data_path, "genres", line)
        wav, fs = sf.read(audio_filename)
```

(continues on next page)

(continued from previous page)

```
        # adjust audio length
        wav = self._adjust_audio_length(wav).astype("float32")

        # data augmentation
        if self.is_augmentation:
            wav_i = self.get_augmentation(wav)
            wav_j = self.get_augmentation(wav)
        else:
            wav_i = wav
            wav_j = wav

        return (wav_i, wav_j), genre_index

    def __len__(self):
        return len(self.song_list)


def get_dataloader(
    data_path=None,
    split="train",
    num_samples=22050 * 29,
    num_chunks=1,
    batch_size=16,
    num_workers=0,
    is_augmentation=False,
):
    is_shuffle = True if (split == "train") else False
    batch_size = batch_size if (split == "train") else (batch_size // num_chunks)
    data_loader = data.DataLoader(
        dataset=GTZANDataset(
            data_path, split, num_samples, num_chunks, is_augmentation
        ),
        batch_size=batch_size,
        shuffle=is_shuffle,
        drop_last=False,
        num_workers=num_workers,
    )
    return data_loader
```

```
args.batch_size = 48


train_loader = get_dataloader(
    data_path="../../codes/split",
    split="train",
    is_augmentation=True,
    num_samples=59049,
    batch_size=args.batch_size,
)
iter_train_loader = iter(train_loader)
(train_wav_i, _), train_genre = next(iter_train_loader)

valid_loader = get_dataloader(
    data_path="../../codes/split",
    split="valid",
```

(continues on next page)

(continued from previous page)

```
    num_samples=args.audio_length,
    batch_size=args.batch_size,
)
test_loader = get_dataloader(
    data_path="../../codes/split",
    split="test",
    num_samples=args.audio_length,
    batch_size=args.batch_size,
)
iter_test_loader = iter(test_loader)
(test_wav_i, _), test_genre = next(iter_test_loader)
print("training data shape: %s" % str(train_wav_i.shape))
print("validation/test data shape: %s" % str(test_wav_i.shape))
print(train_genre)
```

### 15.1.1 Audio Data Augmentations

Now, let's apply a series of transformations, each applied with an independent probability:

- Crop
- Filter
- Reverb
- Polarity
- Noise
- Pitch
- Gain
- Delay

```
import torchaudio
from torchaudio_augmentations import (
    RandomApply,
    ComposeMany,
    RandomResizedCrop,
    PolarityInversion,
    Noise,
    Gain,
    HighLowPass,
    Delay,
    PitchShift,
    Reverb,
)
```

```
args.transforms_polarity = 0.8
args.transforms_filters = 0.6
args.transforms_noise = 0.1
args.transforms_gain = 0.3
args.transforms_delay = 0.4
args.transforms_pitch = 0.4
args.transforms_reverb = 0.4
```

(continues on next page)

(continued from previous page)

```
train_transform = [
    RandomResizedCrop(n_samples=args.audio_length),
    RandomApply([PolarityInversion()], p=args.transforms_polarity),
    RandomApply([Noise()], p=args.transforms_noise),
    RandomApply([Gain()], p=args.transforms_gain),
    RandomApply([HighLowPass(sample_rate=args.sample_rate)], p=args.transforms_
↪filters),
    RandomApply([Delay(sample_rate=args.sample_rate)], p=args.transforms_delay),
    RandomApply([PitchShift(n_samples=args.audio_length, sample_rate=args.sample_
↪rate)], p=args.transforms_pitch),
    RandomApply([Reverb(sample_rate=args.sample_rate)], p=args.transforms_reverb),
]
train_loader.augmentation = Compose(train_transform)
```

Remember, always take a moment to listen to the data that you will give to your model! Let's listen to three examples from our dataset, on which a series of stochastic audio data augmentations are applied:

```
from IPython.display import Audio

for idx in range(3):
    print(f"Iteration: {idx}")
    (x_i, x_j), y = train_loader.dataset[0]

    print("Positive pair: (x_i, x_j):")
    display(Audio(x_i, rate=args.sample_rate))
    display(Audio(x_j, rate=args.sample_rate))
```

```
Iteration: 0
Positive pair: (x_i, x_j):
```

```
<IPython.lib.display.Audio object>
```

```
<IPython.lib.display.Audio object>
```

```
Iteration: 1
Positive pair: (x_i, x_j):
```

```
<IPython.lib.display.Audio object>
```

```
<IPython.lib.display.Audio object>
```

```
Iteration: 2
Positive pair: (x_i, x_j):
```

```
<IPython.lib.display.Audio object>
```

```
<IPython.lib.display.Audio object>
```

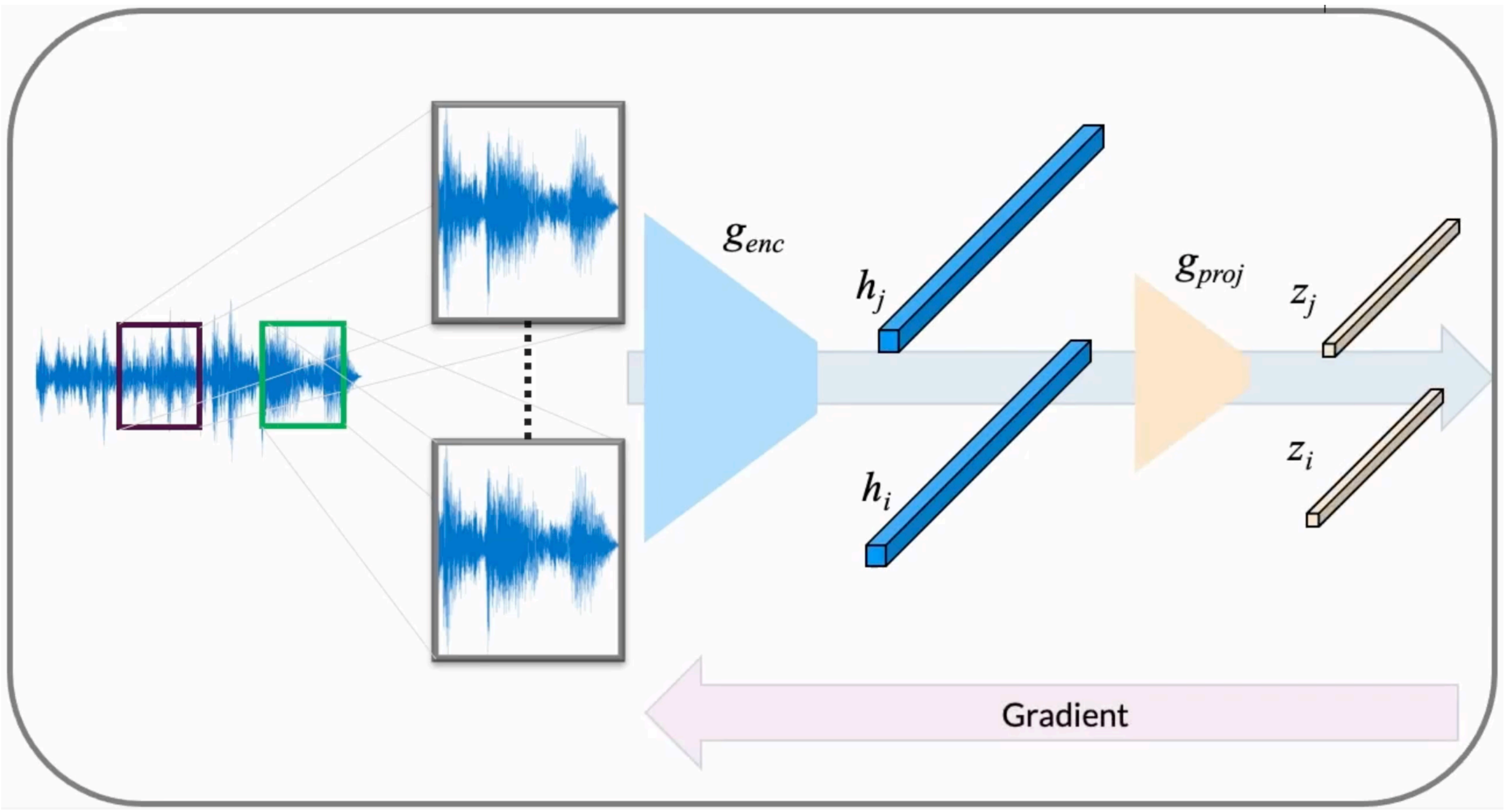

## 15.2 SampleCNN Encoder

First, let us begin with initializing our feature extractor. In CLMR, we chose the SampleCNN encoder to learn high-level features from raw pieces of audio. In the above figure, this encoder is denoted as $g_{enc}$. The last fully connected layer will be removed, so that we obtain an expressive final vector on which we can compute our contrastive loss.

```
import torch.nn as nn


class SampleCNN(nn.Module):
    def __init__(self, strides, supervised, out_dim):
        super(SampleCNN, self).__init__()

        self.strides = strides
        self.supervised = supervised
        self.sequential = [
            nn.Sequential(
                nn.Conv1d(1, 128, kernel_size=3, stride=3, padding=0),
                nn.BatchNorm1d(128),
                nn.ReLU(),
            )
        ]

        self.hidden = [
            [128, 128],
            [128, 128],
            [128, 256],
            [256, 256],
```

(continues on next page)

(continued from previous page)

```
            [256, 256],
            [256, 256],
            [256, 256],
            [256, 256],
            [256, 512],
        ]

        assert len(self.hidden) == len(
            self.strides
        ), "Number of hidden layers and strides are not equal"
        for stride, (h_in, h_out) in zip(self.strides, self.hidden):
            self.sequential.append(
                nn.Sequential(
                    nn.Conv1d(h_in, h_out, kernel_size=stride, stride=1, padding=1),
                    nn.BatchNorm1d(h_out),
                    nn.ReLU(),
                    nn.MaxPool1d(stride, stride=stride),
                )
            )

        # 1 x 512
        self.sequential.append(
            nn.Sequential(
                nn.Conv1d(512, 512, kernel_size=3, stride=1, padding=1),
                nn.BatchNorm1d(512),
                nn.ReLU(),
            )
        )

        self.sequential = nn.Sequential(*self.sequential)

        if self.supervised:
            self.dropout = nn.Dropout(0.5)
        self.fc = nn.Linear(512, out_dim)

    def initialize(self, m):
        if isinstance(m, (nn.Conv1d)):
            nn.init.kaiming_uniform_(m.weight, mode="fan_in", nonlinearity="relu")

    def forward(self, x):
        x = x.unsqueeze(dim=1)  # here, we add a dimension for our convolution.
        out = self.sequential(x)
        if self.supervised:
            out = self.dropout(out)

        out = out.reshape(x.shape[0], out.size(1) * out.size(2))
        logit = self.fc(out)
        return logit
```

Let's have a look at a printed version of SampleCNN:

```
# in the GTZAN dataset, we have 10 genre labels
args.n_classes = 10

encoder = SampleCNN(
    strides=[3, 3, 3, 3, 3, 3, 3, 3, 3],
```

(continues on next page)

(continued from previous page)

```
    supervised=False,
    out_dim=args.n_classes,
).to(device)
print(encoder)
```

```
SampleCNN(
  (sequential): Sequential(
    (0): Sequential(
      (0): Conv1d(1, 128, kernel_size=(3,), stride=(3,))
      (1): BatchNorm1d(128, eps=1e-05, momentum=0.1, affine=True, track_running_
↪stats=True)
      (2): ReLU()
    )
    (1): Sequential(
      (0): Conv1d(128, 128, kernel_size=(3,), stride=(1,), padding=(1,))
      (1): BatchNorm1d(128, eps=1e-05, momentum=0.1, affine=True, track_running_
↪stats=True)
      (2): ReLU()
      (3): MaxPool1d(kernel_size=3, stride=3, padding=0, dilation=1, ceil_
↪mode=False)
    )
    (2): Sequential(
      (0): Conv1d(128, 128, kernel_size=(3,), stride=(1,), padding=(1,))
      (1): BatchNorm1d(128, eps=1e-05, momentum=0.1, affine=True, track_running_
↪stats=True)
      (2): ReLU()
      (3): MaxPool1d(kernel_size=3, stride=3, padding=0, dilation=1, ceil_
↪mode=False)
    )
    (3): Sequential(
      (0): Conv1d(128, 256, kernel_size=(3,), stride=(1,), padding=(1,))
      (1): BatchNorm1d(256, eps=1e-05, momentum=0.1, affine=True, track_running_
↪stats=True)
      (2): ReLU()
      (3): MaxPool1d(kernel_size=3, stride=3, padding=0, dilation=1, ceil_
↪mode=False)
    )
    (4): Sequential(
      (0): Conv1d(256, 256, kernel_size=(3,), stride=(1,), padding=(1,))
      (1): BatchNorm1d(256, eps=1e-05, momentum=0.1, affine=True, track_running_
↪stats=True)
      (2): ReLU()
      (3): MaxPool1d(kernel_size=3, stride=3, padding=0, dilation=1, ceil_
↪mode=False)
    )
    (5): Sequential(
      (0): Conv1d(256, 256, kernel_size=(3,), stride=(1,), padding=(1,))
      (1): BatchNorm1d(256, eps=1e-05, momentum=0.1, affine=True, track_running_
↪stats=True)
      (2): ReLU()
      (3): MaxPool1d(kernel_size=3, stride=3, padding=0, dilation=1, ceil_
↪mode=False)
    )
    (6): Sequential(
      (0): Conv1d(256, 256, kernel_size=(3,), stride=(1,), padding=(1,))
```

(continues on next page)

(continued from previous page)

```
      (1): BatchNorm1d(256, eps=1e-05, momentum=0.1, affine=True, track_running_
↪stats=True)
      (2): ReLU()
      (3): MaxPool1d(kernel_size=3, stride=3, padding=0, dilation=1, ceil_
↪mode=False)
    )
    (7): Sequential(
      (0): Conv1d(256, 256, kernel_size=(3,), stride=(1,), padding=(1,))
      (1): BatchNorm1d(256, eps=1e-05, momentum=0.1, affine=True, track_running_
↪stats=True)
      (2): ReLU()
      (3): MaxPool1d(kernel_size=3, stride=3, padding=0, dilation=1, ceil_
↪mode=False)
    )
    (8): Sequential(
      (0): Conv1d(256, 256, kernel_size=(3,), stride=(1,), padding=(1,))
      (1): BatchNorm1d(256, eps=1e-05, momentum=0.1, affine=True, track_running_
↪stats=True)
      (2): ReLU()
      (3): MaxPool1d(kernel_size=3, stride=3, padding=0, dilation=1, ceil_
↪mode=False)
    )
    (9): Sequential(
      (0): Conv1d(256, 512, kernel_size=(3,), stride=(1,), padding=(1,))
      (1): BatchNorm1d(512, eps=1e-05, momentum=0.1, affine=True, track_running_
↪stats=True)
      (2): ReLU()
      (3): MaxPool1d(kernel_size=3, stride=3, padding=0, dilation=1, ceil_
↪mode=False)
    )
    (10): Sequential(
      (0): Conv1d(512, 512, kernel_size=(3,), stride=(1,), padding=(1,))
      (1): BatchNorm1d(512, eps=1e-05, momentum=0.1, affine=True, track_running_
↪stats=True)
      (2): ReLU()
    )
  )
  (fc): Linear(in_features=512, out_features=10, bias=True)
)
```

## 15.3 SimCLR

Since we removed the last fully connected layer, we are left with a $512$-dimensional feature vector. We would like to use this vector in our contrastive learning task. Therefore, we wrap our encoder in the objective as introduced by SimCLR: we project the final hidden layer of the encoder to a different latent space using a small MLP projector network. In the `forward` pass, we extract both the final hidden representation of our `SampleCNN` encoder (`h_i` and `h_j`), and the projected vectors (`z_i` and `z_j`).

```
class SimCLR(nn.Module):
    def __init__(self, encoder, projection_dim, n_features):
        super(SimCLR, self).__init__()

        self.encoder = encoder
```

(continues on next page)

(continued from previous page)

```
        self.n_features = n_features

        # Replace the fc layer with an Identity function
        self.encoder.fc = Identity()

        # We use a MLP with one hidden layer to obtain z_i = g(h_i) = W(2)σ(W(1)h_i)␣
↪where σ is a ReLU non-linearity.
        self.projector = nn.Sequential(
            nn.Linear(self.n_features, self.n_features, bias=False),
            nn.ReLU(),
            nn.Linear(self.n_features, projection_dim, bias=False),
        )

    def forward(self, x_i, x_j):
        h_i = self.encoder(x_i)
        h_j = self.encoder(x_j)

        z_i = self.projector(h_i)
        z_j = self.projector(h_j)
        return h_i, h_j, z_i, z_j


class Identity(nn.Module):
    def __init__(self):
        super(Identity, self).__init__()

    def forward(self, x):
        return x
```

## 15.4 Loss

Here, we apply an InfoNCE loss, as proposed by van den Oord et al. (2018) for contrastive learning. InfoNCE loss compares the similarity of our representations $z_i$ and $z_j$, to the similarity of $z_i$ to any other representation in our batch, and applies a softmax over the obtained similarity values. We can write this loss more formally as follows:

$$\ell_{i,j} = -\log \frac{\exp\left(\left(z_i, z_j\right)/\tau\right)}{\sum_{k=1}^{2N} \mathbb{1}_{[k \neq i]} \exp\left(\left(z_i, z_k\right)/\tau\right)}$$

The similarity metric is the cosine similarity between our representations:

$$\left(z_i, z_j\right) = \frac{z_i^\top \cdot z_j}{\parallel z_i \parallel \cdot \parallel z_j \parallel}$$

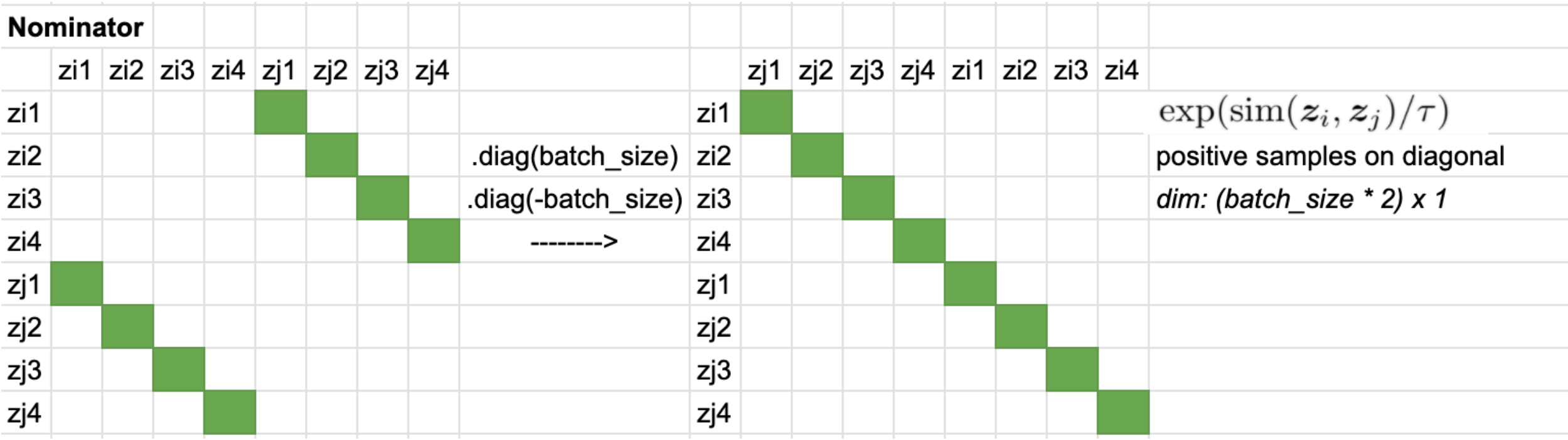

```
z = torch.cat((z_i, z_j), dim=0)`
```

```
sim = self.similarity_f(z.unsqueeze(1), z.unsqueeze(0)) / self.temperature
```

```
sim_i_j = torch.diag(sim, self.batch_size * self.world_size)
sim_j_i = torch.diag(sim, -self.batch_size * self.world_size)
```

```
positive_samples = torch.cat((sim_i_j, sim_j_i), dim=0).reshape(N, 1)
```

```
negative_samples = sim[self.mask].reshape(N, -1)
```

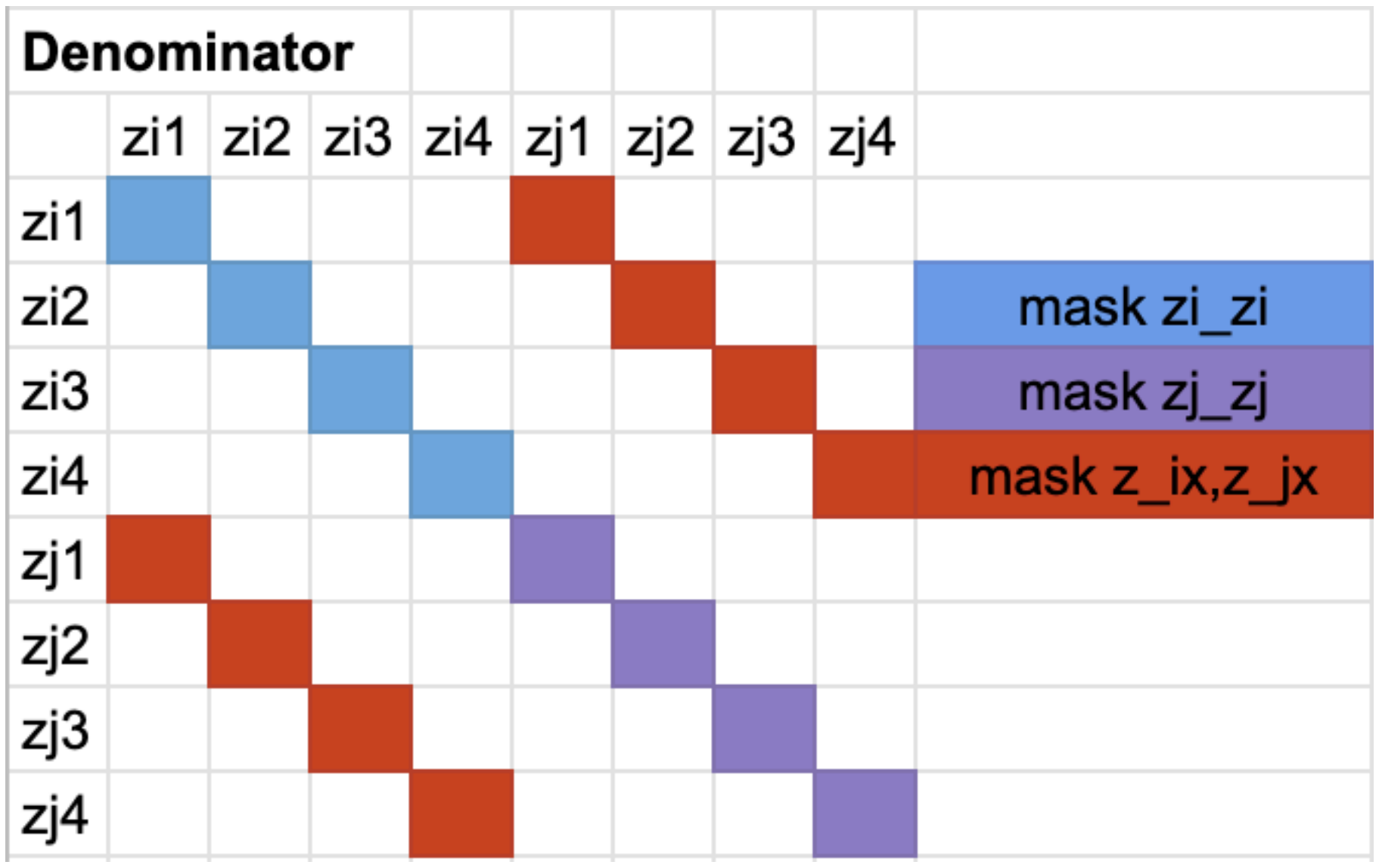

```
import torch
import torch.nn as nn


class NT_Xent(nn.Module):
    def __init__(self, batch_size, temperature, world_size):
        super(NT_Xent, self).__init__()
        self.batch_size = batch_size
        self.temperature = temperature
        self.world_size = world_size

        self.mask = self.mask_correlated_samples(batch_size, world_size)
        self.criterion = nn.CrossEntropyLoss(reduction="sum")
        self.similarity_f = nn.CosineSimilarity(dim=2)

    def mask_correlated_samples(self, batch_size, world_size):
        N = 2 * batch_size * world_size
        mask = torch.ones((N, N), dtype=bool)
        mask = mask.fill_diagonal_(0)
        for i in range(batch_size * world_size):
            mask[i, batch_size * world_size + i] = 0
            mask[batch_size * world_size + i, i] = 0
```

(continues on next page)

(continued from previous page)

```
        return mask

    def forward(self, z_i, z_j):
        """
        We do not sample negative examples explicitly.
        Instead, given a positive pair, similar to (Chen et al., 2017), we treat the␣
↪other 2(N − 1) augmented examples within a minibatch as negative examples.
        """
        N = 2 * self.batch_size * self.world_size

        z = torch.cat((z_i, z_j), dim=0)

        sim = self.similarity_f(z.unsqueeze(1), z.unsqueeze(0)) / self.temperature

        sim_i_j = torch.diag(sim, self.batch_size * self.world_size)
        sim_j_i = torch.diag(sim, -self.batch_size * self.world_size)

        # We have 2N samples, but with Distributed training every GPU gets N examples␣
↪too, resulting in: 2xNxN
        positive_samples = torch.cat((sim_i_j, sim_j_i), dim=0).reshape(N, 1)
        negative_samples = sim[self.mask].reshape(N, -1)

        labels = torch.zeros(N).to(positive_samples.device).long()
        logits = torch.cat((positive_samples, negative_samples), dim=1)
        loss = self.criterion(logits, labels)
        loss /= N
        return loss
```

## 15.5 Pre-training CLMR

**Note:** The following code will pre-train our SampleCNN encoder using our contrastive loss. This needs to run on a machine with a GPU to accelerate training, otherwise it will take a very long time.

```
args.temperature = 0.5  # the temperature scaling parameter in our NT-Xent los


encoder = SampleCNN(
    strides=[3, 3, 3, 3, 3, 3, 3, 3, 3],
    supervised=False,
    out_dim=0,
).to(device)


# get dimensions of last fully-connected layer
n_features = encoder.fc.in_features
print(f"Dimension of our h_i, h_j vectors: {n_features}")

model = SimCLR(encoder, projection_dim=64, n_features=n_features).to(device)

temperature = 0.5
optimizer = torch.optim.Adam(model.parameters(), lr=3e-4)
```

(continues on next page)

(continued from previous page)

```
criterion = NT_Xent(args.batch_size, args.temperature, world_size=1)

epochs = 100
losses = []

for e in range(epochs):

    for (x_i, x_j), y in train_loader:
        optimizer.zero_grad()

        x_i = x_i.to(device)
        x_j = x_j.to(device)

        # here, we extract the latent representations, and the projected vectors,
        # from the positive pairs:
        h_i, h_j, z_i, z_j = model(x_i, x_j)

        # here, we calculate the NT-Xent loss on the projected vectors:
        loss = criterion(z_i, z_j)

        # backpropagation:
        loss.backward()
        optimizer.step()

        print(f"Loss: {loss}")
        losses.append(loss.detach().item())

        break  # we are only running a single pass for demonstration purposes.

    print(f"Mean loss: {np.array(losses).mean()}")
    break  # we are only running a single epoch for demonstration purposes.
```

```
Dimension of our h_i, h_j vectors: 512
Loss: 2.5802388191223145
Mean loss: 2.5802388191223145
```

## 15.6 Linear Evaluation

Now, we would like to evaluate the versatility of our learned representations. We will train a linear classifier on the representations extracted from our pre-trained SampleCNN encoder.

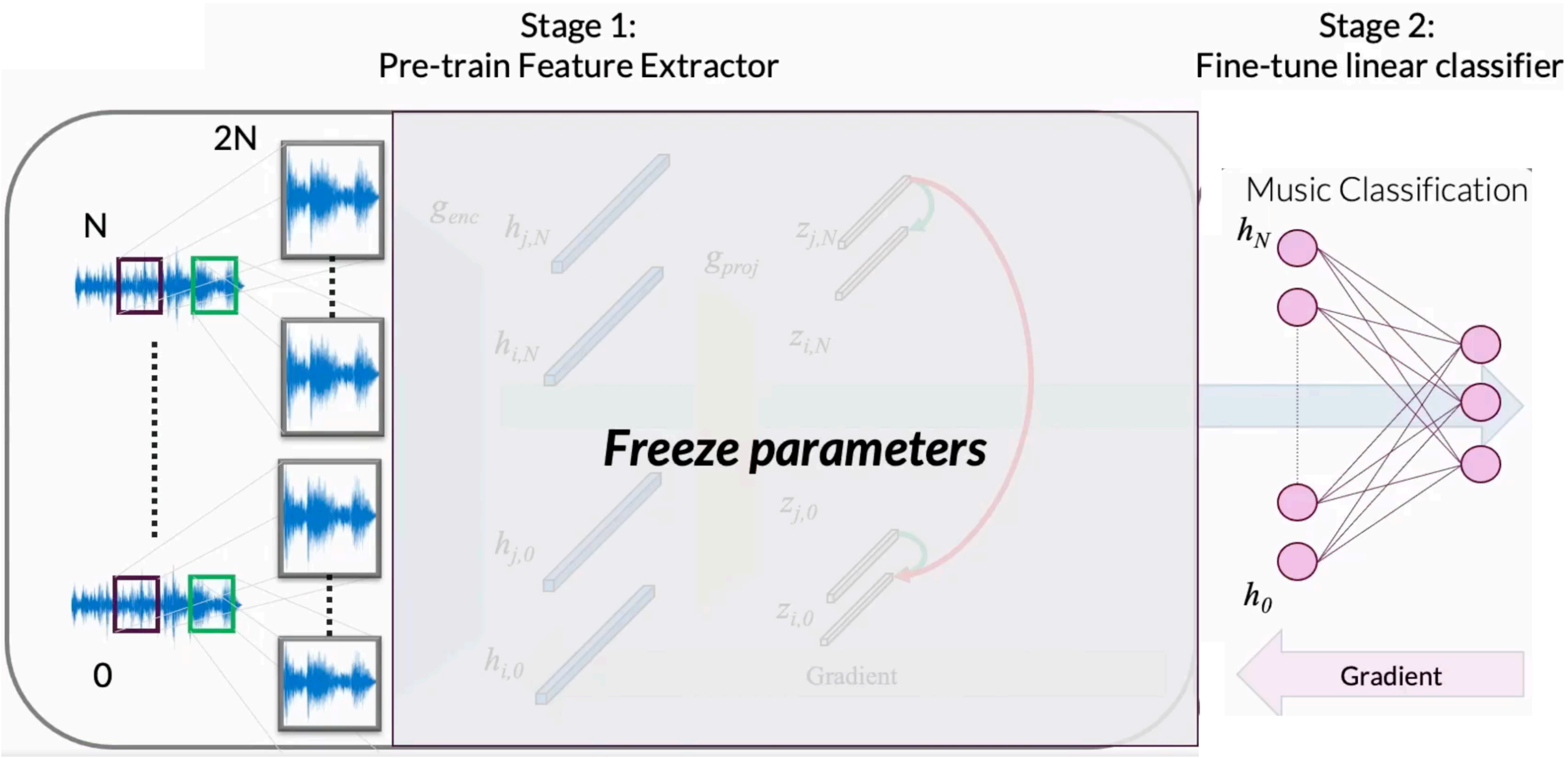

**Warning:** Note that we will not be using data augmentations during linear evaluation.

```
args.batch_size = 16


train_loader = get_dataloader(
    data_path="../../codes/split",
    split="train",
    is_augmentation=False,
    num_samples=args.audio_length,
    batch_size=args.batch_size,
)
valid_loader = get_dataloader(
    data_path="../../codes/split",
    split="valid",
    num_samples=args.audio_length,
    batch_size=args.batch_size,
)
test_loader = get_dataloader(
    data_path="../../codes/split",
    split="test",
    num_samples=args.audio_length,
    batch_size=args.batch_size,
)

iter_train_loader = iter(train_loader)
iter_test_loader = iter(test_loader)


for idx in range(3):
    (train_wav, _), train_genre = train_loader.dataset[idx]
    display(Audio(train_wav, rate=args.sample_rate))
```

```
<IPython.lib.display.Audio object>
```

```
<IPython.lib.display.Audio object>
```

```
<IPython.lib.display.Audio object>
```

Our linear classifier has a single hidden layer and a softmax output (which is already included in the `torch.nn.CrossEntropy` loss function, hence we omit it here).

```
class LinearModel(nn.Module):
    def __init__(self, hidden_dim, output_dim):
        super().__init__()
        self.hidden_dim = hidden_dim
        self.output_dim = output_dim
        self.model = nn.Linear(self.hidden_dim, self.output_dim)

    def forward(self, x):
        return self.model(x)
```

```
def train_linear_model(encoder, linear_model, epochs, learning_rate):

    # we now use a regular CrossEntropy loss to compare our predicted genre labels␣
↪with the ground truth labels.
    criterion = nn.CrossEntropyLoss()

    # the Adam optimizer is used here as our optimization algorithm
    optimizer = torch.optim.Adam(
        linear_model.parameters(),
        lr=learning_rate,
    )

    losses = []
    for e in range(epochs):
        epoch_losses = []
        for (x, _), y in tqdm(train_loader):

            optimizer.zero_grad()

            # we will not be backpropagating the gradients of the SampleCNN encoder:
            with torch.no_grad():
                h = encoder(x)

            p = linear_model(h)

            loss = criterion(p, y)

            loss.backward()
            optimizer.step()

            # print(f"Loss: {loss}")
            epoch_losses.append(loss.detach().item())

        mean_loss = np.array(epoch_losses).mean()
        losses.append(mean_loss)
```

(continues on next page)

(continued from previous page)

```
        print(f"Epoch: {e}\tMean loss: {mean_loss}")
    return losses
```

```
args.linear_learning_rate = 1e-4
args.linear_epochs = 15
print(f"We will train for {args.linear_epochs} epochs during linear evaluation")


# First, we freeze SampleCNN encoder weights
encoder.eval()
for param in encoder.parameters():
    param.requires_grad = False

print(
    f"Dimension of the last layer of our SampleCNN feature extractor network: {n_
↪features}"
)


# initialize our linear model, with dimensions:
# n_features x n_classes

linear_model = LinearModel(n_features, args.n_classes)
print(linear_model)

losses = train_linear_model(
    encoder,
    linear_model,
    epochs=args.linear_epochs,
    learning_rate=args.linear_learning_rate,
)
```

In the `evaluate` function, we perform a full pass of the test dataset and extract the predictions from our linear classifier, given the representations of our pre-trained encoder.

```
# Run evaluation


def evaluate(encoder, linear_model=None, test_loader=None):
    encoder.eval()
    if linear_model is not None:
        linear_model.eval()

    y_true = []
    y_pred = []

    with torch.no_grad():
        for (wav, _), genre_index in tqdm(test_loader):
            wav = wav.to(device)
            genre_index = genre_index.to(device)

            # reshape and aggregate chunk-level predictions
            b, c, t = wav.size()

            with torch.no_grad():
```

(continues on next page)

(continued from previous page)

```
                if linear_model is None:
                    logits = encoder(wav.squeeze(1))
                else:
                    h = encoder(wav.squeeze(1))
                    logits = linear_model(h)

            logits = logits.view(b, c, -1).mean(dim=1)
            _, pred = torch.max(logits.data, 1)

            # append labels and predictions
            y_true.extend(genre_index.tolist())
            y_pred.extend(pred.tolist())

    return y_true, y_pred
```

```
import seaborn as sns
from sklearn.metrics import accuracy_score, confusion_matrix

y_true, y_pred = evaluate(encoder, linear_model, test_loader)

accuracy = accuracy_score(y_true, y_pred)
cm = confusion_matrix(y_true, y_pred)
sns.heatmap(
    cm, annot=True, xticklabels=GTZAN_GENRES, yticklabels=GTZAN_GENRES, cmap="YlGnBu"
)
print("Accuracy: %.4f" % accuracy)
```

```
100
 ↪%|████████████████████████████████████████████████████████████████████████████████
 ↪19/19 [00:26<00:00,  1.39s/it]
```

```
Accuracy: 0.1069
```

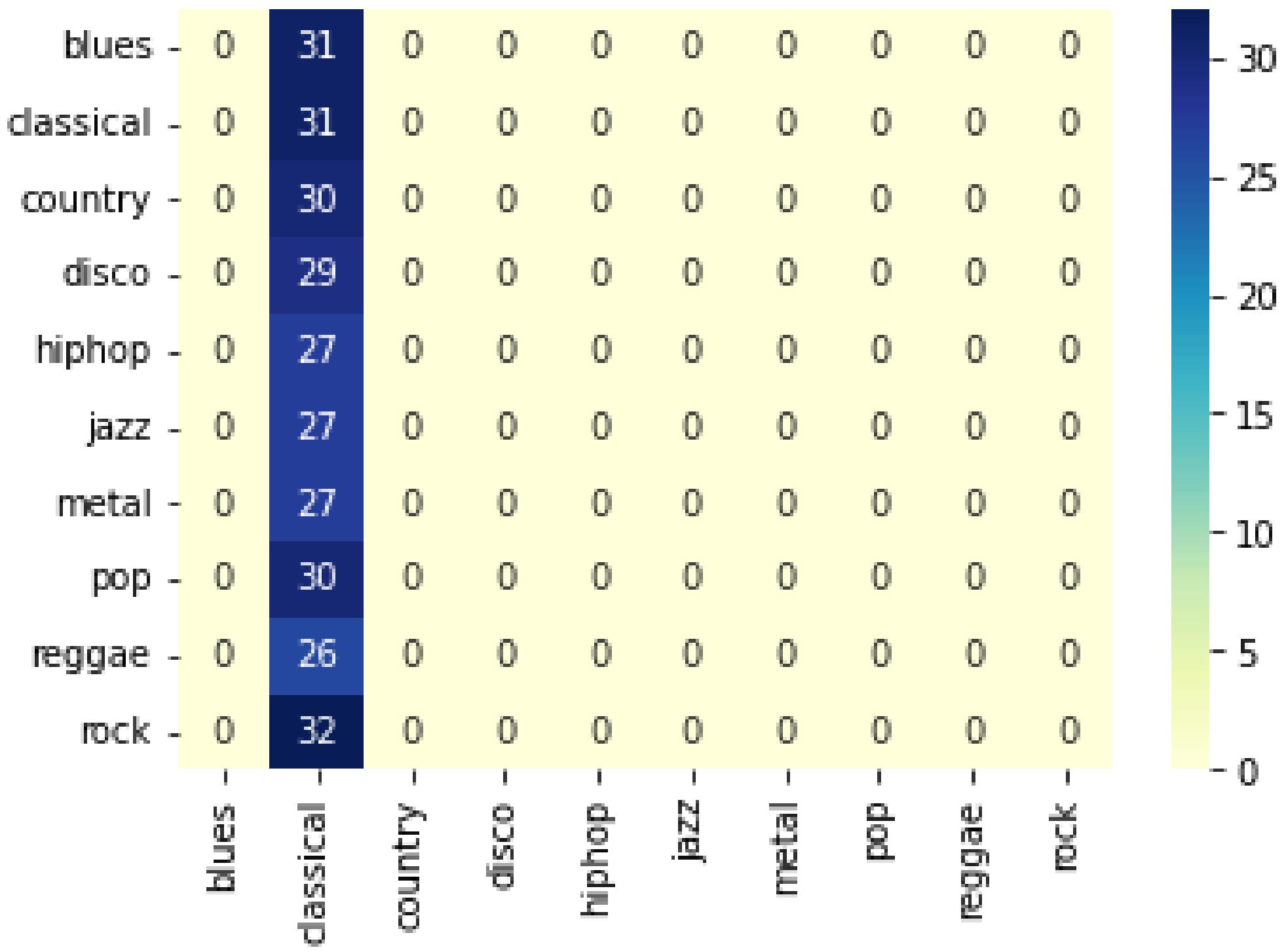

We can also load the weights of a fully pre-trained CLMR model to our `SampleCNN` encoder. The pre-trained representations will hopefully be more expressive for the linear classifier to solve the problem of music classification.

**Note:** Pre-training the encoder takes a while to complete, so let's load the pre-trained weights into our encoder now to speed this up:

```
from collections import OrderedDict

pre_trained_weights = torch.load("./clmr_pretrained.ckpt", map_location=device)

# this dictionary contains a few parameters we do not need  in this tutorial, so we␣
↪discard them here:
pre_trained_weights = OrderedDict(
    {
        k.replace("encoder.", ""): v
        for k, v in pre_trained_weights.items()
        if "encoder" in k
    }
)


# let's load the weights into our encoder:
encoder = SampleCNN(
    strides=[3, 3, 3, 3, 3, 3, 3, 3, 3],
    supervised=False,
    out_dim=0,
).to(device)
encoder.fc = Identity()
encoder.load_state_dict(pre_trained_weights)
```

(continues on next page)

(continued from previous page)

```
encoder.eval()

# we re-initialize our linear model here to discard the previously learned parameters.
linear_model = LinearModel(n_features, args.n_classes)
losses_with_clmr = train_linear_model(
    encoder,
    linear_model,
    epochs=args.linear_epochs,
    learning_rate=args.linear_learning_rate,
)
```

### 15.6.1 Get ROC-AUC and PR-AUC scores on test set

Let's now compute the accuracy of a *linear* classifier, trained on the representations from a pre-trained CLMR model.

```
y_true_pretrained, y_pred_pretrained = evaluate(encoder, linear_model, test_loader)

pretrained_accuracy = accuracy_score(y_true_pretrained, y_pred_pretrained)
cm = confusion_matrix(y_true_pretrained, y_pred_pretrained)
sns.heatmap(
    cm, annot=True, xticklabels=GTZAN_GENRES, yticklabels=GTZAN_GENRES, cmap="YlGnBu"
)
print("Accuracy: %.4f" % pretrained_accuracy)
```

```
100
 ↪%|████████████████████████████████████████████████████████████████████████████████████████████████████████████████
 ↪19/19 [00:25<00:00,  1.33s/it]
```

```
Accuracy: 0.5517
```

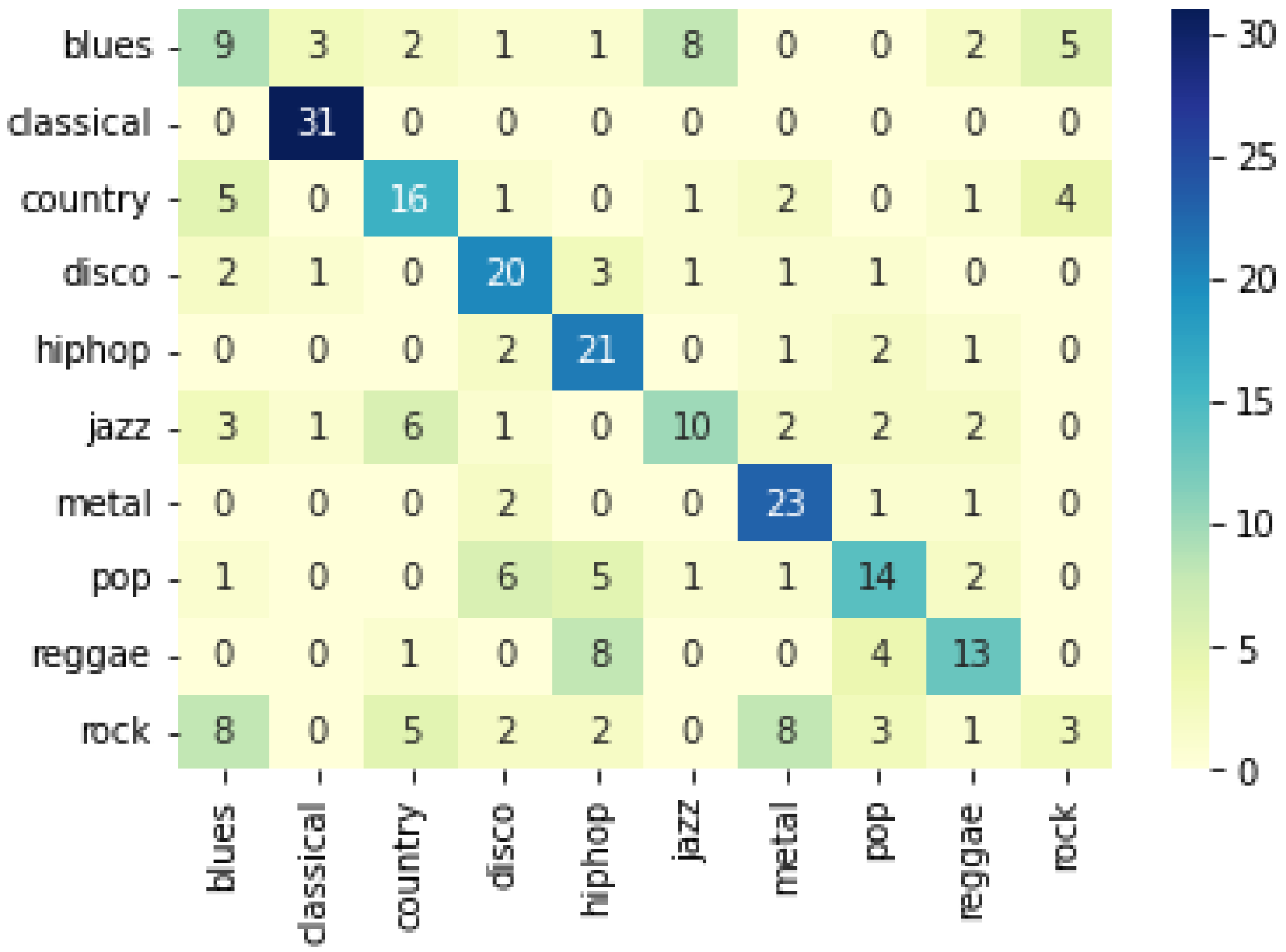

```
import matplotlib.pyplot as plt

plt.plot(losses, label="Losses")
plt.plot(losses_with_clmr, label="With pre-trained CLMR network")
plt.legend()
plt.show()
```

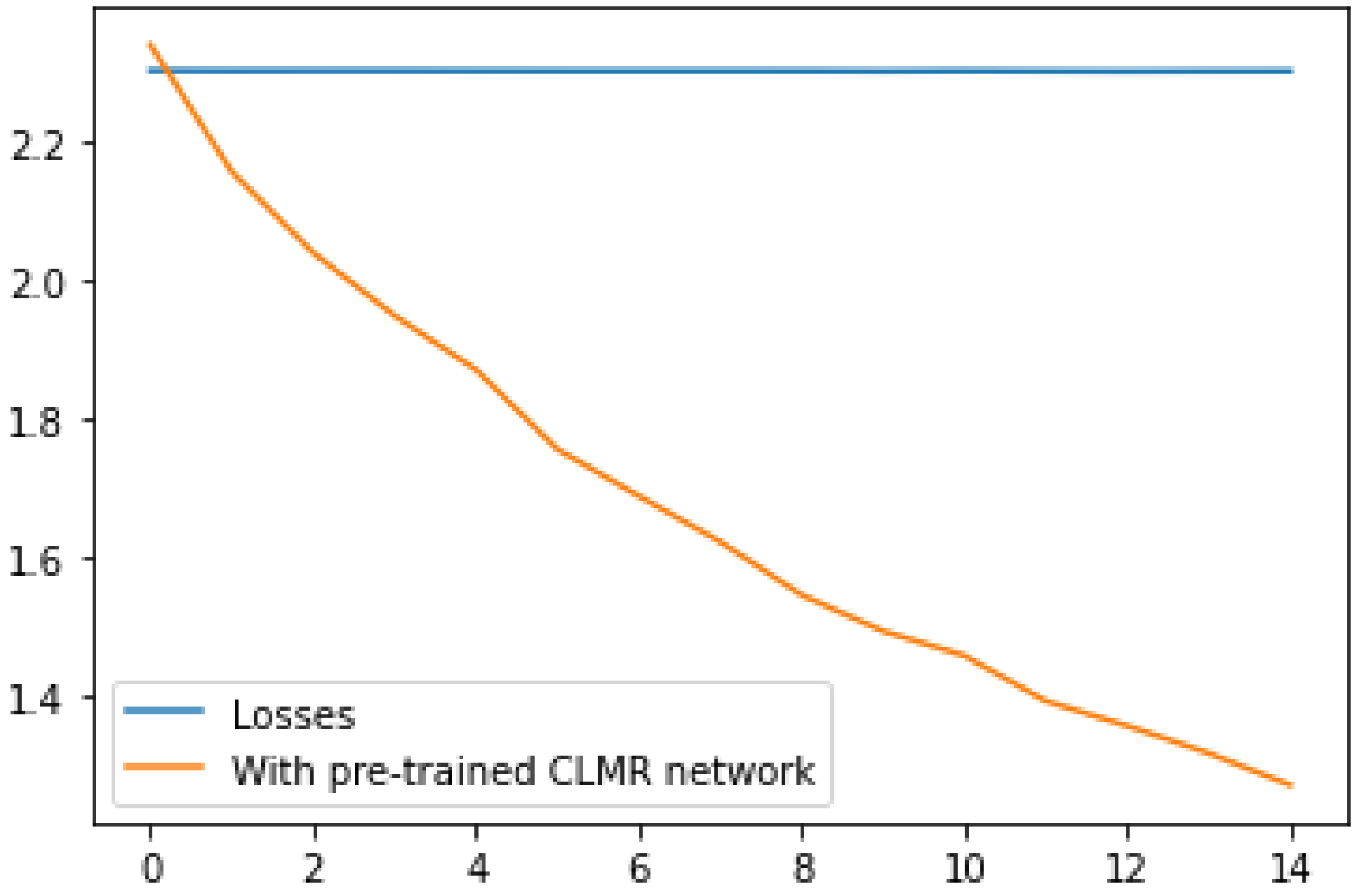

## 15.7 How does a supervised SampleCNN model compare?

```
# let's load the weights into our encoder:
supervised_samplecnn = SampleCNN(
    strides=[3, 3, 3, 3, 3, 3, 3, 3, 3],
    supervised=True,
    out_dim=args.n_classes,
).to(device)

optimizer = torch.optim.Adam(supervised_samplecnn.parameters(), lr=3e-4)
criterion = nn.CrossEntropyLoss()

epochs = 15
losses = []
for e in range(epochs):
    epoch_losses = []
    for (x, _), y in tqdm(train_loader):
        optimizer.zero_grad()
        logits = supervised_samplecnn(x)

        # here, we calculate the NT-Xent loss on the projected vectors:
        loss = criterion(logits, y)

        # backpropagation:
        loss.backward()
        optimizer.step()

        # print(f"Loss: {loss}")
        epoch_losses.append(loss.detach().item())

    mean_loss = np.array(epoch_losses).mean()
    losses.append(mean_loss)
    print(f"Epoch: {e}\tMean loss: {mean_loss}")
```

```
plt.plot(losses, label="Supervised losses")
plt.plot(losses_with_clmr, label="With pre-trained CLMR network")
plt.legend()
plt.show()
```

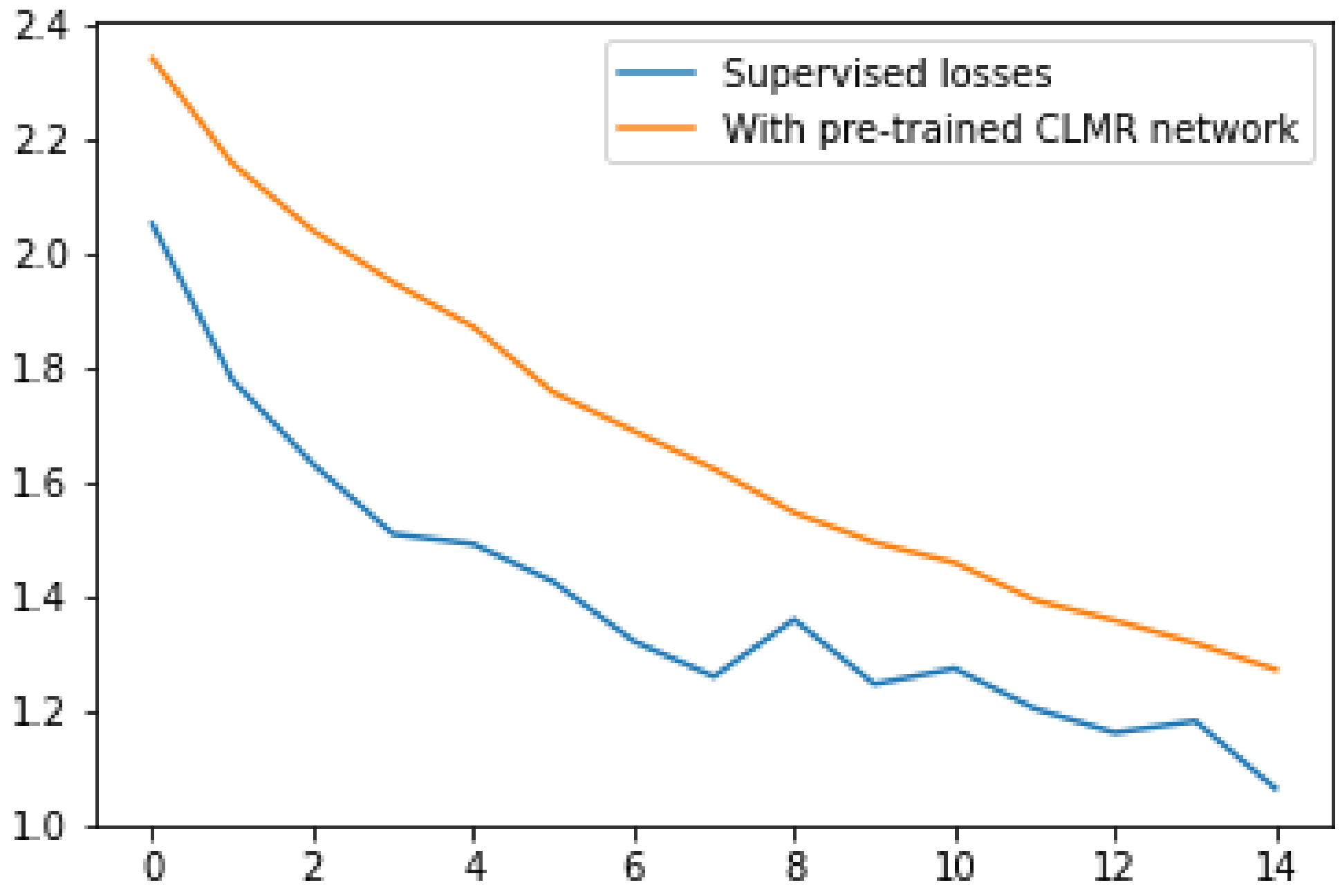

```
y_true_supervised, y_pred_supervised = evaluate(
    supervised_samplecnn, linear_model=None, test_loader=test_loader
)

supervised_accuracy = accuracy_score(y_true_supervised, y_pred_supervised)
cm = confusion_matrix(y_true_supervised, y_pred_supervised)
sns.heatmap(
    cm, annot=True, xticklabels=GTZAN_GENRES, yticklabels=GTZAN_GENRES, cmap="YlGnBu"
)
print("Accuracy: %.4f" % supervised_accuracy)
```

```
100
↪%|███████████████████████████████████████████████████████████████████████████
↪19/19 [00:26<00:00,  1.41s/it]
```

```
Accuracy: 0.4966
```

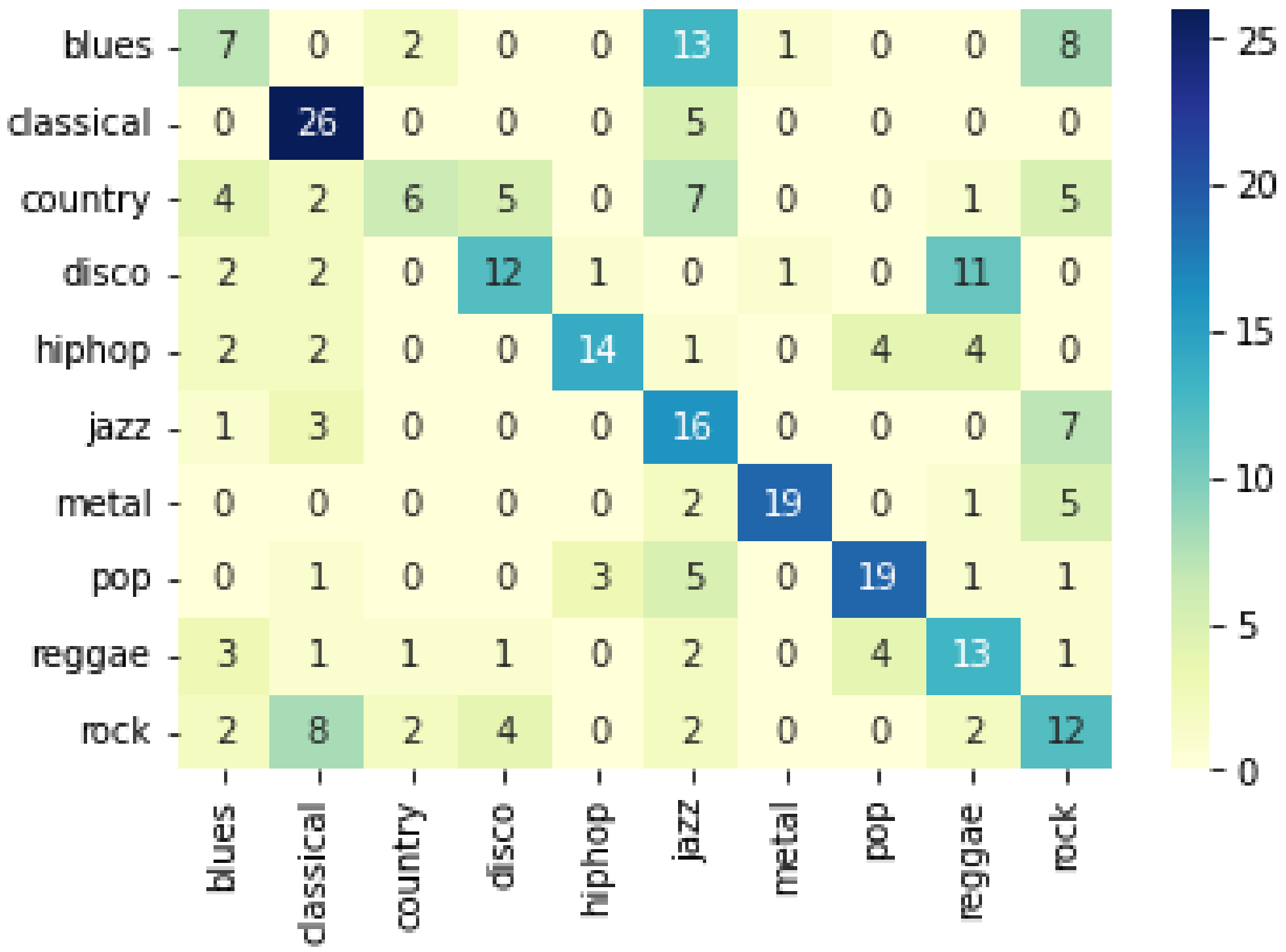

## 15.8 Conclusion

In conclusion, we observed how a self-supervised model, that pre-trains its network by way of leveraging the underlying structure of the data, can learn strong representations for the downstream task of music classification.

We have:

1. Pre-trained a SampleCNN encoder with CLMR.
2. Evaluated the representations with a linear classifier.
3. Loaded pre-trained weights from CLMR trained on the MagnaTagATune dataset.
4. Trained a linear classifier on these representations.

Our final accuracy when training the linear classifier for 15 epochs is **~55.2%** on the downstream task of music classification on the GTZAN dataset.

It is important compare against an equivalent network that is trained in a supervised manner. Therefore, we also trained a supervised SampleCNN model from scratch, which reached an accuracy of **~49.6%**.

| Model | Accuracy |
|---|---|
| Supervised SampleCNN | 49.6% |
| Self-supervised CLMR + SampleCNN | 55.2% |

**Note:** Note that in these experiments, the supervised model may have well overfitted on the GTZAN training data. This tutorial is by no means an exhaustive search for an optimal set of model and training parameters.

In conclusion, it is exciting to see that a linear classifier reaches a comparable performance, compared to a fully optimized encoder, using representations that were learned in a task-agnostic manner by way of self-supervised learning.

# Part V

# Towards Real-world Applications

CHAPTER

SIXTEEN

# MLOPS

Academia and industry have different goals and focuses for good reasons. But it's useful to learn what is happening on the other side.

---

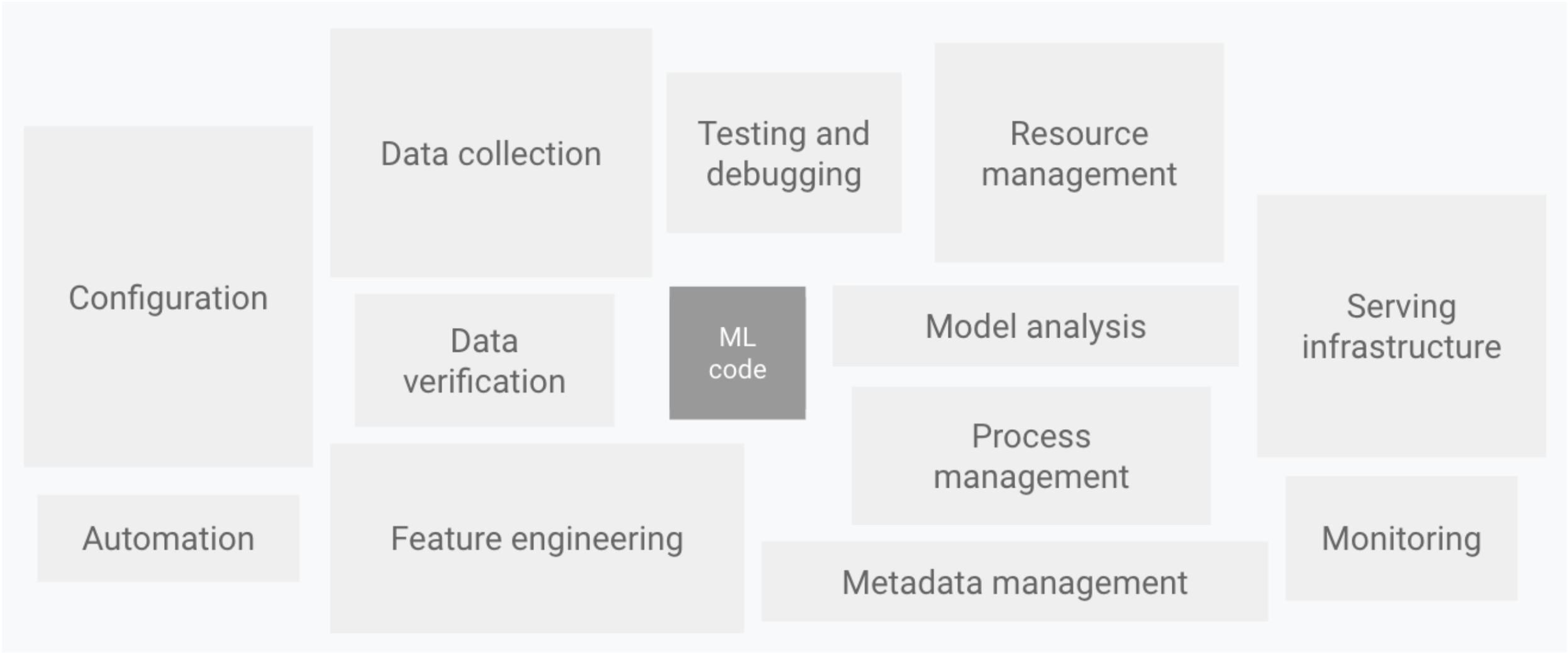

This famous image from Machine Learning: The High Interest Credit Card of Technical Debt, [SHG+14] shows various kinds of modules that are needed, on top of ML code, to build an ML system. There are so many of them that people even coined a new word, "MLOps". Among these, we will discuss the creation and management of datasets, evaluation of an ML model, and deployment in this chapter.

## 16.1 Dataset Creation

### 16.1.1 1. Item Sampling

To create a dataset, one first needs to collect the items. If the source of an item is limited in some sense, we might want to simply maximize the number of items. Otherwise, we need to sample the items from a bigger set, which might be the whole population (e.g., the whole catalog of streaming service).

In general, the goal of this sampling is to statistically represent the target set. But how? Here are some tips.

- Use metadata (year, artist, language, genre) to ensure diversity
- Make sure the items at extreme ends are included (e.g., the oldest and the newest songs)

OpenMIC-2018 [HDM18] set a high bar on this. The source of the dataset is FMA [DBVB16] which has full tracks, and the goal was to sample 10-second segments so that each instrument class is represented well enough. The authors first trained an instrument recognition model using an existing dataset. The model was used to approximate class occurrence. Based on this, the authors sampled the (assumed to be) positive 10-second segments of an instrument – from least to most likely instrument class. Finally, then applied a rule so that no two clips share a source track.

### 16.1.2 2. Annotation

Once collecting the items, you'll go through an annotation process.

- Defining a taxonomy can be an overwhelmingly difficult task! Ask experts, see if there's existing taxonomy you can use (e.g., WordNet).
- Educate the annotators so that everyone has the same understanding of the task and the labels.
- Be skeptical about the quality - Have multiple annotators and use the agreement to validate the labels.
- After labeling, if you're lack items with some labels, manually add the items. I.e., do some preemptive active learning

### 16.1.3 3. Postprocessing

All done? Yes – almost. But there are a few more things to do.

- Mistakes in the split leads to mistakes in training and evaluation!
  - Make sure the (label and/or any other) distributions of training/validation/testing sets are similar.
  - Allocate enough items to validation and testing sets.
    - For example, when the dataset is not that large – Which is better? 90:5:5 vs 70:20:10. Of course, it all depends. But I'd prefer the latter since i) a 22% decrease of the training set is probably not critical while a 400% increase in validation set means our model selection will be significantly more reliable.

## 16.2 Dataset Management

In the ideal world, you have nothing to do once a dataset is created. In the world we're living in, it might be just the beginning!

- Version your dataset. It can be like versioning software (1.0, 1.0.1, 1.1.0, ..) which has nice rules about, for example, semantic versioning. Or maybe simply the dates and some explanation.
  - Save the version of the dataset with your models.
- You may add new data samples. Why? □♀ Anyway, it happens.
  - Be aware of the distribution of the new subset and the result
  - Be consistent on the data pipeline (software and parameters used during audio processing)
- Keep your data samples up-to-date. Add recent samples!
- Keep your labels up-to-date.
  - E.g., genre labels (new genres may emerge), labels from language models

## 16.3 Evaluation: It is more than a single number

### 16.3.1 Choice of metric(s)

- In papers, conciseness is a virtue!
  - We have to compare models (to show my model beats yours!). We prefer a single number such as AUC-ROC, PR-AUC, or average F-1 score, one that summarizes the performance of tags / with various metrics (e.g., precision and recall).
- In industry, it might be a bit different.
  - You would want to evaluate the model in more detail. For example, the performance of each tag would matter.
  - Depending on use cases, people may focus on either precision or recall.

### 16.3.2 Optimize for your target metric

- High precision? or High recall?
  - Know your application!
- Thresholding or not?
  - Even with softmax, if the target is high precision, simply thresholding with value works.
  - Confidence estimation can be done in various way.

## 16.4 Deployment

### 16.4.1 Notes

- Ensure the reproducibility of:
  - Software/your code!
  - Model
  - Data processing pipeline
    * Decoding mp3, or if it's mp3 vs wav input, resampling algorithm, how to downmix, ..
- Is your model actually useful for the whole catalog you have? E.g.,:
  - If album cover images are used in the model, are they going to be available for all the music tracks?
  - If you used lyrics, would it work for all the languages you need to support?

### 16.4.2 Food for thought: Aggregating segment level predictions

We often avoid this issue when using public datasets (e.g., MSD) as they come with 30-second preview only. If you have access to full tracks, congratulations! It's an opportunity to improve the performance! This is where some ideas of multiple instance learning can help you.

Assuming the model is working properly, simply averaging the predictions/probabilities/logits would result in a better performance. This is because when misclassifications are rare, they are ignored as we choose the majority.

One would want to go further and let a model handle the aggregation. For example, a Transformer can be used on top of segment level predictions. This could solve some corner cases where averaging fails. Imagine an instrumental detector that is deployed for track-level classification but trained at a segment-level. Averaging (or majority voting) would result in incorrect classification if more than 50% of segments have no vocal; even if the track is not instrumental.

# CHAPTER SEVENTEEN

# UNDER-EXPLORED PROBLEMS IN ACADEMIA

Your choices as a researcher are affected by the circumstance such as

| \ | Has more freedom in | Has Bias towards |
|---|---|---|
| Academia | Choosing the topics | Publishable topics |
| Industry | Using resources (time, budget, work force) | Profitable topics |

As a result, we'll see each entity uses [A] to specialize in [B]. And that's great! But, for the same reason, some topics are getting little attention in academia.

## 17.1 What makes a topic difficult to work on in academia?

- When it feels like it's solved → You can't write a paper about it anymore!
- When it's hard to create a dataset → In this data-driven era, it's a deal-breaker.
- When the problem is too new / there's no dataset for it → No way for sure.
- When it's difficult to evaluate → Don't feed Reviewer 2 a reason for rejection!

## 17.2 Let's talk about research topics

Disclaimer - This section is meant to be subjective. Also, as the content is based on the diagnosis of the current research field, it will expire as time goes by.

### 17.2.1 Speech/music classification

**Although**

- It seems easy
- Many methods have achieved 100% accuracy in Gtzan speech/music dataset [Tza99],

**It is an interesting problem because**

- The model is needed anyway and there's no reliable public model since Gtzan speech/music dataset [Tza99] is pretty small
- The problem can be defined further such as:
  - Clip-level decision → short segment-level decision (say, 1 second)

- More than binary decision - {100% Music – many levels in between – 100% speech} + {something neither music or speech} (e.g., [MelendezCatalanMGomez19], [HWW+21])

### 17.2.2 Language classification

**Although**

- We were not doing it (nearly at all) because there was no public dataset

**It is an interesting problem because**

- It's one of the main components in music recommendation systems.
- It is popular in Industry - According to publication records, ByteDance [CW21] / Spotify [Rox19] / YouTube [CSR11] have done it.
- There is a public dataset now [SPD+20]

### 17.2.3 Mood recognition

**Although**

- It has lost popularity for these reasons:
  - Tagging tasks sort of overshadowed it
  - Hard to get large-scale data // while we have to write deep learning papers!
  - Hard to evaluate (fundamentally, completely subjective)
  - Maybe a lot about lyrics, which are also hard to get.

**It is an interesting problem because**

- Users still want to find some songs by mood.
  - Mood-based playlists/radio stations are popular!
  - Check out this repo[GCCE+21] for a comprehensive list of mood-related datasets

### 17.2.4 Year/decade/era

**Although**

- No one does it explicitly
- Metadata is supposed to cover this pretty well
- MSD includes it and it works pretty okay [BMEWL11]

**It is an interesting problem because**

- And yes, there is demand! Metadata is NOT always there or correct
- Relevant to user's musical preference

### 17.2.5 Audio codec quality (mp3, wav, etc)

**Although**

- Music services are supposed to always have high-quality audio

**It is an interesting problem because**

- There are many fake CD-quality/fake HD audio files
- Indie music/Directly publishing + sample-based music producing = Increase of audio quality issue

### 17.2.6 Hierarchical Classification

**Although**

- There are little datasets that have hierarchical taxonomies

**It is an interesting problem because**

- We can do a better job by learning the knowledge in the hierarchy
- The users of your model may want it! Even if they did not explicitly want a label hierarchy, it might make more sense to have one based on the labels in demand.

# Part VI

# Conclusion

CHAPTER

# EIGHTEEN

# CONCLUSION

Congratulations! You finished the book, executed every code we typed, and read every line we wrote!

In the first chapter, The Basics, we defined music classification and introduced its applications. We then looked into input representations with a special focus on biological plausibility. We also looked into music classification datasets with a special focus on the secrets of how to use some popular datasets correctly. In the evaluation section, we showed the concepts of important metrics such as precision and recall as well as code demo to compute them. After finishing this chapter, we hope you're ready to start working on your music classification model.

In the second chapter, Supervised Learning, we reviewed popular architectures - their definitions, pros, and cons. We also demonstrated data augmentation methods for music audio - the code, spectrograms, and audio signals you can play. At the end of the chapter, we showed a full example of data preparation, model training, and evaluation on Pytorch. After this chapter, you can implement a majority of music classification models that were introduced during the deep learning era.

In the third chapter, Semi-Supervised Learning, we covered transfer learning and semi-supervised learning – approaches that became popular, recently, due to annotation cost. Both are strategies one can consider when there is only a small number of labeled items. These approaches can be useful in many real-world situations where you only have, for example, less than a thousand labeled items.

In the fourth chapter, Self-Supervised Learning, an even more radical approach. The goal of self-supervised learning is to learn useful representations without any labels. To achieve the goal, researchers assume some structural/internal patterns purely within input and design loss functions to predict the patterns. We covered a wide range of self-supervised learning methods introduced in music, speech, and computer vision. The lesson of this chapter liberates you from the worry of getting annotations.

In the fifth chapter, Towards Real-world Applications, we introduce you to what people care about in industry. After finishing this chapter, you can understand the procedures and tasks researchers and engineers in industry spend time on.

We're delighted that you have studied music classification with us. Did you achieve your goal while reading it? Are your questions solved now? We hope we also achieved our goals - lowering the barrier of music classification to the newcomers, providing methods to cope with data issues, and narrowing the gap between academia and industry. Please feel free to reach out to us if you have any questions or feedback.

Best wishes,

Minz, Janne, and Keunwoo.

# Part VII

# Resources

CHAPTER

# NINETEEN